\documentclass[a4paper,11pt,preprintnumbers]{article}

\usepackage[T1]{fontenc}
\usepackage[utf8]{inputenc}

\usepackage{fontawesome5}

\usepackage{amsmath,amssymb,mathtools,bm,slashed}
\usepackage{graphicx}
\usepackage{booktabs,array}
\usepackage{microtype}
\usepackage[dvipsnames,x11names]{xcolor}
\usepackage{tcolorbox}
\usepackage[compat=1.0.0]{tikz-feynman}

\definecolor{forestgreen}{rgb}{0.13, 0.55, 0.13}

\usepackage{pifont}
\newcommand{\xmark}{\ding{55}}%
\usepackage{wasysym}

\usepackage{lmodern}

\usepackage{xcolor}
\usepackage{listings}

\definecolor{codebg}{RGB}{247,247,247}
\definecolor{codecomment}{RGB}{0,120,0}
\definecolor{codekeyword}{RGB}{0,0,180}
\definecolor{codestring}{RGB}{160,30,30}
\definecolor{codenumber}{RGB}{100,100,100}

\lstdefinestyle{baryonetpython}{
  language=Python,
  basicstyle=\ttfamily\small,
  backgroundcolor=\color{codebg},
  frame=single,
  rulecolor=\color{black},
  framerule=0.4pt,
  framesep=4pt,
  numbers=left,
  numberstyle=\tiny\color{codenumber},
  numbersep=8pt,
  keywordstyle=\color{codekeyword},
  commentstyle=\color{codecomment},
  stringstyle=\color{codestring},
  showstringspaces=false,
  breaklines=true,
  breakatwhitespace=true,
  columns=fullflexible,
  keepspaces=true,
  tabsize=2,
  xleftmargin=1.5em,
  xrightmargin=0.5em
}

\usepackage{float}
\usepackage[section]{placeins} 
\usepackage[labelfont=bf]{caption}

\def\linkcolor{blue!90!black}
\usepackage[
  colorlinks=true,
  urlcolor=\linkcolor,
  anchorcolor=\linkcolor,
  citecolor=\linkcolor,
  filecolor=\linkcolor,
  linkcolor=\linkcolor,
  menucolor=\linkcolor,
  linktocpage=true
]{hyperref}

\usepackage[nameinlink,capitalize]{cleveref}

\usepackage[numbers,sort&compress]{natbib}

\usepackage{tcolorbox}
\tcbuselibrary{most} 

\newcommand{\boxref}[1]{Box~\ref{#1}}

\newtcolorbox[auto counter, number within=section]{WideBox}[2][]{%
  enhanced, breakable,
  float, floatplacement=tp,
  colback=blue!2!white,
  colframe=blue!40!black,
  fonttitle=\bfseries, coltitle=white,
  boxrule=0.8pt,
  enlarge left by=0cm,
  enlarge right by=0cm,
  width=\dimexpr\textwidth+0cm\relax,
  title={Box~\thetcbcounter:~#2},
  #1
}

\newcommand{\avg}[1]{\left\langle #1 \right\rangle}

\newcommand{\dd}{\mathrm{d}}

\newcommand{\vev}[1]{\avg{#1}}

\allowdisplaybreaks

\usepackage[margin=2cm]{geometry}
\begin{document}

\begin{titlepage}

\begin{flushright}
IFT-UAM/CSIC-121 
\end{flushright}

\vspace{0.3cm}

\begin{center}
{\LARGE\bf
Electroweak Baryogenesis with \texttt{BARYONET}: \\a self-contained review of the WKB approach\\[2pt]
{\small \emph{Do Not Get Lost: A Survival Guide}}
}

\vspace{1.0cm}

{\large Giulio Barni\footnote{\href{mailto:giulio.barni@ift.csic.es}{giulio.barni@ift.csic.es}}}

\vspace{0.4cm}

\textit{Instituto de F\'isica Te\'orica IFT-UAM/CSIC, Cantoblanco, E-28049, Madrid, Spain}\\
October 28, 2025

\vspace{0.8cm}

\begin{abstract}
\noindent
We present a comprehensive, self-contained pedagogical computation of the baryon asymmetry of the Universe within electroweak baryogenesis (EWBG), from the derivation of the semiclassical, CP-dependent force to the formulation and solution of the transport equations obtained from the Boltzmann equations—all implemented in the open-source code \texttt{BARYONET}\footnote{\emph{electroweak BARYOgenesis with Numerical Evaluation of the Transport equations}.}. Our analysis follows the semiclassical WKB approach, where spatially varying complex masses across expanding bubble walls feel CP-violating forces that bias plasma transport. Starting from the stationary Boltzmann equation in the wall frame and projecting onto a hierarchy of velocity moments, we derive a compact, fluid-like system of coupled differential equations for chemical potentials and velocity perturbations. After obtaining the solutions, one can define the left-handed baryon chemical potential, which acts as the source term for the weak sphalerons. These processes generate the baryon asymmetry in front of the wall, which is subsequently frozen once it passes through it.

We validate the framework against established formalisms and provide benchmarks in representative scenarios, including singlet extensions of the Standard Model, two-Higgs-doublet models, and Higgs–$\phi^6$ constructions. The resulting \texttt{BARYONET} implementation delivers an automated, reproducible pipeline for WKB-based baryogenesis studies, connecting formal derivations with phenomenological applications. 

In parallel, we revisit standard EWBG ingredients—\emph{diffusion constants}, \emph{Yukawa/helicity-flip rates}, and \emph{strong/weak sphaleron rates}—to clarify conventions, update numerical inputs, and present a pedagogical derivation, ensuring transparent reproducibility.

\end{abstract}

\end{center}

\vspace{-1.8cm}

\begin{center}
    \includegraphics[scale=1.2]{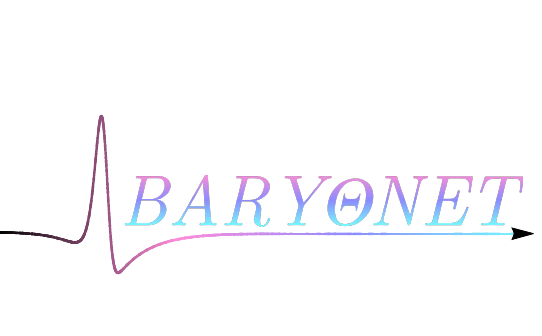}
\end{center}

\end{titlepage}

\tableofcontents

\section{Introduction}

Electroweak baryogenesis (EWBG) is an appealing, testable explanation for the origin of the cosmic matter–antimatter asymmetry at the weak scale. In its modern form it realizes Sakharov’s three conditions—baryon number violation, $C$ and $CP$ violation, and departure from equilibrium—within a hot, expanding plasma undergoing the electroweak phase transition (EWPT): anomalous sphaleron transitions violate $B$, $CP$–violating interactions bias particle–antiparticle transport across the bubble wall, and the out–of–equilibrium environment provided by a first–order phase transition (FOPT) allows the bias to be converted into net baryon number and subsequently preserved inside the broken phase~\cite{Sakharov:1967dj,KUZMIN198536,Shaposhnikov:1986jp,SHAPOSHNIKOV1987757,COHEN1990561,Cohen:1993nk}.

The Standard Model (SM) by itself falls short on both dynamical fronts. Lattice and perturbative studies show that for $m_h\simeq125$~GeV the EW symmetry breaking proceeds through a smooth crossover rather than first order~\cite{Kajantie:1995kf,Kajantie:1996mn,Kajantie:1996qd,Gurtler:1997hr,Csikor:1998eu,Laine:1998qk,Aoki:1999fi,DOnofrio:2015gop}, precluding efficient departure from equilibrium; moreover, Cabibbo–Kobayashi–Maskawa $CP$ violation yields sources many orders of magnitude too small to explain the observed baryon asymmetry~\cite{Gavela:1994dt,Huet:1994jb,Kapusta:2007xjq}. These facts motivate extensions of the SM that (i) strengthen the EWPT to a FOPT and (ii) supply additional $CP$–violating phases compatible with laboratory bounds. Representative possibilities include singlet extensions of the Higgs sector, two–Higgs–doublet models (2HDMs), supersymmetric and composite frameworks, and effective descriptions with higher–dimensional operators or dynamical Yukawa couplings across the phase transition; each provides concrete wall profiles (masses and phases) and interaction rates that feed the transport network~\cite{Carena:1996wj,Carena:1997gx,Cline:1998hy,Fromme:2006cm,Profumo:2007wc,Chung:2012vg,Grojean:2004xa,deVries:2017ncy,Vaskonen:2016yiu,Su:2020pjw,Bian:2019kmg,Bruggisser:2017lhc,Bruggisser:2018mus,Bruggisser:2018mrt,Bruggisser:2022ofg,Bruggisser:2022rdm}.

Historically, the modern electroweak baryogenesis picture emerged in two conceptually distinct steps. First, it was realised that the Standard Model already contains a concrete source of baryon number violation at high temperature: anomalous electroweak transitions become unsuppressed in the symmetric phase, so that baryon number can be efficiently violated around the electroweak epoch~\cite{KUZMIN198536,Shaposhnikov:1986jp,SHAPOSHNIKOV1987757}. This immediately made the electroweak scale an attractive arena for baryogenesis. If, in addition, the electroweak phase transition is first order, the expanding bubble walls provide the required departure from equilibrium, while the symmetric phase in front of the wall remains the region where weak sphalerons are active~\cite{COHEN1990561,Cohen:1993nk,Trodden:1998ym,Riotto:1999yt,Morrissey:2012db}.

A second conceptual step was the shift from attempting to bias topology directly to phrasing the problem in terms of an effective charge potential. In spontaneous baryogenesis, a slowly varying CP--odd background with derivative coupling to a current acts as a thermodynamic bias, shifting particle and antiparticle energies in a way analogous to a chemical potential~\cite{CohenKaplan:1987,CohenKaplan:1988}. Cohen, Kaplan and Nelson showed how this logic can be implemented at the electroweak phase transition: a spacetime--dependent Higgs-sector phase can be removed from the Yukawa interactions by an anomaly--free field redefinition, after which it reappears as a derivative coupling to a fermionic current. The resulting charge potential biases the plasma, while weak sphalerons in the symmetric phase convert that bias into baryon number~\cite{COHEN1990561,CohenKaplanNelson:1991EWPT,CohenKaplanNelson:1991SB,Nelson:1991ab,Cohen:1994ss}. In this way, the scalar background provides the CP--odd bias, the electroweak anomaly provides baryon number violation, and the moving wall provides the out--of--equilibrium stage. Modern transport treatments can be viewed as increasingly quantitative implementations of this same organising principle~\cite{Cohen:1993nk,Cline:2000nw,Fromme:2006wx,Cline:2020jre,Kainulainen:2024qpm,vandeVis:2025efm}.

On the theory side, two complementary approaches underlie contemporary EWBG computations. The first is the real–time Schwinger–Keldysh (Kadanoff–Baym) formalism, which yields quantum kinetic equations that are gradient–expanded to obtain semiclassical dispersion relations, force terms, and collision integrals for out–of–equilibrium distribution functions~\cite{kadanoff1962quantum,Schwinger1961,Keldysh:1964ud,Kainulainen:2001cn,Kainulainen:2002th,Prokopec:2003pj,Prokopec:2004ic,Konstandin:2005cd}. The second is the Wentzel–Kramers–Brillouin (WKB)/semiclassical method, particularly suitable for thick walls, where $CP$–odd corrections to single–particle dispersion relations induce opposite forces on particles and antiparticles traversing the wall; these forces seed left–handed charge in front of the wall, which weak sphalerons partially convert into baryon number~\cite{Joyce:1994zn,Joyce:1994zt,Huet:1995sh,Funakubo:1996dw,Davoudiasl:1997jh,Cline:2000nw}. Careful comparisons have clarified the correspondence of observable sources in the two pictures, the role of elastic ($W$) scatterings, and the fluid closure used to reduce the kinetic system to a tractable diffusion network~\cite{Lee:2004we,Cirigliano:2009yt,Cirigliano:2011di,Herranen:2008hi,Herranen:2008hu,Herranen:2008di,Herranen:2010mh,Balazs:2004ae}. In particular, the semiclassical force can be derived consistently within both frameworks: the original CTP–based derivations are due to Kainulainen, Prokopec, Schmidt and collaborators~\cite{Kainulainen:2001cn,Kainulainen:2002th}, with subsequent reviews and reformulations in~\cite{Fromme:2006wx,Cirigliano:2009yt}. Later works~\cite{Prokopec:2003pj,Prokopec:2004ic,Konstandin:2005cd} introduced a new collisional semiclassical source, which was shown to be spurious in~\cite{Kainulainen:2021oqs}.

Across models, the phenomenological pipeline has crystallised into a sequence of decidedly nontrivial steps. First, one determines the finite–temperature effective potential and constructs—analytically or parametrically—the bubble–wall profiles for the relevant fields. On this foundation, one derives the $CP$–even and $CP$–odd sources together with the linearised collision integrals in a boosted, near–equilibrium plasma. The next stage—often the numerically dominant challenge, amounting to a stiff, multiscale boundary–value problem—is to solve the coupled fluid/transport system. Finally, the resulting left–handed baryon chemical potential must be folded with the weak–sphaleron profile to predict the frozen–in asymmetry~$\eta_B$~\cite{Joyce:1994zn,Joyce:1994zt,Joyce:1994fu,Huet:1995sh,Cline:2000nw,Bodeker:2004ws,Fromme:2006wx,Fromme:2006cm,Espinosa:2011eu,Konstandin:2013caa,Cline:2020jre,Cline:2021dkf,Kainulainen:2024qpm,Li:2025kyo,Branchina:2025adj}.

In this work, as already done in the literature, we will focus on the computation of the $CP$--odd sector of the transport equations,
assuming a stationary bubble--wall background with a given steady--state velocity $v_w$.
This procedure, standard in semiclassical treatments of electroweak baryogenesis,
treats $v_w$ as an external input parameter. The determination of the steady--state bubble--wall velocity has been the subject of extensive investigation
(see, e.g.,~\cite{Moore:1995si,John:2000zq,Moore_2000,Bodeker:2009qy,Konstandin:2014zta,Huber:2013kj,Kozaczuk:2015owa,
Balaji:2020yrx,BarrosoMancha:2020fay,Wang:2020zlf,Laurent:2020gpg,Vanvlasselaer:2020niz,Leitao:2014pda,
Gouttenoire:2021kjv,Dorsch:2021nje,DeCurtis:2022hlx,Azatov:2022tii,Laurent:2022jrs,DeCurtis:2023hil,
DeCurtis:2024hvh,vandeVis:2025plm,Ekstedt:2024fyq,Branchina:2025adj}),
since, as we will show, the predicted baryon asymmetry can vary drastically with~$v_w$. Recent work explored the impact of out--of--equilibrium corrections
to the wall dynamics~\cite{Branchina:2025adj},
demonstrating that such effects can modify the predicted wall velocity by up to $\mathcal{O}(60\%)$ relative to the near--equilibrium treatment.
Consequently, even moderate corrections to the steady--state wall propagation
can induce significant variations in the generated baryon asymmetry,
highlighting the importance of a consistent treatment of wall dynamics
in future electroweak baryogenesis studies.

\medskip\noindent\textit{What this review provides.}
Despite many excellent reviews of EWBG over the past decades~\cite{Trodden:1998ym,Riotto:1998bt,Riotto:1999yt, Dine:2003ax,Cline:2006ts, Morrissey:2012db,10.1088/978-1-6817-4457-5,Garbrecht:2018mrp,Bodeker:2020ghk}, including very recent overviews~\cite{vandeVis:2025efm} (to which we refer for a complete list of references), a practical gap persists: a \emph{public, end-to-end, reproducible} implementation that unifies transport formalisms and conventions across the literature in the semiclassical approach. This work closes that gap by offering a self-contained account of the semiclassical transport approach and, crucially, by packaging it into a single framework. We assemble the standard ingredients—WKB kinematics, the Liouville operator in the wall frame, CP projection, a controlled moment expansion, and a model-independent closure of collision terms—into a compact \emph{fluid network} written in terms of \emph{dimensionless thermal kernels}, cleanly factorised from model inputs. We also provide an operational, precise definition of the left-baryon chemical potential \(\mu_{B_L}\) that sources weak sphalerons, together with the exact integrating-factor expression used to compute \(\eta_B\). 

Alongside the narrative review, we release \texttt{BARYONET}, an open and modular Python code that delivers a reproducible pipeline \emph{from model to \(\eta_B\)}: it ingests wall profiles (masses, phases, and their derivatives), assembles the transport system in multiple established frameworks, solves the linear boundary-value problem, and outputs \(\eta_B\) with figure scripts and tests. In contrast to recent reviews that summarise formal developments but do not provide a unified, maintained reference implementation or cross-validated pipeline, our package offers a single workflow with benchmark overlays and documented choices (diffusion sets, thermal masses, sphaleron rates), enabling direct comparison across frameworks. The code and examples are available at 
\href{https://github.com/GiulioBarni/BARYONETv1}{\faGithub\ BARYONET}.

\medskip\noindent\textit{Scope and organisation.} 
Section~\ref{sec:from-ewbg-to-baryonet} frames the central problem of the paper: how the observed baryon asymmetry can be reduced, within the semiclassical EWBG setup, to a concrete transport calculation across an expanding electroweak bubble wall.
Section~\ref{sec:WKB} develops the semiclassical (WKB) framework: we set up the background, derive the dispersion relations and group velocities, and obtain the $CP$--odd semiclassical force. Section~\ref{sec:boltzmann} builds the stationary Boltzmann equation in the wall frame, performs the $CP$ decomposition and the moment expansion, and introduces the universal thermal kernels together with a model--independent closure of collision terms. Section~\ref{sec:pheno} then specifies the minimal matter content commonly used in EWBG, writes the full fluid network explicitly, and defines the left--baryon chemical potential that sources weak sphalerons. Section~\ref{sec:BAU} derives the weak--sphaleron evolution equation for $n_B(z)$ and the resulting baryon asymmetry $\eta_B$. Section~\ref{sec:analytic_transport} extracts analytic information from the diffusion network in controlled limits, clarifying how the source profile is mapped into the chemical potentials and, in particular, into $\mu_{B_L}(z)$. Section~\ref{sec:Code} outlines the symbolic--to--numeric pipeline implemented in \texttt{BARYONET} and summarises the numerical inputs used for validation, providing updated values recommended by this work. Section~\ref{sec:Results} presents results and comparisons across representative scenarios (a model--independent benchmark, real--singlet extensions, a Higgs--$\phi^6$ setup, and the 2HDM), highlighting agreements and residuals with previous studies. Finally, Section~\ref{sec:higher-moments} examines the fluid network with an arbitrary number of perturbation modes and highlights the convergence of the moment expansion; we conclude in Section~\ref{sec:Conclusions} with an outlook on theoretical refinements and planned updates to the code.

\begin{figure}
    \centering
    \includegraphics[width=.85\linewidth]{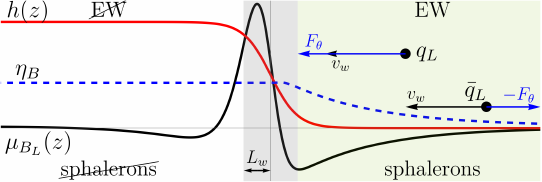}
    \caption{Schematic illustration of EWBG transport across a planar bubble wall. 
The red curve is the Higgs background \(h(z)\), interpolating from the broken phase (left, \(\vev h \neq 0\)) to the symmetric EW phase (right, $\vev h =0$). 
The shaded strip marks the wall of width \(\pm L_w\); the particles in the wall frame are moving with velocity \(v_w\) to the left. 
CP–violating semiclassical forces \(F_\theta\) (blue arrows) act with opposite sign on particles and antiparticles, biasing left–handed quarks \(q_L\) and antiquarks \(\bar q_L\). 
This generates a left–baryon chemical potential \(\mu_{B_L}(z)\) (black), peaked near the wall, which diffuses into the symmetric phase where weak sphalerons are active (grey + green region) and is converted into baryon number. 
The resulting baryon density \(\eta_B\) (blue dashed) accumulates ahead of the wall and is preserved once the wall passes and sphalerons are quenched in the broken phase. Curves are illustrative and not to scale.}
    \label{fig:sketch}
\end{figure}

\medskip\noindent\textit{Notation for frameworks.}
Throughout the review, we use the shorthand
\textbf{FH04}~\cite{Bodeker:2004ws}, \textbf{FH06}~\cite{Fromme:2006wx},
\textbf{CK}~\cite{Cline:2020jre}, and \textbf{KV24}~\cite{Kainulainen:2024qpm}
to denote \emph{our implementation of the fluid networks as defined in the cited papers}.
When we say “FH04/FH06/CK/KV24” below, we refer to those concrete implementations (species content,
moment structure, and collision closure) as summarised in Table~\ref{tab:schemes-notation}.

\begin{table}[t]
  \centering
  \setlength{\tabcolsep}{6pt}
  \renewcommand{\arraystretch}{1.18}
  \begin{tabular}{l p{0.2\linewidth} p{0.2\linewidth} p{0.2\linewidth} c}
    \hline
    \textbf{Scheme} &
    \textbf{Tracked species} &
    \textbf{Velocity range} &
     \textbf{Closure} &
    \(\boldsymbol{N_{\rm PERT,\max}}\) \\
    \hline
    FH04~\cite{Bodeker:2004ws} &
    \(q_3\), \(t^c\) &
    small$-v_w$ & constant $R=-v_w$&
    \(2\) \\
    FH06~\cite{Fromme:2006wx} &
    \(t,\ b,\ t^c,\ h\)~~~~~~~~ {\color{white}{a}} $+W$ scattering &
    small$-v_w$ & constant $R=-v_w$&
    \(2\) \\
    CK~\cite{Cline:2020jre} &
     \(t_L,\ b_L,\ t_R,\ h\)~~~~~~~~ {\color{white}{a}} $+W$ scattering &
    large$-v_w$ & constant $R=-v_w$&
    \(2\)\\
    KV24~\cite{Kainulainen:2024qpm} &
     \(t_L,\ b_L,\ t_R,\ h\)~~~~~~~~ {\color{white}{a}} $+W\ $scattering {\color{white}{a}} ~ +improved treatment of rates &
     large$-v_w$ &
     constant $R=-v_w$ or variance&
    \(\leq 50\) \\
    \hline
  \end{tabular}
  \caption{Summary of the frameworks used as \emph{notation} in this review. 
  The first column lists the frameworks and their original paper. The second column indicates which species is tracked in that implementation.
  The third column states the range of velocity that is suited for that implementation.
  The last column gives the maximum number of perturbations (chemical potential + velocity perturbation) retained per species.}
  \label{tab:schemes-notation}
\end{table}

\medskip\noindent\textit{Additional conventions.}
Unless stated otherwise, we work in the wall frame. Spatially, the \emph{broken} phase ($\vev h \neq 0$) is on the left and the \emph{symmetric} EW phase on the right; i.e.\ $z<0$ is broken and $z>0$ is symmetric. In the plasma frame, the bubble wall moves to the right (positive $z$) with velocity $v_w$; equivalently, in the wall frame, the plasma and particle flux drift to the left (negative $z$). The symbol $q_3$ denotes the third–generation left–handed quark doublet, $t_R$ the right–handed top, and $h$ the Higgs. By “perturbations’’ we mean velocity moments in the fluid description: chemical potentials $\mu_i$ are always included, while $u_i^{(\ell+1)}$ with $\ell=0,1,\dots,N_{\rm PERT, \max}-1$ are the retained velocity modes ($N_{\rm PERT, \max}=2$ corresponds to $\mu_i$ plus a single velocity mode $u_i^{(1)}$). Differences due to diffusion sets, thermal mass choices, and sphaleron profiling are called out explicitly when relevant.

\noindent\emph{Blue boxes} offer concise, pedagogical clarifications of technical points encountered in this review; expert readers may safely skip them without loss of continuity.

\section{From EWBG to \texttt{BARYONET}}
\label{sec:from-ewbg-to-baryonet}

The starting point is the observed absence of antimatter on cosmological scales. In terms of the
entropy-normalised net baryon number, the measured asymmetry is
\begin{equation}
  \eta_B^{\rm obs}
  \equiv
  \frac{n_B-n_{\bar B}}{s}
  \simeq
  8.7\times 10^{-11}.
\end{equation}
This value is inferred with high
precision from the CMB baryon density, for instance by Planck, and is consistent with the baryon
density required by light-element abundances in Big-Bang nucleosynthesis (BBN)
\cite{Planck:2018vyg,Fields:2019pfx}. Thus the asymmetry must already have been present by the
time of BBN. It cannot be interpreted as a statistical fluctuation of an initially symmetric universe:
a random fluctuation
would give a relative asymmetry of many orders of magnitude
below the observed value. Moreover, in a cosmology with an inflationary stage, any pre-existing
conserved number density generated before inflation would be diluted by the accelerated expansion
unless it is regenerated afterwards \cite{Guth:1980zm}. The baryon asymmetry therefore calls for a
dynamical mechanism operating after the last inflationary dilution and before BBN.

Sakharov identified the three necessary ingredients for such a dynamical generation of baryon number:
baryon-number violation, violation of \(C\) and \(CP\), and departure from thermal equilibrium
\cite{Sakharov:1967dj}. Baryon-number violation is needed to change the net baryon charge from its
initial value. \(C\) and \(CP\) violation are needed so that baryon-producing and antibaryon-producing
processes do not occur at identical rates. Finally, departure from equilibrium is required because, in
thermal equilibrium and under the usual assumptions of CPT invariance, any generated asymmetry
would be erased by detailed balance. In addition, once an asymmetry has been produced, the
baryon-violating processes responsible for its generation must either freeze out or become sufficiently
suppressed, otherwise they wash out the final result.

The Standard Model contains part of this structure, but not enough for successful baryogenesis.
Electroweak anomalous transitions violate \(B+L\) at high temperature and are active in the symmetric
phase \cite{KUZMIN198536,Shaposhnikov:1986jp,SHAPOSHNIKOV1987757}. The CKM phase provides
\(CP\) violation, but the corresponding bias is far too small for the observed baryon asymmetry
\cite{Jarlskog:1985ht,Gavela:1993ts,Gavela:1994dt,Huet:1994jb}. In addition, for the measured Higgs
mass the electroweak transition is a smooth crossover rather than first order, so the Standard Model
does not provide the required out-of-equilibrium environment at the electroweak scale
\cite{Kajantie:1995kf,Kajantie:1996mn,Kajantie:1996qd,DOnofrio:2014rug,DOnofrio:2015gop}. This
motivates electroweak baryogenesis: new physics near the electroweak scale can make electroweak
symmetry breaking proceed through a first-order phase transition and can introduce additional
\(CP\)-violating phases. In this way the same dynamics that modifies the electroweak phase transition
also supplies the out-of-equilibrium environment and the new \(CP\)-odd bias needed to generate
\(\eta_B\).

In principle, the relevant problem is a fully quantum, real-time, out-of-equilibrium calculation. The
natural language for such a treatment is the Schwinger--Keldysh formalism and the corresponding
Kadanoff--Baym equations for non-equilibrium Green's functions
\cite{Schwinger1961,Keldysh:1964ud,kadanoff1962quantum}. In practice, this problem is too complex
to solve in full generality for phenomenological scans. Moreover, the plasma surrounding an expanding
electroweak bubble is not expected to be arbitrarily far from equilibrium, unless the transition is
extremely strong or the microscopic dynamics invalidates a quasiparticle description. 
One can then work in the semiclassical regime in which the real-time quantum kinetic equations admit
a gradient expansion and an on-shell quasiparticle description. In the Schwinger--Keldysh language,
this corresponds to the quasiparticle, gradient-expanded limit of the Wigner-transformed
Kadanoff--Baym equations, which leads to Boltzmann equations for phase-space distribution functions.
Equivalently, the same single-particle ingredients can be obtained by a WKB expansion in the slowly
varying wall background: quasiparticles acquire local dispersion relations, group velocities and
semiclassical forces, which then enter the Boltzmann transport equations
\cite{Kainulainen:2001cn,Kainulainen:2002th,Prokopec:2003pj,Prokopec:2004ic}.

The standard EWBG transport problem is then the following. One considers the plasma in the
background of an expanding bubble wall, usually approximated as planar and moving at a constant
steady-state velocity \(v_w\). Across the wall, particle masses and \(CP\)-violating phases vary with
position. Particles and antiparticles crossing the wall therefore experience different semiclassical
forces, or equivalently different source terms in the transport equations. Solving the Boltzmann
equations for the active species gives the local chemical potentials and velocity perturbations induced
by the wall. These profiles are then combined into the baryon-left chemical potential
\(\mu_{B_L}(z)\), the weighted left-handed charge bias seen by weak sphalerons. In front of the wall,
where the electroweak symmetry is unbroken and weak sphalerons are active, this bias is partially
converted into net baryon number. Once the wall passes, the produced baryon number enters the broken
phase, where sphalerons must be sufficiently suppressed so that the asymmetry freezes in rather than
being washed out.

This is the standard semiclassical charge-transport problem addressed in the EWBG literature. The
rest of this paper fixes the conventions used to formulate this problem, derives the WKB sources and
the fluid transport equations, and then implements the resulting pipeline in \texttt{BARYONET}.

\paragraph{Input and output.}
The starting point is not a first-principles computation of the phase transition, but a reduced wall
background and a set of interaction rates,
\begin{equation}
  \left\{
  m_i(z),\theta_i(z),v_w,L_w,\Gamma_a,\Gamma_{\rm ws}(z)
  \right\}.
\end{equation}
Here $m_i(z)$ and $\theta_i(z)$ are the mass and CP-violating phase profiles of the species crossing
the wall, $v_w$ and $L_w$ are the wall velocity and width, $\Gamma_a$ denotes the relaxation and
interaction rates entering the transport network, and $\Gamma_{\rm ws}(z)$ is the weak-sphaleron rate
profile. The quantities to be solved for are the CP-odd departures from equilibrium, parametrised by
chemical potentials and velocity moments. The final observable is the entropy-normalised baryon
asymmetry in the broken phase.

With these conventions, the semiclassical transport calculation can be summarised as
\begin{equation}
  \left\{
  m_i,\theta_i,\Gamma_a,v_w,L_w
  \right\}
  \quad \longrightarrow \quad
  S_i(z)
  \quad \longrightarrow \quad
  Y_{\mathcal F}(z)
  \quad \longrightarrow \quad
  \mu_{B_L}(z)
  \quad \longrightarrow \quad
  \eta_B .
  \label{eq:baryonet-pipeline}
\end{equation}
The first step constructs the CP-violating WKB sources from gradients of the wall background. The
second solves the transport equations in a chosen framework $\mathcal F$, where $\mathcal F$ fixes the
species basis, the retained velocity moments and the collision closure. The third projects the solution
onto the baryon-left chemical potential $\mu_{B_L}$, namely the combination of left-handed charge seen
by weak sphalerons. The last step evolves the baryon density and extracts the frozen value of
$\eta_B$.

The framework labels used below do not denote new transport formalisms introduced in this work.
They refer to specific implementations of existing calculations in the literature, denoted FH04, FH06,
CK and KV24 in Table~\ref{tab:schemes-notation}. The differences among them are the species basis,
the moment truncation, the treatment of collision terms and the velocity regime for which the
approximation is intended \cite{Bodeker:2004ws,Fromme:2006wx,Cline:2020jre,Kainulainen:2024qpm}.
Using a common notation for these frameworks is useful because the same physical pipeline,
Eq.~\eqref{eq:baryonet-pipeline}, can then be discussed without repeatedly translating conventions
between papers.

\paragraph{Approximations and limitations.}
The calculation rests on the standard hierarchy of approximations of the WKB treatment of EWBG
transport \cite{Joyce:1994fu,Joyce:1994zn,Joyce:1994zt,Kainulainen:2001cn,Kainulainen:2002th,
Prokopec:2003pj,Prokopec:2004ic,Konstandin:2013caa}. The wall is taken to be stationary and planar
in its rest frame, so curvature effects, wall acceleration and the distribution of local wall velocities
during bubble growth are neglected. The wall is assumed to be thick compared with the microscopic
de Broglie wavelengths of the particles that dominate transport, i.e. $L_wT > 1$ where $T$ is the temperature of the plasma, allowing a gradient expansion in
$m_i(z)$ and $\theta_i(z)$. The plasma is expanded around boosted local thermal equilibrium, and the
CP-odd departures from equilibrium are treated in linear response. The Boltzmann equation is then
projected onto a finite number of velocity moments, replacing the full momentum-dependent
perturbation by chemical potentials and velocity perturbations. Collision terms are encoded through
thermally averaged rates, rather than through the full momentum-dependent collision operator.
Finally, weak sphalerons are treated in a second step: the transport network is first solved for
$\mu_{B_L}(z)$, which is then inserted into the weak-sphaleron evolution equation for baryon number.

These assumptions define the domain of validity of the calculation implemented in
\texttt{BARYONET}. The code does not determine the finite-temperature effective potential, the bubble
nucleation action, the wall profile or the steady-state wall velocity from first principles. These
quantities are external physical inputs. In particular, the determination of $v_w$ requires the CP-even
out-of-equilibrium plasma response, the hydrodynamic matching across the interface, and the balance
between the driving pressure and the friction exerted by the plasma on the wall
\cite{Moore:1995ua,Moore_2000,Konstandin:2014zta,Dorsch:2018pat,Hoeche:2020rsg,
Wang:2020zlf,Laurent:2022jrs,Ekstedt:2024fyq}. By contrast, the baryogenesis computation performed
here uses $v_w$ as an input and solves the CP-odd transport problem. Likewise, the moment expansion
is not a substitute for a fully momentum-resolved Boltzmann solution: its accuracy depends on the
truncation and closure, especially at large wall velocities or in regimes where the plasma perturbation
develops significant angular structure
\cite{Laurent:2020gpg,Cline:2020jre,Dorsch:2024jjl,Kainulainen:2024qpm}.

\paragraph{What is, and is not, new.}
The purpose of the derivation in the following sections is not to propose a new semiclassical force, a
new sphaleron mechanism or a new moment expansion. These ingredients are taken from the existing
EWBG literature. The role of the review material is instead to fix in one place the conventions,
normalisations, source definitions, collision closures and weak-sphaleron normalisation that enter a
numerical prediction for $\eta_B$. This is necessary because small differences in signs, wall-frame
conventions, thermal weights or rate normalisations can lead to sizeable changes in the final baryon
asymmetry.

The original contribution of this work begins from this convention-fixed formulation. First,
\texttt{BARYONET} provides a public implementation of the WKB transport calculation, exposing the
FH04, FH06, CK and KV24 prescriptions through a common model interface and a common numerical
pipeline. Second, we reassess the numerical inputs entering the transport equations: diffusion
constants, Yukawa and helicity-flip relaxation rates, and strong- and weak-sphaleron rates are
recomputed or updated where possible and translated into a common normalisation. Third, the
benchmark calculations in the literature are reproduced inside the same code base. This both validates
the implementation and provides an independent check of technically involved results that are
otherwise difficult to compare across papers.

With this setup fixed, Sec.~\ref{sec:WKB} derives the WKB source, Sec.~\ref{sec:boltzmann} builds
the fluid transport network, Sec.~\ref{sec:pheno} defines the baryon-left combination $\mu_{B_L}$,
and Sec.~\ref{sec:BAU} derives the weak-sphaleron evolution used to compute $\eta_B$. Only after
these ingredients have been specified do we turn, in Sec.~\ref{sec:Code}, to the numerical
implementation in \texttt{BARYONET}.

\section{WKB formalism}\label{sec:WKB}
In this section, we review the WKB (semiclassical) treatment of a Dirac fermion that crosses a planar bubble wall with a space–dependent complex mass. The following derivation closely follows the approach developed in \cite{Cline:2000nw, Kainulainen:2001cn, Kainulainen:2002th} later reformulated in \cite{Bodeker:2004ws,Fromme:2006wx,Cline:2020jre, Kainulainen:2024qpm}, while making explicit some intermediate steps and conventions. No new derivation of the semiclassical force is claimed here. The purpose of this section is to fix conventions and to define the quantities implemented in the public code presented in Sec.~\ref{sec:Code}.

The guiding idea of WKB is that when the background varies slowly on the scale of the particle’s de-Broglie wavelength, the evolution can be organised as an expansion in spatial gradients of the background. One writes the spinor as a locally plane–wave state with a slowly varying amplitude and solves the Dirac equation order by order in derivatives of the background. At leading order, one recovers a local dispersion relation; at the next orders, one obtains transport equations and small $CP$–odd shifts that encode how the slowly varying complex phase of the mass biases particle and antiparticle propagation. This is the mechanism behind the semiclassical force used in electroweak baryogenesis, as schematically shown in Fig.~\ref{fig:sketch}.

Here, by “background fields” we specifically mean the classical scalar profile across the wall that sources the fermion’s complex mass. In practice, the space dependence of the mass is entirely inherited from this scalar background, so it is convenient to phrase the WKB expansion in terms of the slowly varying quantities $m(z)$ and $\theta(z)$ that are proportional to the scalar profile. In what follows, whenever we refer to expanding in the background, we will do so by expanding in spatial gradients and in the slowly varying amplitudes $m$ and $\theta$ themselves, which faithfully capture the influence of the scalar background on the fermion dynamics.

We work in the wall rest frame and model the wall as static and planar, with normal along the $z$–axis.  The fermion mass is taken to be
\begin{equation}
m(z)=|m(z)|\,e^{i\theta(z)}\,,
\end{equation}
so that the only space dependence enters through the modulus $|m|$ and the $CP-$violating (CPV) phase $\theta$.  The calculation is organised as a controlled gradient expansion in $\partial_z |m|$ and $\partial_z \theta$, valid when the wall is thick compared to microscopic wavelengths (equivalently, when $E\,L_w\!\gg\!1$ for the modes that dominate transport).  Unless explicitly stated otherwise, primes denote derivatives with respect to the wall coordinate, $X'\!\equiv\!\partial_z X$.  All subsequent formulae and checks will be presented for this simple but paradigmatic example, which captures the essential physics of a slowly varying $CP$–violating background across the bubble wall.

\subsection{Setup and assumptions}
\label{subsec:wkb}
We consider, as first explicitly done in \cite{Cline:2000nw}, a single Dirac fermion with a spatially varying complex mass
\begin{equation}
  \bigl(i\gamma^\mu \partial_\mu - m\,P_R - m^* P_L\bigr)\psi = 0,
  \qquad
  m(z)=|m(z)|\,e^{i\theta(z)},\qquad
  P_{R,L}=\tfrac12(1\pm \gamma^5).
\end{equation}
We work in the wall frame and, just for simplicity, set the momentum parallel to the wall to zero initially ($p_x=p_y=0$), transverse momenta will be reintroduced later. The spin of the fermion is taken along $z$, $\sigma_3\chi_s=s\,\chi_s$, with $\chi_s$ the spin eigenstates and $s=\pm1$ related to helicity by $s=\lambda\,\mathrm{sign}(p_z)$. We assume a slowly varying background (sometimes called adiabaticity condition)
\begin{equation}
  \frac{|m|'}{E^2}\ll 1,\qquad \frac{|\theta'|}{E}\ll 1, \qquad
  \text{and similarly for higher derivatives,}
\end{equation}
so that a gradient expansion is justified. We emphasise that primes denote spatial derivatives, which therefore carry dimensions of energy.
Looking for positive–energy solutions with plane–wave time dependence, we write
\begin{equation}
  \Psi_s(t,z) = e^{-iEt}
  \begin{pmatrix} L_s(z) \\ R_s(z) \end{pmatrix}\!\otimes \chi_s,
  \label{eq: full form L}
\end{equation}
which reduces the Dirac equation to
\begin{align}
  (E - i s\,\partial_z) L_s &= m\,R_s, \label{eq:LR1}\\
  (E + i s\,\partial_z) R_s &= m^*\,L_s. \label{eq:LR2}
\end{align}
Eliminating $R_s$ yields a second–order equation for $L_s$,
\begin{equation}
  \Bigl[(E + i s\,\partial_z)\frac{1}{m}(E - i s\,\partial_z) - m^*\Bigr]L_s=0.
\end{equation}
To organise the gradient expansion, we adopt the WKB ansatz
\begin{equation}
  L_s(z) = w(z)\,\exp\Bigl\{\,i\!\int^z p_{c z}(z')\,\dd z'\Bigr\},
\end{equation}
with slowly varying amplitude $w$ and canonical momentum $p_{c z}$. Separating real and imaginary parts of Eq. \eqref{eq: full form L}, one finds
\begin{align}
  E^2 - |m|^2 - p_{c z}^2 + (sE+p_{c z})\,\theta'
  - \frac{|m|'}{|m|}\frac{w'}{w} + \frac{w''}{w} &= 0, \label{eq:WKB-real}\\
  2 p_{c z}\,w' + p_{c z}'\,w - \frac{|m|'}{|m|}(sE+p_{c z})\,w - \theta'\,w' &= 0, \label{eq:WKB-imag}
\end{align}
as explicitly obtained in Eqs.~(15)--(16) of \cite{Cline:2000nw}. At zeroth order (dropping gradients) Eq.~\eqref{eq:WKB-real} reproduces the usual dispersion relation,
\begin{equation}
  E^2 = p_{c z}^2 + |m|^2. \label{eq:disp0}
\end{equation}

\paragraph{Power counting.}
We count $\partial_z \sim \mathcal{O}(\epsilon)$, so $|m'|,\,\theta'\sim \mathcal{O}(\epsilon)$, $w'/w\sim\mathcal{O}(\epsilon)$ and $w''/w\sim\mathcal{O}(\epsilon^2)$. When deriving first–order kinematics, we will consistently drop $\mathcal{O}(\epsilon^2)$ terms like $w''/w$.

\subsection{First–order solution for the canonical momentum}
\begin{WideBox}[label={box:canonical-vs-kinetic}]{Canonical vs.\ kinetic momentum}

In the WKB/Hamilton–Jacobi picture one writes $S(z)$ so that $\psi \sim a(z)\,e^{iS(z)}$ and
the \emph{canonical} momentum is $p_{cz}\equiv \partial_z S$.
A local phase redefinition $\psi \to e^{i\alpha(z)}\psi$ is a \emph{canonical transformation}
on phase space: it preserves the equations of motion but shifts the canonical coordinate,
\[
p_{cz}\;\to\;p_{cz}+\alpha'(z).
\]
Thus $p_{cz}$ is gauge-/phase-convention dependent.

\textbf{Physical momentum.}
Observables must be invariant under canonical rephasings. This selects the \emph{kinetic}
momentum
\[
p_z \;\equiv\; p_{cz}-\alpha'(z),
\]
which controls the group velocity, semiclassical force, and enters the Liouville/Boltzmann
transport equations. In short: use $p_z$ for dynamics; $p_{cz}$ is a convenient phase-space
coordinate.

\textit{First-order gradient form.}
In our setup, the phase $\alpha$ includes the CP phase redefinition. To $\mathcal{O}(\partial_z)$,
the explicit relation used in the text reads
\[
p_z \;=\; \big(p_{cz}-\alpha_{\rm CP}'\big)
\!\left(1 - s_{\rm CP}\,s\,\frac{\theta'}{2E_{0z}}\right),
\]
i.e.\ the pure-gauge subtraction $-\alpha_{\rm CP}'$ plus the standard $\mathcal{O}(\theta')$
misalignment between phase and group propagation.
\end{WideBox}

Solving Eq.~\eqref{eq:WKB-real} to first order and using $p_0\equiv \mathrm{sign}(p_{c z})\sqrt{E^2-|m|^2}$ from~\eqref{eq:disp0}, one finds
\begin{equation}
  p_{c z}
  = p_0 + s_{CP}\,\frac{sE+p_0}{2p_0}\,\theta' + \alpha',
  \qquad s_{CP}=
  \begin{cases}
    +1 & (\text{particle})\\
    -1 & (\text{antiparticle})
  \end{cases}\!,
  \label{eq:pc-firstorder}
\end{equation}
where $s_{CP}=\pm1$ labels the two CP–conjugate WKB branches.  Under a CP transformation the complex mass phase flips, $\theta\!\to\!-\theta$, which maps the $s_{CP}\!=\!+1$ branch into the $s_{CP}\!=\!-1$ one.  It is therefore convenient to keep a \emph{fixed} phase profile $\theta(z)$ and encode the CP operation entirely through $s_{CP}$.  The additional term $\alpha'(z)$ denotes an arbitrary local rephasing of the spinor field\footnote{Under a local rephasing $\psi\to e^{i\alpha(z)}\psi$ the WKB phase shifts by $\int \alpha' \dd z$, which is equivalent to $p_{c z}\to p_{c z}+\alpha'$. Hence $p_{c z}$ is \emph{not} gauge invariant, while observables built from kinetic quantities (below) are.}, and represents a pure gauge contribution to the \emph{canonical} longitudinal momentum\footnote{We can adopt a fixed phase convention (e.g.\ $\alpha'\!=\!0$) and carry the bookkeeping variable $s_{CP}$; all physical results are expressed in terms of \emph{kinetic} quantities (group velocity, semiclassical force) that are rephasing–invariant.  Or we can also use the “kinetic gauge’’ $\alpha'\!=\!-\tfrac{1}{2}s_{CP}\,\theta'$ so that canonical and kinetic momenta coincide at leading order; either choice leaves observables unchanged. In what follows, we will show that the physical quantities will be, of course, gauge invariant.}.  With this convention, we will always refer to a phase $\theta(z)$ of definite sign and track CPV with $s_{CP}$.

Inverting Eq.~\eqref{eq:pc-firstorder}, the dispersion relation can be written as
\begin{equation}
  E=\sqrt{(p_{c z}-\alpha_{CP})^2+m^2}- s_{CP}\,\frac{s\,\theta'}{2},
  \qquad
  \alpha_{CP} \equiv \alpha' \pm \frac{s_{CP}\,\theta'}{2},
\end{equation}
where the $\pm$ refers to chirality, the difference drops out of kinetic observables. From now on, we write $m$ to mean its modulus $|m|$. Restoring the conserved momentum parallel to the wall, $p_\perp^2\equiv p_x^2+p_y^2$, we define\footnote{The mismatch reported in parts of the literature, see \cite{Cline:2000nw, Bodeker:2004ws}, stems from boosting \emph{after} manipulating the first–order (canonical) dispersion relation.The distinction between the one-dimensional and the full three-dimensional derivations of the 
semiclassical equations, and in particular the need to boost the dispersion relation to a general 
frame prior to any gradient or on-shell manipulation, was already established in 
Refs.~\cite{Prokopec:2003pj,Prokopec:2004ic}, and later emphasised in 
Ref.~\cite{Fromme:2006wx}.
}
\begin{equation}
  E_0 \equiv \sqrt{(p_{c z}-\alpha_{CP})^2+p_\perp^2+m^2},\qquad
  E_{0z}\equiv \sqrt{(p_{c z}-\alpha_{CP})^2+m^2},
\end{equation}
so that, to first order in gradients,
\begin{equation}
  E = E_0 - s_{CP}\,s\,\frac{\theta'\,E_{0z}}{2E_0}.
  \label{eq:E-canonical}
\end{equation}
It is convenient to trade canonical for kinetic variables. Define the kinetic momentum by 
\begin{equation}
  p_z = (p_{c z}-\alpha_{CP})\left(1 - s_{CP}\,s\,\frac{\theta'}{2E_{0z}}\right),
\end{equation}
and the corresponding kinetic energies
\begin{equation}
  E_0 \equiv \sqrt{p_z^2+p_\perp^2+m^2(z)},\qquad
  E_{0z}\equiv \sqrt{p_z^2+m^2(z)}.
  \label{eq: E0 kin}
\end{equation}
Equation~\eqref{eq:E-canonical} then becomes the familiar shifted dispersion relation
\begin{equation}
  E = E_0 \pm \Delta E
  = E_0 - s_{CP}\,s\,\frac{\theta' m^2}{2E_0 E_{0z}},
  \label{eq:disp-kinetic}
\end{equation}
which is manifestly independent of $\alpha'$, i.e.\ gauge invariant. More details on the difference between canonical and kinetic momentum and what is observable are given in Box~\ref{box:canonical-vs-kinetic}.

\subsection{Group velocity and semiclassical force}
Following \cite{Joyce:2000ed,Cline:2000nw}, the group velocity along $z$ is defined as
\begin{equation}
  v_{g z} \equiv \frac{\partial E}{\partial p_{c z}}
  \simeq \frac{p_z}{E_0}\left(1 + s_{CP}\,s\,\frac{\theta' m^2}{2E_0^2 E_{0z}}\right).
  \label{eq:vg}
\end{equation}
Hamilton’s equations, $\dot z=v_{g z}$ and $\dot p_z=-\partial E/\partial z$, give the semiclassical force
\begin{align}
  F_z \equiv \dot p_z
  &\simeq - \frac{(m^2)'}{2E_0}
     + s_{CP}\,s\,\frac{(m^2 \theta')'}{2E_0 E_{0z}}
     - s_{CP}\,s\,\frac{m^2(m^2)'\theta'}{4E_0^3 E_{0z}},
  \label{eq:Fz}
\end{align}
written entirely in kinetic variables. The first term is CP–even and $\mathcal{O}(\partial_z)$, while the last two constitute the CP–odd piece at $\mathcal{O}(\partial_z^2)$, implying opposite forces for particles and antiparticles as they cross the wall. The semiclassical force mechanism was first discussed at the level of the underlying transport picture in 
Refs.~\cite{Joyce:1994fu,Joyce:1994zn,Joyce:1994zt,Cline:2000nw}. 
These works capture the relevant physical effect through modified semiclassical propagation, 
but do not yet provide the final explicit three-dimensional force term. 
A derivation from first principles, including a proper distinction between canonical and physical variables, 
was later presented in \cite{Kainulainen:2001cn}. 
The first complete and consistent three-dimensional semiclassical treatment, including the explicit 
expression for the force $F$, was given in \cite{Kainulainen:2002th} (see also 
\cite{Prokopec:2003pj,Prokopec:2004ic}); importantly, the same 3D force term follows consistently 
both in the WKB approach and within the closed-time-path (CTP) formalism.

Several simple limits provide useful cross–checks. If the phase is constant, $\theta'=0$, all CP–odd terms in Eqs.~\eqref{eq:disp-kinetic} and~\eqref{eq:Fz} vanish, and one recovers the standard dispersion relation and semiclassical force for a real mass background. If the mass is constant, $m'=0$, then $(m^2\theta')'=m^2\theta''$; in the adiabatic regime $\theta''$ is suppressed so the CP–odd effects are correspondingly small, as expected. Finally, a local gauge rephasing $\alpha'\!\to\!\alpha'+\delta\alpha'$ shifts the canonical momentum $p_{c z}$ but leaves the energy in Eq.~\eqref{eq:disp-kinetic}, the group velocity in Eq.~\eqref{eq:vg}, and the force in Eq.~\eqref{eq:Fz} unchanged, confirming rephasing invariance of the kinetic observables. Taken together, Eqs.~\eqref{eq:disp-kinetic}-\eqref{eq:Fz} furnish a gauge–invariant WKB kinematics accurate to first order in gradients, with a clean separation between CP–even and CP–odd pieces; these are the ingredients that enter the Liouville (Boltzmann) operator and the sources used in the transport analysis of electroweak baryogenesis.

\section{Boltzmann equations and fluid network}
\label{sec:boltzmann}

In this section, we turn the single--particle WKB kinematics into macroscopic transport equations, as was first done in \cite{Joyce:1994fu, Joyce:1994zn, Joyce:1994zt, Cline:2000nw, Kainulainen:2001cn, Kainulainen:2002th} and later in \cite{Fromme:2006wx, Cline:2020jre, Cline:2021dkf, Kainulainen:2024qpm}. The strategy is simple in spirit. We start from the semiclassical equations of motion, which tell us how a WKB wave packet moves through the wall, and use them to build the Liouville operator that appears in the Boltzmann equation. We then expand the phase--space distribution around a local equilibrium appropriate to the moving plasma, separate CP--even from CP--odd pieces, and finally reduce the kinetic equation to a set of ordinary differential equations --- the ``fluid network'' --- for a few collective variables, namely the chemical potential and velocity perturbations. Along the way we will make explicit which terms are kept at each order in the gradient expansion, and we will isolate the CP--violating source terms that ultimately seed the baryon asymmetry. In practice, however, we will focus exclusively on the CP--odd part of this network and will not attempt to solve the CP--even equations that control the backreaction on the bubble wall and the determination of its steady--state velocity. The actual computation of the steady--state wall velocity is a technically involved problem, requiring the full CP--even transport machinery and a detailed balance between driving force and friction (see, e.g.,~\cite{Moore:1995si,John:2000zq,Moore_2000,Bodeker:2009qy,Konstandin:2014zta,Huber:2013kj,Kozaczuk:2015owa,Balaji:2020yrx,BarrosoMancha:2020fay,Wang:2020zlf,Laurent:2020gpg,Vanvlasselaer:2020niz,Leitao:2014pda,Gouttenoire:2021kjv,Dorsch:2021nje,DeCurtis:2022hlx,Azatov:2022tii,Laurent:2022jrs,DeCurtis:2023hil,DeCurtis:2024hvh,vandeVis:2025plm,Ekstedt:2024fyq,Branchina:2025adj}). In this work, by contrast, we simply treat $v_w$ as an external input parameter in the steady--state limit, possibly informed by those dedicated studies, and feed it numerically into our baryogenesis calculation rather than determining it self--consistently within the transport system.

\subsection{Assumptions and equilibrium in the wall frame}\label{subsec:equilibrium}

The first ingredient is the choice of equilibrium about which we linearise. We work in the wall frame, where the background is static and depends only on the coordinate $z$ normal to the wall. In collisions, we assume that the \emph{kinetic} longitudinal momentum is conserved: to leading order in gradients, WKB particles scatter elastically in the slowly varying background, so the relevant on–shell energy entering the equilibrium distribution is the kinetic energy \(E_0\). This choice is consistent with the WKB Hamiltonian used in the previous section, where all observable quantities were expressed in terms of the kinetic momentum \(p_z\).

With this in mind, the local equilibrium\footnote{In the context of EWBG, the term \emph{local equilibrium} is often used loosely, but it specifically refers to \emph{local thermal equilibrium}: elastic scatterings are fast enough to maintain isotropic, thermal momentum distributions characterized by a common local velocity $u^\mu(x)$ and temperature $T(x)$, while number–changing processes are slower and allow for small chemical potentials $\mu_a(x)$ between species. In the literature is commonly assume a uniform temperature, $T=\text{const.}$, and treat deviations only through $\mu_a(x)$ and $u^\mu(x)$. A detailed discussion of the distinction between thermal, kinetic, and chemical equilibrium is provided in Appendix~\ref{app:chem_vs_kin}.
} phase–space distribution for a species \(i\) is taken to be the usual Fermi–Dirac (FD) or Bose–Einstein (BE) function, but evaluated at the energy seen by the plasma moving with velocity \(v_w\) relative to the wall. In the wall frame, this amounts to boosting the single–particle energy by the factor \(\gamma_w=(1-v_w^2)^{-1/2}\) and replacing \(E\) with the kinetic expression \(E_0^i=\sqrt{p_z^2+p_\perp^2+m_i^2(z)}\). The result is
\begin{equation}
\label{eqn: boost distribution}
  f_{0w}^i(z,\mathbf p)=\frac{1}{\exp\!\big[\beta\gamma_w\big(E_0^i(z,\mathbf p)+v_wp_z\big)\big] \pm 1}\,,
  \qquad \beta\equiv \frac{1}{T}\,,
\end{equation}
where the upper (lower) sign is for fermions (bosons) and \(p_\perp^2\equiv p_x^2+p_y^2\) is the momentum conserved parallel to the wall\footnote{In local equilibrium one writes $f_{0w}=1/\{\exp[\beta(u\!\cdot\!p)]\pm1\}$.
In the wall frame the plasma drifts along $-z$, so $u^\mu=\gamma_w(1,0,0,-v_w)$ and, with metric $(+,-,-,-)$, $u\!\cdot\!p=\gamma_w(E_0+v_w p_z)$.}. For later manipulations, it is convenient to also introduce the unboosted shorthand \(f_0(E_0)\equiv [\exp(\beta E_0)\pm 1]^{-1}\), which is recovered smoothly in the limit \(v_w\to 0\).

A technical convention will be used repeatedly and is therefore stated here. The derivatives of \(f_{0w}\) that appear upon linearisation are taken with respect to its \emph{argument} \(\gamma_w(E_0+v_w p_z)\); we denote these by primes, e.g.
\(
f'_{0w}\equiv \partial_{\gamma_w(E_0+v_w p_z)} f_{0w}
\)
and similarly for \(f''_{0w}\). 

This specification of the equilibrium ensemble completes the starting point for our transport analysis. In the next subsection we will describe how small, space–dependent departures from \(f_{0w}\) are parametrised by a chemical potential and by an out-of–equilibrium piece \(\delta f\), and how their CP properties are organised so that the CP–violating sources can be identified unambiguously.

\subsection{Fluid ansatz, CP decomposition, and gradient expansion}\label{subsec:ansatz}

To turn the single–particle kinematics into transport, we parametrise small departures from local equilibrium by a chemical potential and by a residual distortion of the phase–space shape. The guiding principle is that collisions in the slowly varying wall background drive each species toward a boosted equilibrium characterised by the kinetic energy \(E_0\), while the wall forces and spatial gradients continuously perturb the distribution.

We therefore write, for a given species, see \cite{Joyce:1994fu, Joyce:1994zt, Cline:2000nw, Bodeker:2004ws}, its distribution function as\footnote{At linear order, hydrodynamic perturbations separate into CP–even and CP–odd components. The temperature and flow distortions $(\delta T,\ \delta u^\mu\!\sim\!v_w)$ are CP–even, while the chemical potentials $\mu_i$—which describe particle–antiparticle asymmetries—are CP–odd. This separation explains why, in electroweak baryogenesis, one usually studies only the perturbations in velocity and chemical potentials when computing the CP–odd transport that sources the baryon asymmetry, treating $T$ as fixed by rapid kinetic equilibration. Conversely, when determining the wall velocity and friction—CP–even quantities controlled by energy–momentum flow—one must also include the temperature perturbation $\delta T$. Couplings between CP–even and CP–odd sectors appear only beyond leading order.
}
\begin{equation}
  f(z,\mathbf p)=\frac{1}{\exp\!\big[\beta\gamma_w\big(E_i(z,\mathbf p)+v_wp_z\big)-\beta\mu(z)\big] \pm 1}+\delta f(z,\mathbf p)\,,
\end{equation}
where \(\mu\) is the chemical potential associated to the species and \(\delta f\) encodes departures from kinetic equilibrium that cannot be absorbed into a local density shift. Consistently with this interpretation, \(\delta f\) is imposed not to carry particle number,
\begin{equation}
  \int \! d^3p\delta f(z,\mathbf p)=0\,, 
\end{equation}
so that number–density deviations are entirely captured by \(\mu(z)\)\footnote{As pointed out in \cite{Dorsch:2021nje}, by a generalised fluid Ansatz we expand the non–equilibrium part of the distribution as a tensor series in momenta,
$\delta f \!=\! w^{(0)} + p_\mu w^{(1)\mu} + p_\mu p_\nu w^{(2)\mu\nu} + \cdots$,
rather than keeping only density and drift (perfect–fluid) modes. This is more general and can encode entropy–producing effects (e.g.\ shear/bulk–like corrections), but it loosens the direct link to physical observables: the coefficients $w^{(n)}$ are not uniquely interpretable (they mix under boosts, depend on chosen weights/normalisation, and their mapping to hydrodynamic fields is model– and scheme–dependent).}. In the wall frame, the kinetic WKB dispersion reads \(E=E_0+\Delta E\), with Eqs. \eqref{eq: E0 kin} and \eqref{eq:disp-kinetic}, so that CP violation enters linearly through \(\Delta E\). It is convenient to expand around the boosted kinetic equilibrium $f_{0w}(z,\mathbf p)$ and to treat both \(\mu\) and \(\Delta E\) as small in the gradient counting. Keeping terms up to second order in gradients, see \cite{Fromme:2006wx, Cline:2020jre, Kainulainen:2024qpm}, the distribution becomes
\begin{equation}\label{eq:f-expanded}
  f \simeq f_{0w}
  + f'_{0w}\,\big(\gamma_w\,\Delta E - \mu\big)
  + \tfrac12\,f''_{0w}\,\big(\gamma_w\,\Delta E - \mu\big)^2
  + \delta f
  + \mathcal O(\partial_z^3)\,.
\end{equation}
Following Refs.~\cite{Cline:2020jre, Kainulainen:2024qpm}, it is natural to split the perturbations into CP–even and CP–odd parts,
\begin{equation}
  \mu=\mu_e+\mu_o,\qquad \delta f=\delta f_e+\delta f_o,
\end{equation}
and to organise the full distribution as\footnote{For any phase–space quantity $X$ (e.g.\ energy shift, chemical potential, or a deformation of the distribution) let $\bar X$ denote its CP–conjugate evaluated at the same $(z,\mathbf p)$ in the wall frame. Define the CP–even/odd projections
$X_e \equiv \tfrac12\,(X+\bar X)$ and $X_o \equiv \tfrac12\,(X-\bar X)$.  We then write
$X = X_e + s_{\rm CP}\,X_o$ with $s_{\rm CP}=+1$ for particles and $s_{\rm CP}=-1$ for antiparticles.  In practice, we keep a fixed CP sign for all the ``$o$" quantities, i.e. +1, and encode the CP flip entirely in $s_{\rm CP}$, so we need not track barred quantities explicitly.}
\begin{equation}
  f \equiv f_e + s_{\rm CP}\,f_o\,.
\end{equation}
Reading off the even and odd pieces from \eqref{eq:f-expanded} gives, to the stated order,
\begin{align}
  f_e &\simeq f_{0w}- f'_{0w}\mu_e
  + \tfrac12 f''_{0w}\gamma_w^2 \Delta E^2 
  + \delta f_e\,,
  \\
  f_o &\simeq f'_{0w}\,\big(\gamma_w\,\Delta E - \mu_o\big)- f''_{0w}\,\gamma_w\,\Delta E\mu_e
  + \delta f_o\,.
\end{align}
This decomposition makes the subsequent CP projection of the Boltzmann equation transparent: the even part collects the density–like response and the quadratic \(\Delta E^2\) correction that survives averaging, while the odd part contains the linear CP–violating response to the wall and the CP–odd chemical potential \(\mu_o\). In the next step we will insert \(f=f_e+s_{\rm CP} f_o\) into the stationary Boltzmann equation \(L[f]=C[f]\), with the Liouville operator built from the WKB group velocity and force, and we will isolate the terms that act as CP–even and CP–odd \emph{sources} for the fluid variables.

At this stage, before moving on, it is useful to emphasise that the \(\mu_o\) entering the following expressions are local CP--odd asymmetry parameters rather than exact equilibrium chemical potentials; their physical interpretation is summarised in \boxref{box:mu_interpretation}.

\begin{WideBox}{Interpretation of the chemical potentials in the transport network}
	\label{box:mu_interpretation}
	
	The chemical potentials appearing in the transport network should be understood as \emph{local linear-response parameters} that encode the particle--antiparticle asymmetry carried by each plasma species, rather than as equilibrium chemical potentials imposed by an exactly conserved charge. In the near-equilibrium regime relevant for fluid transport, the distribution functions are expanded around thermal equilibrium, and the CP--odd perturbation is parametrised by a small quantity \(\mu_{o,i}(z)\) for each species \(i\). To leading order, this quantity is proportional to the corresponding local charge-density asymmetry,
	\begin{equation}
		n_i(z)-\bar n_i(z)=\chi_i\,\mu_{o,i}(z),
	\end{equation}
	with \(\chi_i>0\) the appropriate susceptibility.
	
	The sign of \(\mu_{o,i}\) therefore has a direct physical meaning. A positive \(\mu_{o,i}\) corresponds to a positive asymmetry of the charge carried by that species, namely an excess of particles over antiparticles in the corresponding channel, while a negative \(\mu_{o,i}\) corresponds to an excess of antiparticles over particles. Equivalently, \(\mu_{o,i}\) measures the local thermodynamic bias associated with that charge: a positive value favours excitations carrying the corresponding positive charge, whereas a negative value favours the opposite charge.
	
	We will refer to the \(\mu_{o,i}\) as chemical potentials, although this is not literally what they are. In the transport network, they are better understood as linear-response parameters describing the slowly varying CP--odd departure from equilibrium induced by the wall background. Their role is to encode how the CP--violating source is converted into local charge asymmetries, which are then redistributed by diffusion and interactions and eventually bias the weak-sphaleron processes through the left-handed combination \(\mu_{B_L}\).
\end{WideBox}

\subsection{Liouville operator and the stationary Boltzmann equation}\label{subsec:liouville}

With the kinematics in hand, the transport problem starts from the Boltzmann equation\footnote{We use a different convention of signs for the collision operator than the one used in \cite{Dorsch:2024jjl, Cline:2020jre}, where we define the collision operator to have a $-$ in the definition in App. \ref{app: collision operators}.},
\begin{equation}
  L[f]\equiv(\dot z\,\partial_z+\dot p_z\,\partial_{p_z})\,f = C[f]\,,
\end{equation}
written in the wall frame and evaluated in a steady state, so that there is no explicit time derivative. The Liouville operator is built from the WKB group velocity and force derived earlier in Eqs. \eqref{eq:vg} and \eqref{eq:Fz}. Two simple chain–rule identities make the action of \(L\) transparent once we expand around the boosted kinetic equilibrium \(f_{0w}(z,\mathbf p)\). Since \(f_{0w}\) depends on \(z\) and \(p_z\) only through its argument \(\gamma_w(E_0+v_w p_z)\), differentiation gives
\begin{equation}
  \partial_z f_{0w} = \gamma_w\,f_{0w}'\,\frac{(m^2)'}{2E_0}\,,
  \qquad
  \partial_{p_z} f_{0w} = \gamma_w\,f_{0w}'\Bigl(\frac{p_z}{E_0}+v_w\Bigr),
  \label{eq:chain}
\end{equation}
where we stress that \(f_{0w}'\) denotes the derivative of \(f_{0w}\) with respect to its argument. Substituting \eqref{eq:chain} and, at this stage, keeping only the leading pieces \(v_{g z}\simeq p_z/E_0\) and \(F_z\simeq -(m^2)'/(2E_0)\), one immediately finds
\begin{align}
  L[f_{0w}] &= v_{g z}\,\partial_z f_{0w}+F_z\,\partial_{p_z} f_{0w}
  \simeq \frac{p_z}{E_0}\,\gamma_w f_{0w}'\,\frac{(m^2)'}{2E_0}
  -\frac{(m^2)'}{2E_0}\,\gamma_w f_{0w}'\Bigl(\frac{p_z}{E_0}+v_w\Bigr)\notag\\
  &= -v_w\,\gamma_w\,\frac{(m^2)'}{2E_0}\,f_{0w}'\,.
  \label{eq:Se-from-Lf0}
\end{align}
This is the CP–even driving term that appears later as the source in the fluid equations. It has a simple physical meaning: in the wall frame, a nonzero wall velocity \(v_w\) biases the flow in the presence of a spatially varying mass. 

Going beyond the leading order is equally systematic. The CP–odd corrections to \(v_{g z}\) and \(F_z\), together with the CP–odd piece of the distribution arising from \(\Delta E\), generate the familiar structures proportional to \((m^2\theta')'\) and to \(m^2(m^2)'\theta'\). They enter through the same two terms, but are suppressed by one extra gradient, as anticipated by the counting. In practice, one now substitutes the expanded distribution \(f=f_e+s_{\rm CP}f_o\) into \(L[f]\), uses the identities \eqref{eq:chain}, and keeps terms consistently up to first order in gradients for CP–even and up to second order for CP–odd. The result is a linear combination of derivatives of the chemical potentials \(\mu_{e,o}\), derivatives of the out-of-equilibrium piece \(\delta f_{e,o}\), and a remainder that does not depend on perturbations. We shall refer to this remainder as the \emph{source}, because it persists even if \(\mu\) and \(\delta f\) are set to zero and therefore initiates the departure from equilibrium.

The algebra is straightforward but not particularly illuminating, so we postpone the fully expanded expressions to the next subsection, where we project \(L[f]\) onto its CP–even and CP–odd parts. There we will see explicitly how the even source \eqref{eq:Se-from-Lf0} sits at \(\mathcal O(\partial_z)\) while the odd source enters at \(\mathcal O(\partial_z^2)\), and how the various gradient structures organize themselves in terms that multiply \(\mu_{e,o}\), \(\delta f_{e,o}\), or no perturbation at all.
\subsection{CP projection and identification of sources}\label{subsec:CPproj}

Armed with the Liouville operator, the next step is to separate the kinetic equation into its CP–even and CP–odd parts, \cite{Fromme:2006wx,Cline:2020jre, Kainulainen:2024qpm}. This is more than a bookkeeping choice: once the distribution is expanded as \(f=f_e+s_{\rm CP}f_o\), with \(f_e\) even and \(f_o\) odd under CP, the Boltzmann equation \(L[f]=C[f]\) cleanly splits into two independent equations at the orders of interest. Operationally, we define the projections by combining the particle equation with its CP conjugate,\vspace{-0.4em}
\[
  L[f]\Big|_{\rm CP\text{-}odd} \equiv \tfrac12\!\left(L[f]-\overline{L[\bar f\,]}\right), 
  \qquad
  L[f]\Big|_{\rm CP\text{-}even} \equiv \tfrac12\!\left(L[f]+\overline{L[\bar f\,]}\right),
\]
where the overline denotes flipping the CP–odd ingredients (\(s_{\rm CP}\to -s_{\rm CP}\)). This projection isolates the pieces that must change (or not) sign between particles and antiparticles.

For the CP–odd projection, the result can be written, retaining terms up to \(\mathcal O(\partial_z^2)\) and linear in the perturbations, as
\begin{align}
  L[f]\Big|_{\rm CP\text{-}odd}
  &= -\,\frac{p_z}{E_0}\,f'_{0w}\,\mu_o'
     + v_w\gamma_w\,\frac{(m^2)'}{2E_0}\,f''_{0w}\,\mu_o
     -\frac{(m^2)'}{2E_0}\,\partial_{p_z}\delta f_o
     +\frac{p_z}{E_0}\,\partial_z \delta f_o \notag\\[0.3em]
  &\quad
     +\, s\,v_w\gamma_w\,\frac{m^2(m^2)'\theta'}{4E_0^2E_{0z}}
        \Big(\frac{f''_{0w}}{E_0}-\gamma_w f'''_{0w}\Big)\mu_e
     +\gamma_w v_w\,\frac{s\,(m^2\theta')'}{2E_0E_{0z}}
         f''_{0w}\,\mu_e \notag\\[0.3em]
  &\quad
     +\,\frac{s\,p_z m^2\theta'}{2E_0^2E_{0z}}
        \Big(\gamma_w f''_{0w}-\frac{f'_{0w}}{E_0}\Big)\mu_e'
     +\frac{s}{2E_0E_{0z}}
        \Big[(m^2\theta')'-\frac{m^2(m^2)'\theta'}{2E_0^2}\Big]\partial_{p_z}\delta f_e
     +\frac{s\,p_z m^2\theta'}{2E_0^3E_{0z}}\,\partial_z \delta f_e \notag\\[0.3em]
  &\quad
     +\,\gamma_w v_w\,\frac{s\,(m^2\theta')'}{2E_0E_{0z}}
         f'_{0w}
     + s\,v_w\gamma_w\,\frac{m^2(m^2)'\theta'}{4E_0^2E_{0z}}
        \Big(\gamma_w f''_{0w}-\frac{f'_{0w}}{E_0}\Big)
     + \mathcal O(\partial_z^3,\mu^2)\,.
  \label{eq:Lodd}
\end{align}
The final line contains the CP–violating \emph{source} terms, i.e.\ the pieces independent of \(\mu\) and \(\delta f\). Collecting them explicitly,
\begin{equation}
  S_o \equiv
 - \gamma_w v_w\,\frac{s\,(m^2 \theta')'}{2 E_0 E_{0z}}
  f'_{0w}
  -
  s\,v_w\gamma_w\,\frac{m^2(m^2)' \theta'}{4 E_0^2 E_{0z}}
  \Big(\gamma_w f''_{0w}-\frac{f'_{0w}}{E_0}\Big).
  \label{eq:So}
\end{equation}
A feature that is worth emphasising is that all CP–odd structures are of order \(\partial_z^2\), as anticipated by the WKB counting at the end of section~\ref{subsec:wkb}.

The CP–even projection proceeds in the same way, and one finds
\begin{align}
  L[f]\Big|_{\rm CP\text{-}even}
  &= -\,\frac{p_z}{E_0}\,f'_{0w}\,\mu_e'
     + v_w\gamma_w\,\frac{(m^2)'}{2E_0}\,f''_{0w}\,\mu_e
     -\frac{(m^2)'}{2E_0}\,\partial_{p_z}\delta f_e
     +\frac{p_z}{E_0}\,\partial_z \delta f_e \notag\\[0.3em]
  &\quad
     -\,\frac{s\,p_z m^2\theta'}{2E_0^3E_{0z}}\,f'_{0w}\,\mu_o'
     -\frac{s\,v_w\gamma_w}{2E_0E_{0z}}
        \Big[(m^2\theta')'\Big(1+{p_z \over v_w E_0 }\Big
        )-\frac{m^2(m^2)'\theta'}{2E_0^2}\Big] f''_{0w}\,\mu_o \notag\\[0.3em]
  &\quad
     +\,\frac{s}{2E_0E_{0z}}
        \Big[(m^2\theta')'-\frac{m^2(m^2)'\theta'}{2E_0^2}\Big]\partial_{p_z}\delta f_o
     +\frac{s\,p_z m^2\theta'}{2E_0^3E_{0z}}\,\partial_z \delta f_o \notag\\[0.3em]
  &\quad
     -\, v_w \gamma_w\,\frac{(m^2)'}{2E_0}\,f'_{0w}
     + \mathcal O(\partial_z^2,\mu^2)\,.
  \label{eq:Leven}
\end{align}
The CP–even \emph{source} is simply the last term,
\begin{equation}
  S_e \equiv v_w \gamma_w\,\frac{(m^2)'}{2E_0}\,f'_{0w}\,,
  \label{eq:Se}
\end{equation}
which already appeared in the warm–up computation \(L[f_{0w}]\). Finally Eqs.\eqref{eq:Lodd}--\eqref{eq:Leven} match Eq. (18) of \cite{Fromme:2006wx} and Eq. (16) of \cite{Cline:2020jre}. 

Solving directly for $\delta f_{e,o}(z,\mathbf p)$ and $\mu(z)$ in Eqs.~\eqref{eq:Lodd}-\eqref{eq:Leven} is a hard task (see \cite{Laurent:2020gpg, DeCurtis:2022hlx,DeCurtis:2023hil,DeCurtis:2024hvh,Dorsch:2024jjl} for solving the CP even part): the unknown $\delta f$ depends on the full three–momentum, the collision operator couples angles and energies with nonlocal kernels, and the resulting integro–differential system is stiff across the wall. Here, to obtain a simpler description, we \emph{project} the kinetic equations onto a small set of macroscopic moments—e.g.\ density and longitudinal flux—by averaging over momentum with suitable weights. In practice, replacing $\delta f_{e,o}$ by its weighted moments and integrating \eqref{eq:Lodd}-\eqref{eq:Leven} yields a closed, first–order linear system in $z$ for a pair of collective variables $(\mu,u)$ in each CP sector, at the order in perturbations considered here, where the sources $S_o$ and $S_e$ from \eqref{eq:So}-\eqref{eq:Se} enter as inhomogeneous terms and the interaction rates specify the collision closure. This moment (fluid) reduction retains the physically relevant charges and currents while avoiding the full momentum–resolved kinetics. However, for the CP–even sector it is now well established that low–order moment closures are unreliable: energy–momentum flow is sensitive to the angular structure and high–energy tails of $\delta f$, so truncations can violate exact conservation, misestimate friction and wall acceleration, and even produce spurious instabilities near the wall. Consequently, recent analyses abandon the moment projection for the CP–even problem and solve the Boltzmann equation directly for $\delta f_e(z,\mathbf p)$ using modern numerical techniques, achieving stable and accurate determinations of the friction and wall velocity~\cite{Laurent:2020gpg,DeCurtis:2022hlx,DeCurtis:2023hil,DeCurtis:2024hvh,Dorsch:2024jjl}.

\subsection{Moment expansion and normalisation}\label{subsec:moments}

As anticipated in the previous subsection, the Boltzmann equation is an integro–differential equation in phase space, and to turn it into a simpler set of ordinary differential equations, we project it onto a few low–order moments in momentum, first done in \cite{Cline:2000nw} and then later reused in \cite{Fromme:2006wx}, although the first systematic and organised projection on moments has been done in \cite{Kainulainen:2024qpm}. The idea is to weight the kinetic equation with simple functions of \(p_z/E_0\), integrate over momentum, and thereby trade the detailed shape of \(\delta f\) for a handful of “fluid” variables that capture the slow evolution driven by the wall. This procedure is effective because collisions quickly relax most angular structures, while the wall sources act coherently on low moments\footnote{Physically, the wall drives a coherent drift mainly along \(z\) (building density and flux), while collisions rapidly randomise directions, washing out higher angular distortions. Thus, the slow dynamics are captured by the lowest ``fluid–like'' moments (chemical potentials and longitudinal flux). A useful image is a stirred coffee cup: the spoon sets a bulk swirl, while molecular collisions kill small eddies. This picture breaks down as \(v_w\!\to\!1\), where angular modes are not efficiently damped and one must retain the full moment hierarchy.}
.

A convenient choice is to use the longitudinal velocity \(p_z/E_0\) as the building block of the weights. For each non–negative integer \(\ell\) we define a weighted phase–space average of any function \(X(z,\mathbf p)\) by
\begin{equation}
  \big\langle X \big\rangle \equiv \frac{1}{N_1}\int \! d^3p\,X(z,\mathbf p)\,,
  \qquad
  N_1 \equiv \int \! d^3p\, f'_{0w, {\rm FD}}(m=0)=-\gamma_w{2\pi^3 \over 3}T^2 \,,
\end{equation}
where the overall factor \(N_1\) fixes dimensions and is chosen so that the coefficients that appear below are \emph{universal} functions of \(x\equiv m/T\) and \(v_w\) (fermions and bosons differ only through the sign in \(f_{0w}\) that might appear in \(X(z,\mathbf p)\)). With this normalisation in place, the velocity moments of the out-of-equilibrium part are introduced as
\begin{equation}
  u_\ell(z) \equiv \Big\langle \Big(\frac{p_z}{E_0}\Big)^\ell \,\delta f(z,\mathbf p)\Big\rangle .
\end{equation}
The zeroth moment of \(\delta f\) vanishes by construction, because \(\delta f\) does not carry particle number; the first two moments, \(u_1\) and \(u_2\), will become the protagonists of the two–equation fluid system.

Projecting the stationary Boltzmann equation \(L[f]=C[f]\) with the weight \((p_z/E_0)^\ell\) produces the \(\ell\)-th moment equation
\begin{equation}
  \Big\langle \Big(\frac{p_z}{E_0}\Big)^\ell L \Big\rangle
  =
  \Big\langle \Big(\frac{p_z}{E_0}\Big)^\ell \delta C \Big\rangle ,
\end{equation}
where \(\delta C\) is the linearised collision term that we will properly define in the next subsections. The left–hand side can be written entirely in terms of \(\mu\), \(u_\ell\), and a set of coefficient functions by using the chain–rule identities established earlier for derivatives of \(f_{0w}\) and by substituting the WKB expressions for \(v_{gz}\) and \(F_z\). After a straightforward manipulation one finds a structure that repeats for each \(\ell\): a term proportional to the gradient of the chemical potential, a term proportional to \(\mu\) itself, two terms involving \(\partial_z\delta f\) and \(\partial_{p_z}\delta f\), and, in the CP–odd channel, the source pieces coming from \((m^2\theta')'\) and \(m^2(m^2)'\theta'\). To keep the notation compact, to make the physical content transparent and to align with the literature we follow closely \cite{Cline:2020jre, Kainulainen:2024qpm}, therefore, it is useful to name the two universal kernels that multiply \(\mu'\) and \(\mu\),
\begin{equation}
  D_\ell \equiv \Big\langle \Big(\frac{p_z}{E_0}\Big)^\ell f'_{0w} \Big\rangle,
  \qquad
  Q_\ell \equiv \Big\langle \frac{p_z^{\ell-1}}{2E_0^\ell}\, f''_{0w} \Big\rangle .
  \label{eq: D and Q}
\end{equation}
With these definitions, and keeping terms to the same gradient order used in the WKB section, the even part of the \(\ell\)-th moment reads schematically
\begin{equation}
  \Big\langle \Big(\frac{p_z}{E_0}\Big)^\ell L \Big\rangle_{e/o}
  =
  -\,D_{\ell+1}\,\mu'
  + v_w\gamma_w\,(m^2)'\,Q_{\ell+1}\,\mu
  + \Big\langle \Big(\frac{p_z}{E_0}\Big)^{\ell+1}\partial_z\delta f\Big\rangle
  - (m^2)'\Big\langle \frac{p_z^\ell}{2E_0^{\ell+1}}\,\partial_{p_z}\delta f\Big\rangle
  - \mathcal S_\ell \, .
\end{equation}
The explicit expressions of \(\mathcal S_\ell^{e/o}\) follow directly from the CP–even and CP–odd sources identified at the distribution level and amount to simple averages of those terms with the same weight \((p_z/E_0)^\ell\). At this stage, two remnants still link the moment equations to the detailed shape of \(\delta f\): the averages of \(\partial_z\delta f\) and \(\partial_{p_z}\delta f\). In the next subsection we show how they can be rewritten in terms of the velocity moments \(u_\ell\) and their derivatives by using a mild factorization assumption together with an integration–by–parts identity. This is the only additional closure needed to arrive at a closed fluid system for \(\mu\) and the velocity perturbations \(u_\ell\).
\subsection{Closure: factorization and truncation}\label{subsec:closure}

Projecting the Boltzmann equation onto moments has reduced the kinetic problem to a few phase–space averages. Two kinds of terms still prevent the system from closing on a small set of variables: averages that contain derivatives of the shape distortion, such as
\(\big\langle (p_z/E_0)^{\ell+1}\partial_z\delta f\big\rangle\) and
\(\big\langle p_z^\ell/(2E_0^{\ell+1})\,\partial_{p_z}\delta f\big\rangle\),
and the tower of higher moments \(u_\ell\) generated by the weights \((p_z/E_0)^\ell\).
The first ingredient is a mild factorisation rule for averages that contain \(\delta f\). We parametrise the first velocity moment as
\[
  u_1(z)\equiv u(z)=\Big\langle \frac{p_z}{E_0}\,\delta f\Big\rangle,
\]
and approximate higher moments by pulling out the momentum dependence into a weight built on the equilibrium distribution. Concretely, for any function \(X\) we write
\begin{equation}
  \big\langle X\delta f\big\rangle \simeq
  \Big[X\frac{E_0}{p_z}\Big]u,
  \qquad
  [X]\equiv\frac{1}{N_0}\int d^3p\,Xf_{0w},\qquad
  N_0\equiv \int d^3p\,f_{0w}(m)=\gamma_w\,\hat N_0,
  \label{eq:factorization}
\end{equation}
where the square brackets denote a \emph{different} normalisation from the \(N_1\)–weighted averages used for \(D_\ell\) and \(Q_\ell\). This factorisation, first introduced in \cite{Fromme:2006wx} and later used also in \cite{Cline:2020jre, Cline:2021dkf, Kainulainen:2024qpm}, preserves the exact number–sum rule \(\int d^3p\,\delta f=0\) and isolates the single amplitude \(u\) as the carrier of the out-of-equilibrium contribution. Applying \eqref{eq:factorization} to the second velocity moment gives
\begin{equation}
  u_2=\Big\langle \Big(\frac{p_z}{E_0}\Big)^2\delta f\Big\rangle
  \simeq\Big[\frac{p_z}{E_0}\Big]u \equiv Ru.
\end{equation}
The quantity \(R\) has a simple meaning: it is the average longitudinal velocity in the boosted equilibrium. Since \(f_{0w}\) is a boost of an isotropic distribution, its average velocity along the boost direction is fixed by kinematics to be the negative of the wall speed. Boosting back to the plasma frame makes this transparent and yields the exact identity
\begin{equation}
  R=\Big[\frac{p_z}{E_0}\Big]=-v_w\,.
  \label{eq:R}
\end{equation}
Thus the two–moment truncation closes the tower at \(u_2=Ru\) with no further modeling\footnote{A constant–\(R\), ``factorization'' closure has been used since the early literature, often with \(R\) set by the boosted equilibrium, \(R=-v_w\) (see e.g.\ \cite{Cline:2000nw}, and the discussion in \cite{Kainulainen:2024qpm}). Other ad hoc choices appear, such as setting the last moment to zero (\(R=0\)) or equal to the previous one (\(R=1\)). All constant–\(R\) prescriptions imply the same gradient relation \(u'_{\ell}=R\,u'_{\ell}\), i.e.\ derivatives propagate down the hierarchy with geometric weight \(R\). The choice \(R=-v_w\) is singled out by covariance: it reproduces the exact drift of \(f_{0w}\) in the wall frame and introduces no extra parameters.}.

The second ingredient is an integration–by–parts identity that trades the remaining averages of \(\partial_z\delta f\) and \(\partial_{p_z}\delta f\) for derivatives of the velocity moments and a single additional coefficient. As explicitly noted in \cite{Kainulainen:2024qpm}, using Eq. \eqref{eq:disp-kinetic} and keeping the dependence on \(m(z)\) explicit, one finds
\begin{equation}
  \Big\langle \frac{p_z^{\ell+1}}{E_0^{\ell+1}}\partial_z \delta f\Big\rangle
  - (m^2)'\Big\langle \frac{p_z^\ell}{2E_0^{\ell+1}}\,\partial_{p_z}\delta f\Big\rangle
  = u'_{\ell+1} + \ell(m^2)'\bar R u_{\ell},
  \label{eq:IBP}
\end{equation}
where boundary terms vanish because we assume that \(\delta f\) dies off sufficiently fast in momentum space. The new coefficient
\[
  \bar R \equiv \Big[\,\frac{1}{2p_z E_0}\,\Big] ,
\]
is well defined even though the integrand is divided by \(p_z\): instead of taking a principal value, as already considered in the literature \cite{Cline:2020jre, Kainulainen:2024qpm}, it is easier to evaluate \(\bar R\) after boosting the integrand to the plasma frame, where \(f_0\) is isotropic and \(d^3p/E_0\) is Lorentz invariant. Performing the angular integral then leads to a one–dimensional representation,
\begin{equation}
  \bar R = \frac{\pi}{\gamma_w^2\,\hat N_0}\int_{m}^{\infty}\! dE
  \ln\!\left|\frac{p - v_w E}{p + v_w E}\right|\, f_0(E),
  \qquad p\equiv\sqrt{E^2-m^2},
  \label{eq:Rbar}
\end{equation}
where the statistic is included through \(f_0\).

Equations \eqref{eq:R} and \eqref{eq:IBP} are precisely what is needed to close the moment hierarchy. The factorization fixes the relation \(u_2=R\,u\), while the integration–by–parts identity eliminates the explicit derivatives of \(\delta f\) in favor of \(u'_{\ell}\) and \(u_{\ell}\) with a single extra kernel \(\bar R\). Specialising to the lowest two moments \(\ell=0,1\) then produces a closed, first–order system for the pair \((\mu(z),u(z))\) in each CP sector. In the next subsection we will write this system explicitly in matrix form, identify the corresponding sources by averaging the CP–even and CP–odd driving terms, and discuss the role of collisions in coupling different species.
\paragraph{Variance truncation.}
In the moment expansion of the longitudinal Boltzmann equation the hierarchy is unclosed: the equation for $u_{\ell}\!\equiv\!\langle (p_z/E_0)^{\ell}\,\delta f\rangle$ contains $u_{\ell+1}$.  Besides the constant–$R$ ``factorization'' (which assumes $u_{\ell+1}'=R\,u_{\ell}'$), as was introduced and explained in \cite{Kainulainen:2024qpm}, a physically motivated closure is to set the $\ell$-th \emph{central} moment of the perturbation to zero,
\begin{equation}
\Big\langle \Big(\frac{p_z}{E_0}-u_{1}\Big)^{\!\ell}\,\delta f\Big\rangle=0.
\label{eq:var-zero}
\end{equation}
Expanding the binomial gives an algebraic relation that expresses the highest raw moment in terms of lower ones,
\begin{equation}
u_{\ell}\;=\;\sum_{k=1}^{\ell-1}\binom{\ell}{k}\,(-u_1)^{\,\ell-k}\,u_k \;-\;(-u_1)^{\,\ell}\,u_0,
\qquad
u_0\equiv\langle\delta f\rangle,
\label{eq:poly-closure}
\end{equation}
which, in the standard setup where the density mode $u_0$ is handled by the chemical potential (hence $u_0=0$ for $\delta f$), simplifies to
\begin{equation}
u_{\ell}\;=\;\sum_{k=1}^{\ell-1}\binom{\ell}{k}\,(-u_1)^{\,\ell-k}\,u_k.
\label{eq:poly-closure-u0zero}
\end{equation}
Differentiating \eqref{eq:poly-closure-u0zero} yields a linear relation for the derivatives,
\begin{equation}
u_{\ell}' \;=\; \sum_{i=1}^{\ell-1} R_i\,u_{i}'\,,
\qquad
R_i=\binom{\ell}{i}\,(-u_1)^{\,\ell-i}\ \ (i\ge2),\quad
R_1=\binom{\ell}{1}(-u_1)^{\ell-1}-\sum_{k=1}^{\ell-1}\binom{\ell}{k}(\ell-k)(-u_1)^{\ell-k-1}u_k,
\label{eq:R-coeffs}
\end{equation}
so that the closure adapts smoothly to the local drift $u_1$ and introduces no extra parameters\footnote{In \cite{Kainulainen:2024qpm} a slightly different implementation of the ``variance'' closure is used (they keep a nonzero right–hand side for the central moment). The authors report that switching between these variants has no visible numerical impact on $\eta_B$ within their benchmarks. We consider the RHS to be zero since with $\ell=0$ it has to vanish by construction.}.

\subsection{From moments to a two–equation fluid network}\label{subsec:fluid-system}

With these ingredients, the \(\ell=0\) and \(\ell=1\) moment equations become, respectively,
\begin{tcolorbox}
\vspace{-0.5cm}
\begin{align}
  -D_1\mu'(z) + u'(z) + v_w\gamma_w(m^2)'(z)Q_1\mu(z)
  = \mathcal S_1(z) + \delta \mathcal C_1(z), \label{eq:fluid-eq-1}\\[0.2em]
  -D_2\mu'(z) + Ru'(z) + v_w\gamma_w(m^2)'(z)Q_2\mu(z)
  + (m^2)'(z)\bar Ru(z)
  = \mathcal S_2(z) + \delta \mathcal C_2(z), \label{eq:fluid-eq-2}
\end{align}
\end{tcolorbox}
\noindent where \(\mathcal S_{1,2}\) are the weighted sources obtained by averaging the driving terms identified at the CP–projection stage, and \(\delta \mathcal C_{1,2}\) are the corresponding moments of the linearised collision integral. These two equations are written once and for all; what distinguishes the CP–even and CP–odd sectors is solely the choice of \(\mathcal S_{1,2}\) appropriate to each case. For the even sector, the sources are proportional to \(v_w\gamma_w(m^2)'\). For the odd sector, they are proportional to the two gradient structures \((m^2\theta')'\) and \(m^2(m^2)'\theta'\) that carry the CP–violating physics of the wall. In both sectors, the coefficients \(D_\ell, Q_\ell, R,\bar R\) are universal functions of the dimensionless mass \(x=m/T\) and of the wall speed \(v_w\), differing only by statistics through the equilibrium distribution.

It is convenient to combine Eqs.~\eqref{eq:fluid-eq-1}-\eqref{eq:fluid-eq-2} into a compact block–matrix form by introducing the state vector \(w(z)\equiv(\mu(z),\,u(z))^{\mathsf T}\). This notation, used in \cite{Cline:2020jre}, makes it straightforward to extend the system to an arbitrary number of velocity moments, as developed in Sec.~\ref{sec:higher-moments}, closely following what has been done in \cite{Kainulainen:2024qpm}. The system then reads
\begin{tcolorbox}
\vspace{-0.2cm}
\begin{equation}
  A\,w'(z) + (m^2)'(z)\,B\,w(z) = \mathcal S(z) + \delta \mathcal C(z),
  \label{eq: diffusion system matrix}
\end{equation}
with
\[
A=\begin{pmatrix} -D_1 & 1 \\[0.2em] -D_2 & R \end{pmatrix},
\qquad
B=\begin{pmatrix} v_w\gamma_w\,Q_1 & 0 \\[0.2em] v_w\gamma_w\,Q_2 & \bar R \end{pmatrix},
\qquad
\mathcal S=\begin{pmatrix} \mathcal S_1 \\ \mathcal S_2 \end{pmatrix},
\qquad
\delta \mathcal C=\begin{pmatrix} \delta \mathcal C_1 \\ \delta \mathcal C_2 \end{pmatrix}.
\]
\end{tcolorbox}
\noindent
This is a linear, first–order boundary–value problem along \(z\). Physically, one requires the perturbations to vanish far away from the wall, where the background becomes homogeneous and the sources switch off, so \(\mu,u\to 0\) as \(z\to \pm\infty\). Once these conditions are imposed, the system can be solved independently in the CP–even and CP–odd channels, because the sources do not mix them at the order considered. Hereafter, we restrict to the CP-odd sector, the piece that sources the weak–sphaleron bias and is therefore the component relevant for computing the baryon asymmetry \(\eta_B\). The chemical potential produced in the CP–odd sector is the quantity that will feed the anomalous baryon number violation in front of the wall and determine the final baryon asymmetry.

\subsection{Collision terms and model–independent closure}
\label{subsec:collisions}

To close the two–moment fluid system we must specify how interactions relax small
departures from boosted kinetic equilibrium. We work with the \textit{linearised}
collision operator and separate, at the level of the integrand, inelastic (number/charge–
changing) and elastic (momentum–redistributing) contributions, following derivation and considerations in \cite{Cline:2000nw, Kainulainen:2024qpm}. More details are given in Appendix~\ref{app: collision operators}. Throughout this
subsection, objects \emph{inside} the momentum averages are denoted in plain font by
\(\delta C\); once averaged, we denote them by calligraphic \(\delta\mathcal C\).

\medskip
\noindent For each species, we define the two projected moments
\begin{equation}
  \delta\mathcal C_1 \equiv \big\langle \delta C \big\rangle,
  \qquad
  \delta\mathcal C_2 \equiv \Big\langle \frac{p_z}{E_{0}}\delta C \Big\rangle\,.
  \label{eq:DC-moments-def}
\end{equation}
We split the linearised collision integrand into \emph{inelastic} and \emph{elastic} parts,
\(\delta C \simeq \delta C^{\rm inel}+\delta C^{\rm el}\).
Inelastic reactions drive specific linear combinations of the dimensionless chemical
potentials \(\xi_i\!\equiv\!\mu_i/T\) toward zero, whereas elastic scatterings conserve all
charges and only relax the momentum–space distortion \(\delta f\).
Let \(s_{ij}\in\{-1,0,+1\}\) be the coefficient that specifies how species \(j\) enters
channel \(i\) (negative if it appears in the initial state of \(i\), positive if it appears
in the final state, and zero if absent). Then the wall–frame, linearised integrand can be written as
\begin{align}
  \delta C^{\rm inel}_a &\simeq
  -f^{a}_{0w}\sum_{i,j} \Gamma_{i\to j}^{\rm wf,\,inel}\, s_{ij}\,\xi_j
  -
  \delta f^{a}(p)\sum_{i,j} \Gamma_{i\to j}^{\rm wf,\,inel},
  \label{eq:integrand-raw}
  \\
  \delta C^{\rm el}_a &\simeq -
  \delta f^{a}(p)\sum_{i,j} \Gamma_{i\to j}^{\rm wf,\,el}\,.
  \label{eq:integrand-raw 1}
\end{align}
Here \(i/j\) label, respectively, the reaction channel and the participating species; the sums run over all open processes that involve \(a\) and over all species that participate in each process. As first explicitly noted in \cite{Kainulainen:2024qpm}, “wf” denotes the \emph{wall frame}, and \(\Gamma\) is the microscopic rate for that process in that frame. Inelastic channels include both decays \((1\!\to\!2)\) and inelastic scatterings \((2\!\to\!2)\). For elastic channels the first (chemical–potential) term is absent because no charge is exchanged, while the \(\delta f\) term survives and encodes momentum relaxation.

At this stage the \(\Gamma^{\rm wf}\) retain their full momentum dependence. In the Box~\ref{box:rates} we explain the difference with the \textit{per-particle} rates. Equations~\eqref{eq:integrand-raw}-\eqref{eq:integrand-raw 1} are therefore the most model–agnostic form of the linearised collision integrand; the only inputs are the list of channels and the coefficients \(s_{ij}\).

\begin{WideBox}[label={box:rates}]{Rates convention}
\textbf{Notation: un-averaged vs.\ per–particle rates.}
In Eq.~\eqref{eq:integrand-raw}$-$\eqref{eq:integrand-raw 1} the symbols $\Gamma$ denote momentum–dependent collision rate in the wall–frame Boltzmann operator; they still depend on the test particle energy/momentum $p$ and are \emph{not} per–particle rates. In order to define the per–particle rate, we need to thermally average over initial/final states, then kernels become momentum–independent:
\begin{align*}
\Gamma\;&\equiv\;\text{un-averaged, momentum-dependent rate},\\
\widehat{\Gamma}\;&\equiv\;\text{thermally averaged (per–particle) rate}.    
\end{align*}
From now on, we use a hat to denote averaged rates, reserving $\Gamma$ for unaveraged kernels.
\end{WideBox}

\subsubsection{Averaged rates.}
To match common usage in the literature, see \cite{Joyce:1994fu, Joyce:1994zn, Joyce:1994zt, Huet:1994jb, Cline:2000nw, Kainulainen:2001cn, Kainulainen:2002th, Fromme:2006wx, Fromme:2006cm, Cline:2020jre, Cline:2021dkf, Kainulainen:2024qpm}, we first emphasise that the $\Gamma$’s in Eqs.~\eqref{eq:integrand-raw}--\eqref{eq:integrand-raw 1} are \emph{not yet averaged}: they are momentum–resolved wall–frame kernels that retain full dependence on the test–particle momentum $p$. To embed them into a fluid description, one trades $\Gamma(p)$ for a single \emph{per–particle} rate via a thermal average,
\begin{equation}
    \Gamma(p) \longrightarrow \frac{T^3/12}{n_X}\,\frac{12}{T^3}\int \frac{d^3p}{(2\pi)^3}\, f_{0,X}(p)\, \Gamma(p)\,\equiv  \kappa^X\,\widehat\Gamma \;,
\end{equation}
where $f_{0,X}$ is the chosen equilibrium weight and $n_X\!\equiv\!\int d^3p\, f_{0,X}/(2\pi)^3$ fixes the normalisation. With this convention, $\widehat\Gamma$ coincides with the per–particle rates defined in Appendices~\ref{app:rate Gammay} and~\ref{app:rate GammaM}, and the conversion factor
\begin{equation}
    \kappa^X \;=\; \frac{T^3/12}{n_X} \;=\; 
    \begin{cases}
        \dfrac{\pi^2}{9 \zeta(3)}\simeq 0.912, & \text{FD}\\
        \dfrac{\pi^2}{12 \zeta(3)}\simeq 0.685, & \text{BE}\\
        \dfrac{\pi^2}{12}\simeq 0.822, & \text{MB}\\
    \end{cases}
\end{equation}
is the usual per–degree–of–freedom rescaling between the $(12/T^3)$ convention and the $(1/n_X)$ convention\footnote{We have verified that including the full mass dependence in $n_X$ does not lead to any qualitative change in the results.}. By default, we take $f_{0,X}$ to be the Fermi–Dirac/Bose–Einstein distribution of the tracked species; alternatively, some works (explicitly \cite{Kainulainen:2024qpm}) adopt a Maxwell–Boltzmann (MB) weight to remain statistics–agnostic. In particular, \cite{Kainulainen:2024qpm} normalise with $N_1$ instead of the factor $12/T^3$; in that normalisation $\widehat{\Gamma}$ cannot be \emph{directly} identified with the per–particle rates in the appendices without an additional overall rescaling. The chosen averaging/normalisation coherently fixes the set of universal projection constants that accompany the averaged rates in the moment equations; if one includes spin/colour degeneracies, $n_X$ (and hence $\kappa^X$) must be adjusted consistently.

\subsubsection{Momentum–space average of the collision operator}
Exact collision integrals in Eqs.~\eqref{eq:integrand-raw}--\eqref{eq:integrand-raw 1} retain a residual dependence on the test–particle momentum \(p\).
Within the moment projection, we replace each microscopic, momentum–resolved kernel by a \emph{momentum–space averaged}
(per–particle) rate \(\widehat{\Gamma}_i\), appropriately normalised so as to match the conventions used in the appendices. This quantity captures the
weighted impact of the underlying process on the chosen moment and renders the collision terms momentum–independent in the fluid system. Following \cite{Kainulainen:2024qpm}, we implement this replacement and perform the two projections producing \emph{species–dependent} normalisations, which we collect into
\begin{equation}
  K^{(\ell)} \equiv \Big\langle \Big(\frac{p_z}{E_{0}}\Big)^{\ell}f_{0w}\Big\rangle.
  \label{eq:K-defs}
\end{equation}
These \(K^{(\ell)}\) are \emph{universal} functions of the dimensionless mass \(x_a\!\equiv\!m_a/T\) and of the wall velocity \(v_w\). One can further write
\begin{equation}
    K^{(\ell)}=-K_0 \left[ \left(p_z \over E_0\right)^\ell\, \right], \quad \text{with}\quad  K_0=-{\langle f_{0w}\rangle} \ ,
\end{equation}
which makes contact with the notation of Ref.~\cite{Cline:2020jre}: for \(\ell=0\) it reduces to \(-K_0\), while for \(\ell=1\) it gives \(+v_w K_0\). 

\medskip
\noindent With the averaged rate \(\widehat{\Gamma}\) and the normalisations \(\kappa^X\) and \(K^{(\ell)}\) in place, the two averaged collision moments take the compact form
\begin{align}
  \delta\mathcal C_1^a
  &\supset -K_a^{(0)} \kappa^X\sum_{i,j} \widehat{\Gamma}_{i\to j}^{\rm inel} s_{ij}\,\xi_j,
  \label{eq:DC0-final}\\
  \delta\mathcal C_2^a
  &\supset -K_a^{(1)}\kappa^X \sum_{i,j} \widehat{\Gamma}_{i\to j}^{\rm inel} s_{ij}\,\xi_j
     - \kappa_a^X\,\widehat{\Gamma}^{a}_{\rm tot}\,u_{1,a},
  \qquad
  \widehat{\Gamma}^{a}_{\rm tot}\equiv \sum_{i\in a}\widehat{\Gamma}_i^{\rm el}+\widehat{\Gamma}_i^{\rm inel}\,.
  \label{eq:DC1-final}
\end{align}
Thus, inelastic channels relax the appropriate combinations of \(\xi_j\) (dimensionless chemical potentials) in both moments, while elastic processes
(and the elastic part of inelastic channels) appear only as a \emph{flow–damping} contribution in the second moment via \(\widehat{\Gamma}^{a}_{\rm tot}\) and the first velocity mode \(u_{1,a}\). Then, Eqs.\eqref{eq:DC0-final}--\eqref{eq:DC1-final}, besides different normalisations, are equal in form to Eq. (5.3) of \cite{Kainulainen:2024qpm}.
\medskip
\paragraph{Total rate.}
We define the momentum–averaged total interaction rate of species $a$ as in \eqref{eq:DC1-final}, which acts as the effective damping of the first velocity moment $u_{1,a}$ in the fluid equations. Physically, $\widehat{\Gamma}_{\rm tot}^a$ quantifies the relaxation of momentum perturbations of species $a$ due to collisions with the plasma. However, in the moment formalism a different definition is used, see \cite{Bodeker:2004ws,Fromme:2006wx, Fromme:2006cm, Konstandin:2013caa,Espinosa:2011eu, Cline:2020jre, Cline:2021dkf, Kainulainen:2024qpm}, that is in terms of the moment functions $D_{\ell,i}$ and recast it as
\begin{equation}
\widehat{\Gamma}_{{\rm tot},i} = -\frac{D_{2,i}(x,v_w)}{v_wD_{1,i}(x,v_w) D_i}~~\overset{v_w\ll 1}{\longrightarrow}~~\frac{K^{\rm FH06}_{4,i}}{D_i\,K^{\rm FH06}_{1,i}}\,,
\end{equation}
where $D_{\ell, i}$ are the moment weights defined in \eqref{eq: D and Q}, $K^{\rm FH06}_{\ell,i}$ are the moment weights defined in FH06 and $D_i$ denotes the (species–dependent) diffusion constant. We stress that this identification is \emph{formally circular} if $D_i$ is itself defined using $\widehat{\Gamma}_{{\rm tot},i}$ and then reused to compute it; the microphysically consistent approach is to obtain $\widehat{\Gamma}_{{\rm tot},i}$ from the full interaction width of the particle and only subsequently define the diffusion constants\footnote{We thank Enrique Fernandez-Martinez for this clarifying remark, raised during a discussion.}. In this review we follow the standard conventions adopted in the EWBG literature for these quantities and proceed accordingly.

\medskip
\paragraph{Strong sphalerons and the quark sector.}
Since in this review we are interested in EWBG within the SM augmented by new physics that renders the EWPT first order, we necessarily track the evolution of the top/bottom chiral sector together with the Higgs, i.e.\ \((t_L,b_L,t_R,h)\). In this setting the QCD \emph{strong sphalerons} (SS) must be included: they violate axial (chiral) charge in the quark sector while conserving baryon number, and they rapidly drive the combination \(\sum_{q}^{n_f}(\xi_{q_L}-\xi_{q_R})\) toward zero in front of the wall, see \cite{Huet:1995sh, Cline:2000nw, Fromme:2006wx, Cline:2020jre, Cline:2021dkf, Kainulainen:2024qpm}. At the moment level, as done in \cite{Kainulainen:2024qpm}, we include the SS contribution as follows
\begin{equation}
  \delta\mathcal C_1^a\big|_{ss}
  =
  -K^{(0)}_{a}c_a^{ss}\,{\Gamma}_{ss}\,
    \sum_{q}^{n_f}\!\big(\xi_{q_L}-\xi_{q_R}\big),
  \qquad
  \delta\mathcal C_{\ell \geq 2}^a\big|_{ss}\simeq 0,
  \label{eq:SS-only}
\end{equation}
where ${\Gamma}_{ss}$ is the strong sphaleron rate, \(c_a^{ss}=+1\) for left–handed quarks, \(c_a^{ss}=-1\) for right–handed quarks of any flavour, and \(c_a^{ss}=0\) for non–quark species (e.g.\ Higgs, leptons). Ref.~\cite{Kainulainen:2024qpm} justifies the restriction to the lowest moment following from the fact that SS dynamics is dominated by deep–infrared gauge modes and is known from lattice studies only at the integrated level; any genuine momentum dependence beyond \(\delta\mathcal C_1\) is strongly suppressed\footnote{Reference \cite{Kainulainen:2024qpm} checked that allowing a small nonzero SS contribution in \(\delta\mathcal C_2\) does not affect the results within numerical uncertainties. See App.~\ref{app: collision operators} for discussion and references. This is important since in all the other references it has been considered in the higher momenta as $\delta\mathcal C_2^a=-K_a^{(1)}c_a^{ss}\,{\Gamma}_{ss}\,
    \sum_{q}^{n_f}\!\big(\xi_{q_L}-\xi_{q_R}\big)$. In \texttt{BARYONET} for all the implementations except KV24, this contribution is considered in $\delta \mathcal{C}_2$, while when implementing KV24 it is not.}.

\paragraph{Eliminating light quarks.}
Light flavours enter our fluid network only through QCD strong sphalerons. In the absence of Yukawa mixing, they satisfy
\(\mu_{q_R}=-\mu_{q_L}\) for each light species \(q\), and—on the transport time scale where electroweak sphalerons are slow—baryon number conservation implies that their net baryon density vanishes. One can then solve the light–quark potentials in terms of the heavy sector, as in \cite{Cline:2000nw, Fromme:2006wx, Cline:2020jre, Cline:2021dkf, Kainulainen:2024qpm},
\begin{equation}
  \mu_{q_L} = -\mu_{q_R}
  = D_0^{t}\mu_{t_L} + D_0^{b}\mu_{b_L} + D_0^{t}\mu_{t_R},
  \label{eq:muqL-solution}
\end{equation}
where \(D_0^{\alpha}\) are the \(\ell=0\) thermal kernels introduced earlier. Substituting
\eqref{eq:muqL-solution} into the strong–sphaleron functional gives
\begin{equation}
  \sum_{q} \big(\mu_{q_L}-\mu_{q_R}\big)
  =
  \big(1+9D_0^{t}\big)\mu_{t_L}
  +
  \big(1+9D_0^{b}\big)\mu_{b_L}
  -
  \big(1-9D_0^{t}\big)\mu_{t_R}\equiv \mu_{ss}.
  \label{eq:SS-functional-heavy}
\end{equation}

\medskip
\noindent\textbf{Summary.}
We have (i) linearised the microscopic collision \emph{integrand} \(\delta C\) by separating inelastic, number–/charge–changing channels from purely elastic, momentum–redistributing ones; (ii) compressed the momentum dependence of each channel into a single \emph{average} \(\widehat{\Gamma}\) by a thermal average with the proper normalisation; and (iii) projected the result onto the fluid moments that appear in the truncated Boltzmann hierarchy. The second step introduces universal equilibrium averages \(K_a^{(\ell)}\) and the species–dependent normalisation \(\kappa_a^X\). Stacking all species then yields a compact, reusable closure for the linearised, \emph{averaged} collision operator, reported below.

\begin{tcolorbox}[colback=gray!8,colframe=black,title={linearised averaged collision operator}]
\vspace{-0.3cm}
\begin{equation}
\delta\mathcal C_{\ell+1}^a
  = -K_a^{(\ell)}\kappa_a^X \sum_{i,j} \widehat{\Gamma}_{i\to j}^{\rm inel} s_{ij}\,\mu_j
     -K_a^{(\ell)} \delta_{\ell 0}{\Gamma}_{ss} \mu_{ss}
     -\kappa_a^X\,\widehat{\Gamma}^{a}_{\rm tot}\,u_{\ell,a},
  \qquad
  \widehat{\Gamma}^{a}_{\rm tot}\equiv \sum_{i\in a}\widehat{\Gamma}_i^{\rm el}+\widehat{\Gamma}_i^{\rm inel}\,.
  \label{eq: linearised collision}
\end{equation}
\end{tcolorbox}

\subsection{Weighted sources}\label{subsec:sources}

Once the Boltzmann equation has been projected on moments, the driving terms that remain after setting \(\mu=\delta f=0\) must be averaged with the same weights \((p_z/E_0)^\ell\) used to build the fluid variables. This step is purely kinematic and turns the distribution–level sources found in the CP projection into the inhomogeneities that appear on the right–hand side of the moment equations. In this subsection we adopt the most recent formulation of the semiclassical framework as presented in \cite{Kainulainen:2024qpm}, while acknowledging the extensive line of works that progressively established and refined this approach \cite{Joyce:1994fu, Joyce:1994zn, Joyce:1994zt, Huet:1995sh, Cline:2000nw, Kainulainen:2001cn, Kainulainen:2002th, Fromme:2006wx, Konstandin:2013caa, Cline:2020jre, Cline:2021dkf}.

In the CP–even sector, the source originates from the term proportional to \((m^2)'\). After the \((p_z/E_0)^\ell\) averaging it collapses to a single coefficient,
\begin{equation}
  Q_\ell^e \equiv \Big\langle \frac{p_z^{\ell-1}}{2E_0^\ell}\,f'_{0w}\Big\rangle,
\end{equation}
so that the weighted source reads
\begin{equation}
  \mathcal S_\ell^{\,e}(z) = v_w\,\gamma_w\,(m^2)'(z)\,Q_\ell^e.
\end{equation}
This is the same structure that already appeared unweighted at the distribution level; the averaging only replaces bare factors of \(1/(2E_0)\) by their thermal expectation values with the appropriate weight.

\begin{figure}
    \centering
    \includegraphics[width=\linewidth]{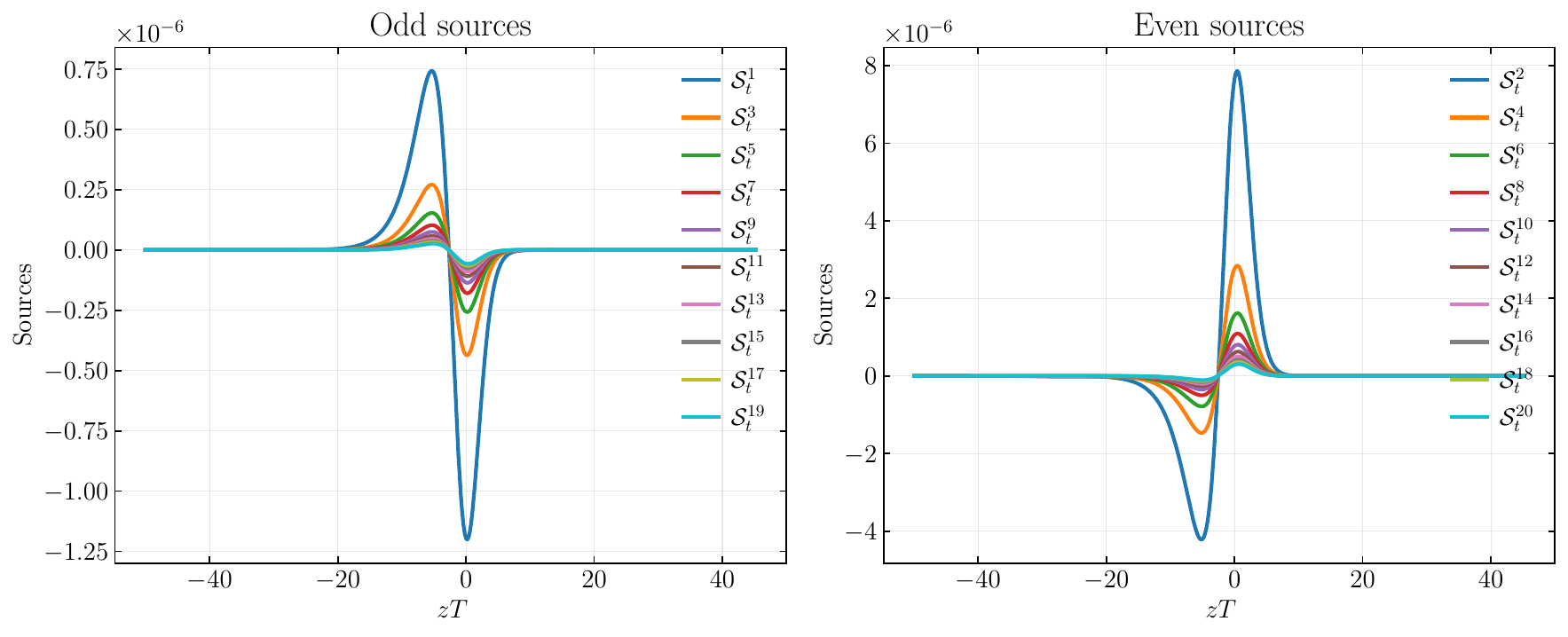}
    \caption{Weighted source multiplets for the top sector as functions of the dimensionless wall coordinate $zT$. The left panel shows the CP–odd sources $\mathcal S^{t}_{\ell}$ for odd $\ell$, while the right panel shows those for even $\ell$. All curves are localised near the interface, and the even sources suffer from velocity suppression in comparison to the
odd ones. This reproduces Fig.~1 of \cite{Kainulainen:2024qpm}.}
    \label{fig:sources}
\end{figure}

The CP–odd sector contains two independent gradient structures, \((m^2\theta')'\) and \(m^2(m^2)'\theta'\). Their weights involve either \(E_{0z}^{-1}\) or an extra power of \(E_0^{-1}\). It is convenient to define
\begin{equation}
  Q_\ell^{\,8o} \equiv \Big\langle \frac{s\,p_z^{\ell-1}}{2E_0^\ell E_{0z}}\,f'_{0w}\Big\rangle,
  \qquad
  Q_\ell^{\,9o} \equiv \Big\langle \frac{s\,p_z^{\ell-1}}{4E_0^{\ell+1}E_{0z}}
  \Big(\frac{f'_{0w}}{E_0}-\gamma_w f''_{0w}\Big)\Big\rangle,
\end{equation}
where the spin label \(s=\pm1\) has been retained explicitly to keep track of the helicity dependence that accompanies the CP–violating energy shift. With these definitions the averaged CP–odd source becomes
\begin{equation}
  \mathcal S_\ell^{\,o}(z)
  =
  -v_w\gamma_w
  \Big[
    (m^2\theta')'Q_\ell^{8o}
    - m^2(m^2)'\theta'Q_\ell^{9o}
  \Big],
\end{equation}
which is nothing but the weighted version of the two CP–odd operators identified earlier. Figure~\ref{fig:sources} visualises the moment–projected CP-odd sources for the top sector, using as a benchmark the \texttt{xSM} model defined below with parameters as in \eqref{eq:BM1}.

\subsection{Dimensionless rescaling and universal kernels}\label{subsec:dimensionless}
To make explicit that the whole framework is temperature–independent—i.e.\ all quantities can be expressed in units of the temperature—it is useful to track how every object scales with \(T\). We measure dimensions in powers of \(T\) and denote this with square brackets \([\,\cdot\,]\).

The starting point is the boosted equilibrium distribution \(f_{0w}\), whose full argument is \(\beta\,\gamma_w (E_0+v_w p_z)\) with \(\beta=T^{-1}\). Derivatives with respect to this argument, as in Sec.~\ref{sec:boltzmann}, therefore, bring down inverse powers of \(T\)
\begin{equation}
    [f_{0w}^{(n)}] = T^{-n}\, .
\end{equation}
From this, the momentum averages used throughout inherit the scalings
\begin{equation}
    [\langle \cdot \rangle] = T, \qquad [[\cdot]] = T^0.
\end{equation}
Consequently, the thermal kernels and kinematic factors appearing in the fluid equations scale as
\begin{align}
    [D_\ell] &= T^0, \qquad [Q_\ell] = T^{-2}, \qquad [R] = T^0, \qquad [\bar{R}] = T^{-2}, \\
    [Q_\ell^e] &= T^{-1}, \qquad [Q_\ell^{8o}] = T^{-2}, \qquad [Q_\ell^{9o}] = T^{-4}, \qquad [K^{(\ell)}] = T^1.
\end{align}
The background profiles contribute gradients with the expected dimensions,
\begin{equation}
    [(m^2)^{(n)}] = T^{2+n}, \qquad [\theta^{(n)}] = T^n.
\end{equation}
With these assignments, the natural dimensionless variables are the reduced chemical potential \(\xi \equiv \mu/T\) and the rescaled wall coordinate \(\tilde z \equiv z\,T\). Written in terms of \(\xi(\tilde z)\), \(x(\tilde z)\equiv |m|/T\), and \(\theta(\tilde z)\), both Eqs.~\eqref{eq:fluid-eq-1} and \eqref{eq:fluid-eq-2} carry an overall and uniform factor \(T^2\). Dividing through by this factor yields a fully dimensionless system in which all coefficients are universal functions of \((x,v_w)\), and any explicit temperature–dependence has been scaled out. This normalisation is not only conceptually clean—cleanly separating physics from units—but also numerically advantageous, as the universal kernels can be pre–tabulated once and interpolated across models.

\section{Phenomenological models}
\label{sec:pheno}

We adopt a minimal, phenomenological setup that captures the structure of EWBG driven by top–Yukawa dynamics without committing to a specific ultraviolet completion, as has been done in \cite{Cline:2020jre, Kainulainen:2024qpm}. The goal is to have a concrete yet model–agnostic playground where the collision operator, the choice of tracked species, and the interplay of fast interactions can be displayed transparently.

\medskip
We keep the four degrees of freedom most tightly coupled by the top Yukawa:
left– and right–handed tops \((t_L,t_R)\), the left–handed bottom \(b_L\), and the Higgs \(h\).
Each species is described by the two–component fluid vector
$
w_i(z)\equiv\begin{pmatrix}\mu_i(z) & u_i(z)\end{pmatrix}$ where $
i\in\{t_L,\;b_L,\;t_R,\;h\},
$
and \(\mu_i\) is the chemical potential and \(u_i\) the longitudinal velocity moment.
With the universal kernels defined earlier, the dynamics is written compactly as the linear, first–order system as in \eqref{eq: diffusion system matrix} with block \(2\times2\) matrices \(A,B\), source vector \(\mathcal S\), and linearised collision vector \(\delta\mathcal C\) specified by the physics below.

\paragraph{Interactions that control equilibration.}
The collision sector includes precisely those processes that efficiently couple \(\{t_L,b_L,t_R,h\}\) in the symmetric phase:
(i) charged–current \(W\) scatterings, which drive \(\mu_{t_L}\simeq\mu_{b_L}\);
(ii) QCD strong sphalerons, which rapidly equilibrate axial quark charge across flavours;
(iii) top helicity flips (``mass insertions''), which damp \(\mu_{t_L}-\mu_{t_R}\) and feed the scalar channel;
(iv) Higgs–number damping in the scalar sector; and
(v) Yukawa exchange among \((t_L,b_L,t_R,h)\).
These ingredients enter only through the components of \(\delta\mathcal C\).

For each species \(i\) we use the general form as in Eq. \eqref{eq: linearised collision}, but with $\ell=0,1$. The first components \(\delta\mathcal C_{1}^a\) are linear in the chemical potentials and contain the rates for the fast
processes listed above. Explicitly, as stated in \cite{Cline:2020jre, Kainulainen:2024qpm}, for the four-species set\footnote{%
The numerical prefactors in Eqs.~\eqref{eq:pheno collision 1}--\eqref{eq:pheno collision 4}
arise from simple counting arguments.
The factor $2$ in front of $\widehat\Gamma_m$ reflects that helicity--flip
processes $t_L\leftrightarrow t_R$ act identically on particles and antiparticles;
when written in terms of charge--asymmetry chemical potentials, both contributions
add, doubling the effective rate.
The factor $3/2$ in the Higgs equation originates from the ratio of internal
degrees of freedom, $N_c/N_h=3/2$, since three quark colours feed a single Higgs
doublet counted as two components.
No additional prefactors appear for $\widehat\Gamma_y$ or $\widehat\Gamma_W$,
because those rates are already defined per particle and include all relevant
multiplicity and symmetry factors in their averaged definitions.%
}

\begin{align}
\delta \mathcal C_1^{t_L} &\propto \widehat\Gamma_y^t \big(\mu_{t_L}-\mu_{t_R}+\mu_h\big)
                            + 2\widehat\Gamma_m^t \big(\mu_{t_L}-\mu_{t_R}\big)
                            + \widehat\Gamma_W \big(\mu_{t_L}-\mu_{b_L}\big)
                            + \widetilde\Gamma_{ss}[\mu],
\label{eq:pheno collision 1}\\[2pt]
\delta \mathcal C_1^{b_L} &\propto \widehat\Gamma_y^t \big(\mu_{b_L}-\mu_{t_R}+\mu_h\big)
                            + \widehat\Gamma_W \big(\mu_{b_L}-\mu_{t_L}\big)
                            + \widetilde\Gamma_{ss}[\mu],
\label{eq:pheno collision 2}\\[2pt]
\delta \mathcal C_1^{t_R} &\propto -\,\widehat\Gamma_y^t \big(\mu_{t_L}+\mu_{b_L}-2\mu_{t_R}+2\mu_h\big)
                            + 2\widehat\Gamma_m^t \big(\mu_{t_R}-\mu_{t_L}\big)
                            - \widetilde\Gamma_{ss}[\mu],
\label{eq:pheno collision 3}\\[2pt]
\delta \mathcal C_1^{h}   &\propto {3\over 2}\widehat\Gamma_y \big(\mu_{t_L}+\mu_{b_L}-2\mu_{t_R}+2\mu_h\big)
                            + \widehat\Gamma_m^h\,\mu_h, 
\label{eq:pheno collision 4}
\end{align}
where \(\widetilde\Gamma_{ss}[\mu]\equiv \Gamma_{ss}\,\mu_{ss}\) with $\mu_{ss}$ defined in Eq. \eqref{eq:SS-functional-heavy}. We recall that $\widehat\Gamma$ are the averaged per-particle rate, and they have dimension $T$, see appendices \ref{app:rate Gammay} and \ref{app:rate GammaM} for full details. We can now stack the four two–component vectors into a single column \(U^T\equiv\big(w_{t_L}^T,w_{b_L}^T,w_{t_R}^T,w_h^T\big)\)
then Eq.~\eqref{eq: diffusion system matrix} applies with
\(A=\mathrm{diag}(A_{t_L},A_{b_L},A_{t_R},A_h)\) block–diagonal,
while \(\mathcal S\) collects the CP–even/odd sources and \(\delta\mathcal C\) the linearised collisions terms just specified.
Far from the wall, where profiles are constant and sources vanish, the boundary conditions are \(U(\pm\infty)=0\).
This closes a compact four–species, two–moment network that faithfully represents the EWBG physics
we aim to test, independent of the ultraviolet details of the CP–violating model.

\subsection{Baryon–left chemical potential \texorpdfstring{$\mu_{B_L}(z)$}{mu_{B_L}(z)}}

Electroweak sphalerons couple to the \emph{left–handed} quark number carried by the $SU(2)_L$ doublets. In the fluid description, this enters through a single linear combination of chemical potentials, the \emph{baryon–left} chemical potential $\mu_{B_L}(z)$. It is the object that sources baryon number in the weak–sphaleron sector (see \cite{PhysRevLett.37.8,1c93e54e-e5c7-3fcb-8a8e-0a3527e35a39}), schematically
\begin{equation}
\frac{d\eta_B}{dz} \;\propto\; \Gamma_{ws}(z)\,\Big(\text{const}\times \mu_{B_L}(z) - A\,\eta_B(z)\Big),
\end{equation}
so once the transport system is solved, $\mu_{B_L}(z)$ is the only input from the charge sector needed by the baryon equation.

By definition,
\begin{equation}
  \mu_{B_L}(z)\equiv\frac{1}{2}\sum_{\text{all quarks }q}\,\mu_{q_L}(z)\,.
\end{equation}
Using the light–quark elimination derived in~\eqref{eq:muqL-solution}, the sum can be written entirely in terms of third–generation variables with the universal thermal kernels \(D_0^{a}\), defined in \eqref{eq: D and Q}, see \cite{Cline:2020jre, Kainulainen:2024qpm},
\begin{tcolorbox}[colback=gray!8,colframe=black,title={Baryon-left chemical potential}]
\begin{equation}
\label{eq:muBL-final}
  \mu_{B_L}(z)
  =
  \dfrac12\!\Big(1+4\,D_0^{t}\Big)\mu_{t_L}(z)
  +
  \dfrac12\!\Big(1+4\,D_0^{b}\Big)\mu_{b_L}(z)
  +
  2\,D_0^{t}\,\mu_{t_R}(z)\,.
\end{equation}
\end{tcolorbox}
\paragraph{Different conventions in FH04 vs.\ FH06.}
The two formalisms define the baryon–left chemical potential in different species bases and with different transport kernels. 
In \textbf{FH04} the diffusion network tracks the third–generation quark doublet $q_3$ and the right–handed top (denoted by its left–handed conjugate $t^c$). The baryon–left combination is
\begin{equation}
\textbf{FH04:}\qquad
\mu_{B_L}(z)
=\frac{1}{2}\Big(1+2\,\kappa_t+\kappa_b\Big)\,\mu_{q_3}(z)
-\frac{1}{2}\,\kappa_t\,\mu_{t^c}(z)\,,
\label{eq:muBL_FH04}
\end{equation}
where the $\kappa_i$ are the standard transport coefficients of \cite{Bodeker:2004ws}.

In \textbf{FH06}, the tracked species are the explicit chiral fields $t_L$, $b_L$, $t_R$ (written as $t^c$ for the left–handed conjugate), and the Higgs $h$ (the subscript $L$ is omitted in their notation). The corresponding definition reads
\begin{equation}
\textbf{FH06:}\qquad
\mu_{B_L}(z)
=\frac{1}{2}\Big(1+4\,K_1^t\Big)\,\mu_t(z)
+\frac{1}{2}\Big(1+4\,K_1^b\Big)\,\mu_b(z)
-2\,K_1^t\,\mu_{t^c}(z)\,,
\label{eq:muBL_FH06}
\end{equation}
with $K_1^a$ the transport kernels of \cite{Fromme:2006wx}. We remind the reader that $\mu_{t_R}=-\mu_{t^c}$ by definition of the conjugated field. 
Equations \eqref{eq:muBL_FH04} and \eqref{eq:muBL_FH06} are equivalent after translating $(\mu_{q_3},\kappa_i)$ into $(\mu_{t},\mu_{b},K_1^a)$ according to the species basis used in each formalism.

\section{Baryon asymmetry}

\begin{WideBox}[label={box:2step}]{Computation of the baryon asymmetry}
The computation of the baryon asymmetry can be approached in two conceptually different ways, depending on how the weak sphaleron processes are treated within the transport system, see \cite{deVries:2017ncy,DeVries:2018aul}.

\medskip
\textbf{Two-step approach (standard method).}
\begin{enumerate}
  \item \emph{Solve the fluid network.}  
  One first determines the steady-state profiles of all particle densities (or chemical potentials) by solving the set of drift–diffusion–reaction equations,
  \[
  0 \;=\; D_i\,n_i''(z) - v_w\,n_i'(z) - \sum_j \Gamma_{ij}\,[\text{linear combinations of }\mu_j(z)] + \mathcal S_i(z),
  \]
  with boundary conditions $n_i(\pm\infty)=0$.  
  In this step, the weak sphaleron processes are neglected under the assumption that they are the slowest reactions in the system.
  \item \emph{Compute the baryon asymmetry.}  
  Once the fluid network is solved, the baryon asymmetry $n_B(z)$ is obtained by solving a separate diffusion equation,
  \[
  0 \;=\; D_B\,n_B''(z) - v_w\,n_B'(z)
  - \Gamma_{ws}(z)\!\left[\mathcal{C}_L\,\frac{\mu_L(z)}{T}
  + \kappa_B\,\frac{n_B(z)}{T^2}\right],
  \]
  with $n_B(+\infty)=0$.  
  The final baryon-to-entropy ratio is extracted in the broken phase as $\eta_B = n_B(z\!\to\!-\infty)/s$.
\end{enumerate}

\medskip
\textbf{One-step approach (fully coupled method)}.  
In this alternative formulation, the baryon density is included as a dynamical variable inside the fluid network itself.  
The system of equations becomes
\[
\begin{cases}
0 = D_i\,n_i'' - v_w\,n_i' - \sum_j \Gamma_{ij}(\mu)\;+\;S_i(z),\\[0.3em]
0 = D_B\,n_B'' - v_w\,n_B' - \Gamma_{ws}(z)\!\left(\mathcal{C}_L \dfrac{\mu_L}{T} + \kappa_B\dfrac{n_B}{T^2}\right),
\end{cases}
\]
In this case, the production and diffusion of the baryon asymmetry are solved \emph{simultaneously} with all other species, rather than in a post-processing step.  
This fully coupled treatment is more demanding numerically but provides a self-consistent picture when the weak sphaleron rate is not hierarchically slower than the other reactions.

\medskip
In this review, as in most of the literature, we adopt the two-step approach.

\end{WideBox}

\label{sec:BAU}
During a first–order electroweak phase transition, bubbles of the broken phase nucleate and sweep through the symmetric plasma. We model a single planar wall moving along $z$ with velocity $v_w$ and width $L_w$, with the broken phase at $z<0$ and the symmetric phase at $z>0$. CP–violating interactions in the wall background create a chiral excess in front of the wall; weak sphalerons convert the left–handed baryon charge carried by quarks into net baryon number. The quantity that couples to weak sphalerons is the baryon–left chemical potential $\mu_{B_L}(z)$ defined earlier as a thermal–weight combination of the heavy–flavour potentials; only its CP–odd part enters.

We treat weak sphalerons as the slowest process in the problem. Accordingly, we first solve the diffusion/transport network for all non-baryonic species (with $\Gamma_{ws}$ switched off) to obtain $\mu_{B_L}(z)$, and only in a second step do we evolve the baryon number using the sphaleron evolution equation to compute $\eta_B$. The rationale and accuracy of this two–step procedure are summarised in Box~\ref{box:2step}. A convenient starting point is the rate equation written in terms of the Chern–Simons (CS) chemical potentials, see \cite{Cline:2000nw, deVries:2017ncy, Bodeker:2004ws, Fromme:2006cm, Cline:2020jre,Kainulainen:2024qpm},
\begin{equation}
\frac{dn_B}{dt}=n_f\Gamma_{ws}\big(-\mu_{\rm CS}+\mu^{0}_{\rm CS}\big),
\end{equation}
with $n_f=3$ are the fermion families, $\Gamma_{ws}\sim\kappa\alpha_W^5 T$ is the weak sphaleron rate in the symmetric phase, and where $\mu_{\rm CS}$ is the response of the plasma to a change in CS number while $\mu^{0}_{\rm CS}$ encodes the chiral charge bias produced by transport. Under the standard hierarchy of rates (sphalerons are the slowest processes in the diffusion network) one can relate $\mu_{\rm CS}\propto\,n_B$ and express $\mu^{0}_{\rm CS}$ as a linear combination of left–handed quark and lepton chemical potentials; recasting the susceptibilities in terms of $\mu_{B_L}$ leads to the compact form used in transport studies,
\begin{equation}
\label{eq:nb_time}
\frac{dn_B}{dt}=-\frac{n_f}{2}\Gamma_{ws}\big(N_c\,\mu_{B_L}\,T^2-A\,n_B\big),
\end{equation}
with $N_c=3$ the number of colours and $A=15/2$. \footnote{A derivation of the coefficient \(A\), together with a discussion of its relation to the reduced transport-network closure, is given in Appendix~\ref{app:A_coeff}.}

Passing to the wall frame and seeking a steady profile, $\partial_t\to -v_w\gamma_w\partial_z$, gives\footnote{In the standard EWBG reduction the baryon–left potential $\mu_{B_L}(z)$ is obtained from a genuine diffusion network for the microscopic species (quarks and Higgs), while the baryon density $n_B(z)$ is evolved by local weak–sphaleron kinetics as in \eqref{eq:nb_time}. A separate diffusion term for $n_B$ (e.g.\ $-\partial_z\!\left[D_B\,\partial_z n_B\right]$) is omitted because: (i) $n_B$, in this two step approach, has no independent carrier/current—its spatial transport is already encoded upstream in $\mu_{B_L}(z)$, so adding $D_B\,n_B''$ would double–count; (ii) the regime is reaction–limited—fast interactions set $\mu_{B_L}$ on wall scales while weak sphalerons act slowly and locally—so a first–order balance in $z$ is the correct leading description; (iii) inside the broken phase $\Gamma_{ws}\!\to 0$, hence diffusion only reshuffles $n_B$ without changing the frozen–in asymmetry $\eta_B$. If one insists on including diffusion consistently, its effect is higher order and negligible when $\Gamma_{ws}\gg D_B/L^2$ over the active layer of width $L$.}
\begin{equation}
\label{eq:baryon_density_z}
\frac{dn_B}{dz}=\frac{n_f}{2v_w\gamma_w}\Gamma_{ws}(z)\Big(N_c\,\mu_{B_L}(z)\,T^2-A\,n_B(z)\Big).
\end{equation}
Sphalerons are suppressed inside the broken phase and active in front of the wall, their profile across the wall is discussed in the next subsection.
\subsection{Weak sphaleron rate profile}
The weak sphaleron rate, $\Gamma_{ws}(z)$, determines the efficiency of baryon number violation across the electroweak bubble wall. Its spatial dependence reflects the rapid change of the Higgs background $h(z)$ between the symmetric and the broken phase.

\medskip
\noindent
\textbf{Symmetric phase.}
In front of the wall, where $h(z)\simeq 0$, the sphaleron transitions are unsuppressed and the rate can be taken as a constant,
\begin{equation}
\Gamma_{ws}(z)\simeq \text{const} \times T\,, \qquad (z\gg0),
\end{equation}
with $\Gamma_{ws}$ obtained from lattice simulations of hot $SU(2)$ gauge theory, including fermionic effects.
The updated value is $\Gamma_{ws}\simeq 8\times10^{-7}\,T$
(see Appendix~\ref{app:sphalerons} for details).

\medskip
\noindent
\textbf{Broken phase.}
Behind the wall, where $h(z)$ grows towards its vacuum expectation value, sphaleron transitions are exponentially suppressed because baryon number violation requires overcoming the sphaleron energy barrier $E_{\rm sph}(T)$. The general expression reads
\begin{equation}
\Gamma_{ws}^{\rm br}(z)\;=\; \kappa\,T^4\,e^{-E_{\rm sph}(z)/T},
\end{equation}
where the exponent is controlled by the classical sphaleron energy~\cite{KLINKHAMER1991245},
\begin{equation}
E_{\rm sph}(T)\;=\;\frac{4\pi}{g}\,B\!\left(\frac{\lambda}{g^2}\right)\,h(z),
\end{equation}
where $B(\lambda/g^2)$ is a dimensionless coefficient that parametrises the energy of the static $SU(2)$ sphaleron solution, obtained from the classical field equations of the electroweak theory for a given ratio $\lambda/g^2$, see~\cite{KLINKHAMER1991245}. For Standard Model parameters ($g\simeq0.65$, $\lambda/g^2\simeq0.3$), the numerical factor
\[
\frac{4\pi}{g}\,B\!\left(\frac{\lambda}{g^2}\right)\simeq\;36.57 \simeq 37,
\]
where we used $B\!\left({\lambda}/{g^2}\right)\simeq 1.89$, and this value has been extracted by interpolating the values that appear in Table II of \cite{KLINKHAMER1991245}. The broken-phase suppression, $E_{\rm sph}(z)/T\simeq 37\,h(z)/T$, thus follows the approximate form $e^{-37\,h(z)/T}$.

\medskip
\noindent
\textbf{Interpolating profile and normalisation.}
To connect the two regimes smoothly, it is convenient to define a dimensionless profile function $f_{\rm sph}(z)$ such that
\begin{equation}
\Gamma_{ws}(z)\;=\;\Gamma_{\rm sph}\,f_{\rm sph}(z), \qquad
f_{\rm sph}(z)\;=\;\min\!\left[1,\,\frac{N\,T}{\Gamma_{\rm sph}}\,
e^{-37\,h(z)/T}\right].
\label{eq:f_sph_profile}
\end{equation}
The prefactor $N$ is fixed phenomenologically by imposing that sphaleron transitions in the broken phase decouple when $v(T_c)/T_c\simeq1$, i.e.
$\Gamma_{ws}^{\rm br}(T_c)\simeq H(T_c)$, with
$H(T)=1.66\sqrt{g_*}\,T^2/M_{\rm Pl}$ the Hubble rate in the radiation era~\cite{Bodeker:2020ghk}.
Taking $g_*\simeq106.75$, $T_c\simeq150~\mathrm{GeV}$, and $M_{\rm Pl}=1.22\times10^{19}~\mathrm{GeV}$ gives
\[
H(T_c)\;\simeq\;1.4\times10^{-18}\,T_c^2\ \text{GeV}^{-1}
\;\simeq\;2.1\times10^{-16}T_c \quad \longrightarrow \qquad N=\frac{H(T_c)/T_c}{e^{-37}} \simeq 1.7 \ .
\]
Equation~\eqref{eq:f_sph_profile} therefore reproduces the standard phenomenological profile,
\begin{equation}
\Gamma_{ws}(z)\;=\;\Gamma_{\rm sph}\,
\min\!\left(1,\frac{1.7\,T}{\Gamma_{\rm sph}}\,
e^{-37\,h(z)/T}\right),
\end{equation}
used in baryogenesis transport equations\footnote{The factor $2.4$ advocated in \cite{Cline:2020jre} might come from the assumption that $H(T_c)/T_c\simeq 10^{-17}$, as explicitly stated in \cite{Bodeker:2020ghk}, and taking the exponent to be $40$, so that $N\simeq 2.4$.}.

\medskip
\noindent
\textbf{Remark on the sphaleron rate below $T_c$.}
The simple parametrisation $\Gamma_{ws}\propto e^{-37\,h/T}$ provides a good approximation to the broken-phase suppression for sufficiently strong transitions, where sphaleron processes are fully frozen. 
Recent analyses within the dimensionally reduced three-dimensional $SU(2)$ gauge–Higgs effective theory, see~\cite{Li:2025kyo}, show that the sphaleron action can be expressed as $\hat S_{3D}=C_{\rm sph}\,v_3(x,y)$ with $C_{\rm sph}\simeq29$, where $v_3$ is the dimensionless Higgs expectation value in the 3D EFT. 
For Standard Model--like couplings and strong transitions, $v_3/(h/T)\simeq1.2$–$1.3$, yielding an effective exponent close to $35$–$38$.
Hence, the profile $e^{-37\,h/T}$ remains numerically consistent with the most recent 3D computation in the ``full-freezing'' regime. More details are given in Appendix~\ref{app:sphaleron3D}.

\subsection{Solving for the BAU}
Now we turn in the solution of the Equation \eqref{eq:baryon_density_z}. Since it is linear, $n_B'(z)+p(z)n_B(z)=q(z)$, with $p(z)=\frac{n_fA}{2v_w\gamma_w}\Gamma_{ws}(z)\ge 0$ and $q(z)=\frac{n_fN_c}{2v_w\gamma_w}\Gamma_{ws}(z)\mu_{B_L}(z)T^2$, introducing the integrating factor $f(z)=\exp\big(\int^z p(z')dz'\big)$, one has $(n_B f)'=q f$. Integrating from $z$ to $+\infty$ yields the exact solution
\begin{align}
\label{eq:nB_full_IF_correct}
n_B(z)&=n_{B}(+\infty)\exp\left[-\frac{A n_f}{2 v_w \gamma_w}\int_{z}^{+\infty}\Gamma_{ws}(z')dz'\right]\\
&\quad-\int_{z}^{+\infty}dz'\,\frac{n_f N_c}{2 v_w \gamma_w}\Gamma_{ws}(z')\mu_{B_L}(z')T^2\exp\left[-\frac{A n_f}{2 v_w \gamma_w}\int_{z}^{z'}\Gamma_{ws}(z'')dz''\right].\nonumber
\end{align}
In EWBG, one imposes $n_B(+\infty)=0$ (no asymmetry far ahead of the wall), so the first term vanishes and
\begin{equation}
\label{eq:nB_simplified}
n_B(z)=-\int_{z}^{+\infty}dz'\,\frac{n_f N_c}{2 v_w \gamma_w}\Gamma_{ws}(z')\mu_{B_L}(z')T^2\exp\left[-\frac{A n_f}{2 v_w \gamma_w}\int_{z}^{z'}\Gamma_{ws}(z'')dz''\right].
\end{equation}
\begin{WideBox}[label={box:signBAU}]{On the sign of the baryon asymmetry}
The literature rarely states this explicitly, but once \emph{conventions} are fixed, the sign of the predicted baryon asymmetry is no longer arbitrary. Conventions to fix include: (i) wall orientation (e.g.\ symmetric phase on the right, broken on the left, wall moving $+\hat z$), (ii) particle vs.\ antiparticle labelling (so that the CP eigenvalue $\mathrm{CP}=\pm1$ refers to what we call ``particle'' or ``antiparticle''), and (iii) the CP--odd vs.\ CP--even decomposition of the kinetic/transport equations such that the CP--violating phase $\theta_{\rm CP}(z)$ has a definite \emph{shape and sign} for particles. With these choices set, the signs of all CP--odd \emph{sources} are fixed, and the predicted baryon asymmetry must be \emph{positive} to match the observed convention $\eta_B>0$. If, under the same fixed conventions, a calculation yields $\eta_B<0$, it describes an ``antimatter universe'' rather than ours. Flipping a model phase (e.g.\ $\theta\!\to\!-\theta$) flips the BAU sign and is \emph{not} equivalent unless one simultaneously flips the baseline conventions (wall direction and particle/antiparticle labelling) everywhere in the pipeline.
\end{WideBox}

The observable baryon asymmetry is the broken–phase value normalised to the entropy density\footnote{We summarise normalisation conventions used in the literature in Box~\ref{box:normalisationBAU}.} $s=\frac{2\pi^2}{45}g_*T^3$,
\begin{tcolorbox}[colback=gray!8,colframe=black,title={Baryon asymmetry generated}]
\begin{equation}
\eta_B\equiv\frac{n_B(-\infty)}{s}=-\frac{45n_f N_c}{4\pi^2 v_w \gamma_wg_* T}\int_{-\infty}^{+\infty}dz\,\Gamma_{ws}(z)\mu_{B_L}(z)\exp\left[-\frac{A n_f}{2 v_w \gamma_w}\int_{-\infty}^{z}\Gamma_{ws}(u)du\right],
\label{eq: baryon asymmetry}
\end{equation}
\end{tcolorbox}
\noindent with $g_*=106.75$ in the SM. The structure of \eqref{eq: baryon asymmetry} makes the physics transparent: the kernel multiplying $\mu_{B_L}$ is the \emph{source} that converts chiral charge into baryon number where sphalerons are active, while the exponential is the \emph{washout} that erases $n_B$ when the wall is too slow. For sufficiently large $v_w$ the washout integral is small and part of the chiral excess diffuses into the broken phase, where $f_{\rm sph}\ll 1$, freezing in as the final $\eta_B$.
\paragraph{Note.}
With the conventions fixed in Box~\ref{box:signBAU} phase assignments, spatial orientation (broken phase to the left, symmetric to the right), wall motion to $+z$ in the plasma frame (and particle drift to $-z$ in the wall frame), and the operational definition of $\mu_{B_L}$ and charge densities—the \emph{sign} of the baryon asymmetry $\eta_B$ is physical and unambiguous. In particular, the overall minus sign in front of Eq.~\eqref{eq: baryon asymmetry}, which is not found in general in the literature, follows from the correct solution. The same sign is found, for example, in Eq.~(49) of~\cite{deVries:2017ncy}.

\begin{WideBox}[label={box:normalisationBAU}]{Photon vs.\ entropy normalisation of the baryon asymmetry}
\textbf{What is being compared?}  Observations often quote the baryon--to--photon ratio 
\(n_B/n_\gamma \approx 6\times 10^{-10}\) (see \cite{Planck:2015fie}), while in this work we use the entropy–normalised
quantity \(\eta_B\equiv n_B/s\) (see Eq.~\eqref{eq: baryon asymmetry}).  They are related by
\[
\eta_B \;=\; \frac{n_\gamma}{s}\;\frac{n_B}{n_\gamma}\,.
\]
For a relativistic plasma in kinetic equilibrium,
\[
n_\gamma \;=\; \frac{2\,\zeta(3)}{\pi^2}\,T^3,
\qquad
s \;=\; \frac{2\pi^2}{45}\,g_{*s}\,T^3,
 \quad \to \quad
\frac{n_\gamma}{s}
\;=\;
\frac{45\,\zeta(3)}{\pi^4}\,\frac{1}{g_{*s}}
\;\simeq\;
\frac{0.555}{g_{*s}}\,.
\]
After $e^+e^-$ annihilation (from CMB decoupling to today), the effective
entropy degrees of freedom are
\[
g_{*s} \;=\; 2 \;+\; \frac{7}{8}\times 3\times 2 \left(\frac{T_\nu}{T_\gamma}\right)^3
\;=\; \frac{43}{11} \;\simeq\; 3.91,
\]
with \(T_\nu/T_\gamma=(4/11)^{1/3}\).  Hence \(\,n_\gamma/s \simeq 0.555/3.91 \simeq 0.142\). Using \(n_B/n_\gamma \simeq 6.1\times 10^{-10}\) one finds
\[
\eta_B \;\simeq\; \Big(\frac{n_\gamma}{s}\Big)\,\frac{n_B}{n_\gamma}
\;\simeq\; 0.142 \times 6.1\times 10^{-10}
\;\simeq\; 8.7\times 10^{-11}.
\]
This is the standard conversion between the two conventions; \(\eta_B\) remains conserved
under entropy–conserving expansion.
\end{WideBox}

\subsection{Beyond a fixed wall velocity: a first–principles view}

Following the idea introduced by \cite{Kainulainen:2024qpm}—where the concept is formulated and the way it should be computed is described qualitatively—we stress that the true baryon asymmetry is generally \emph{not} captured by $\eta_B(v_w)$ evaluated at a single, fixed wall velocity. Since $\eta_B$ can vary strongly with $v_w$, spatial and temporal inhomogeneities must be accounted for. 
Although this is not implemented in \texttt{BARYONET}, here we attempt to make that qualitative proposal operational by outlining a concrete procedure to compute the relevant quantity (e.g. by averaging $\eta_B(v_w)$ over an explicitly defined wall–velocity distribution $P(v_w)$); this is an illustrative implementation of the qualitative framework of \cite{Kainulainen:2024qpm}.

\medskip
\noindent\textbf{First–principles picture.}
Baryon number is produced locally at the moving wall and then transported or damped in the symmetric phase.  
If $v_n(\mathbf{x},t)$ is the \emph{local normal velocity} of the wall element at position $\mathbf{x}$ on the instantaneous wall surface $\Sigma(t)$, and $\eta_B^{\rm loc}(v_n)$ is the 1D yield per unit volume swept by that element, the global asymmetry is
\begin{equation}
\label{eq:etaB_first_principles_simple}
\eta_B = \frac{1}{V_{\rm tot}}\int_{t_{\rm start}}^{t_{\rm end}}\! dt \int_{\Sigma(t)}\! dA\, v_n(\mathbf{x},t)\,
\eta_B^{\rm loc}\!\big(v_n(\mathbf{x},t)\big),
\end{equation}
where $V_{\rm tot}$ is the total converted volume. Defining the \emph{volume–weighted} distribution of wall speeds,
\begin{equation}
P(v)=\frac{1}{V_{\rm tot}}\int_{t_{\rm start}}^{t_{\rm end}}\! dt \int_{\Sigma(t)}\! dA\, 
v_n(\mathbf{x},t)\,\delta\!\big(v-v_n(\mathbf{x},t)\big),
\qquad \int_0^1\! dv\,P(v)=1,
\end{equation}
Eq.~\eqref{eq:etaB_first_principles_simple} becomes
\begin{equation}
\eta_B = \int_0^1\! dv\, P(v)\,\eta_B^{\rm loc}(v),
\end{equation}
showing that the relevant average is over the \emph{volume swept} by the wall.

\medskip
\noindent\textbf{How to obtain $P(v)$.}
A first–principles way is to simulate the coupled scalar–fluid system,
\(\partial_\mu T^{\mu\nu}=-\partial^\nu h\,\mathcal{F}(\partial h)\) and
\(\Box h+\partial V_{\rm eff}(h,T)/\partial h+\mathcal{F}(\partial h)=0\),
with a chosen equation of state and friction model $\mathcal{F}$.  
Tracking the isosurface \(h(\mathbf{x},t)=h_{\rm mid}\) gives \(v_n=-\partial_t h/|\nabla h|\); sampling it in spacetime yields $P(v)$ directly.

\section{Analytic structure of the diffusion network}
\label{sec:analytic_transport}

Before turning to the numerical implementation, it is useful to extract some analytic information from the transport system itself. Even when the full fluid network cannot be solved in closed form, its linear structure is already restrictive enough to clarify several qualitative features of the profiles: why the chemical potentials are localised near the wall, how their tails are controlled by diffusion and damping, under which conditions they approximately track the CP--violating source, and to what extent one can infer the sign structure of $\mu_{B_L}(z)$ from that of the source. Similar analytic limits have long been used in the EWBG literature to build intuition, both in spontaneous--baryogenesis language and in semiclassical diffusion treatments~\cite{Cohen:1994ss,Huet:1995sh,Cline:2000nw,Fromme:2006wx,Cline:2020jre,Cline:2021dkf,Kainulainen:2024qpm}.

The starting point is the linear first--order system in Eq.~\eqref{eq: diffusion system matrix}. For a network of $N$ species, each described by a two--component vector $w_a=(\mu_a,u_a)^{\mathsf T}$, it is convenient to stack all perturbations into a single $2N$--component vector
\begin{equation}
\Psi(z)\equiv \big(w_1(z),w_2(z),\dots,w_N(z)\big)^{\mathsf T}.
\end{equation}
Because the collision terms are linearised around equilibrium, they can always be written schematically as
\begin{equation}
\delta \mathcal C(z)=-\,\Gamma(z)\,\Psi(z),
\end{equation}
where $\Gamma(z)$ is the matrix of relaxation rates and couplings among species. The full transport network then takes the form
\begin{equation}
\Psi'(z)+\mathbb M(z)\,\Psi(z)=\mathbb J(z),
\label{eq:analytic_general_first_order}
\end{equation}
with
\begin{equation}
\mathbb M(z)\equiv \mathbb A^{-1}(z)\Big[\Gamma(z)+(m^2)'(z)\,\mathbb B(z)\Big],
\qquad
\mathbb J(z)\equiv \mathbb A^{-1}(z)\,\mathbb S(z),
\end{equation}
where $\mathbb A$ and $\mathbb B$ are the block matrices obtained by stacking the species-wise matrices appearing in Eq.~\eqref{eq: diffusion system matrix}. Here and below, $(m^2)'(z)\,\mathbb B(z)$ should be understood as shorthand for the appropriate blockwise multiplication by the background gradients of the masses entering each species equation. Equation~\eqref{eq:analytic_general_first_order} makes explicit that the transport problem is linear and nonlocal: the profiles are determined by a source term $\mathbb J(z)$ acted upon by propagation and damping matrices.

The formal solution of Eq.~\eqref{eq:analytic_general_first_order} is written in terms of the evolution operator,
\begin{equation}
\mathcal U(z,z')=
\mathcal P \exp\!\left[-\int_{z'}^{z}\!d\zeta\,\mathbb M(\zeta)\right],
\end{equation}
as
\begin{equation}
\Psi(z)=\mathcal U(z,z_0)\Psi(z_0)+\int_{z_0}^{z}\!dz'\,\mathcal U(z,z')\,\mathbb J(z').
\label{eq:formal_solution_general}
\end{equation}
Imposing the physical boundary conditions $\Psi(\pm\infty)=0$, one obtains the bounded solution by projecting out the exponentially growing modes on either side of the wall. In practice, one often approximates the coefficients as slowly varying or piecewise constant over the wall region. In such a regime, after diagonalising $\mathbb M=V\Lambda V^{-1}$, the normal modes
\begin{equation}
q(z)\equiv V^{-1}\Psi(z),
\qquad
s(z)\equiv V^{-1}\mathbb J(z),
\end{equation}
satisfy decoupled equations
\begin{equation}
q_\alpha'(z)+\lambda_\alpha\,q_\alpha(z)=s_\alpha(z),
\label{eq:modal_first_order}
\end{equation}
with $\lambda_\alpha$ the eigenvalues of $\mathbb M$. The bounded solution is then a sum of Green--function convolutions,
\begin{equation}
q_\alpha(z)=\int_{-\infty}^{+\infty}\!dz'\,G_\alpha(z-z')\,s_\alpha(z'),
\label{eq:modal_convolution}
\end{equation}
where
\begin{equation}
G_\alpha(\Delta z)=
\begin{cases}
e^{-\lambda_\alpha \Delta z}\,\Theta(\Delta z), & \Re \lambda_\alpha>0,\\[4pt]
-\,e^{-\lambda_\alpha \Delta z}\,\Theta(-\Delta z), & \Re \lambda_\alpha<0.
\end{cases}
\label{eq:modal_kernel}
\end{equation}
Therefore every chemical potential, and in particular $\mu_{B_L}(z)$ defined in Eq.~\eqref{eq:muBL-final}, is a linear combination of nonlocal convolutions of the CP--violating sources. This simple observation is already enough to show that $\mu_{B_L}(z)$ need not inherit the zeros of the microscopic source in a one--to--one manner: the network first projects the source onto transport eigenmodes and then smears each mode over a finite diffusion length.

\subsection{Single effective diffusion equation}

The clearest analytic intuition is obtained by reducing the fluid system to a single effective mode. In the present framework, this is not an arbitrary ansatz: it follows directly from the fact that the fluid network in Eq.~\eqref{eq: diffusion system matrix} is linear and first order in the derivatives of the chemical potential and velocity perturbation. For one effective species, or for one suitably projected transport eigenmode, the two equations may always be written in the form
\begin{align}
	-\,D_1\,\mu'(z)+u'(z)+\alpha_1(z)\,\mu(z)+\beta_1(z)\,u(z) &= S_1(z),
	\label{eq:eff_mode_eq1}\\[0.3em]
	-\,D_2\,\mu'(z)+R\,u'(z)+\alpha_2(z)\,\mu(z)+\beta_2(z)\,u(z) &= S_2(z),
	\label{eq:eff_mode_eq2}
\end{align}
where we have absorbed the linearised collision terms into the coefficients multiplying \(\mu\) and \(u\). More explicitly, this means that \(\alpha_{1,2}\) contain both the drift terms proportional to \(v_w\gamma_w(m^2)'Q_{1,2}\) and the collision terms linear in \(\mu\), while \(\beta_{1,2}\) contain the terms linear in \(u\), including the contribution proportional to \((m^2)'\bar R\) in the second equation. In this notation the source terms \(S_{1,2}\) stand for the CP--violating driving terms after projection onto the chosen effective mode.

The key point is that the two equations are first order in \(\mu'\) and \(u'\), so one may eliminate \(u\) and \(u'\) algebraically. A particularly convenient combination is obtained by subtracting \(R\) times Eq.~\eqref{eq:eff_mode_eq1} from Eq.~\eqref{eq:eff_mode_eq2}, which removes \(u'\) exactly:
\begin{equation}
	\bigl(-D_2+R D_1\bigr)\mu'(z)
	+\bigl(\alpha_2-R\alpha_1\bigr)\mu(z)
	+\bigl(\beta_2-R\beta_1\bigr)u(z)
	=
	S_2(z)-R\,S_1(z).
	\label{eq:u_algebraic_pre}
\end{equation}
Provided \(\beta_2-R\beta_1\neq 0\), this gives an algebraic expression for \(u\),
\begin{equation}
	u(z)=
	\frac{
		S_2(z)-R\,S_1(z)
		+\bigl(D_2-RD_1\bigr)\mu'(z)
		-\bigl(\alpha_2-R\alpha_1\bigr)\mu(z)
	}{
		\beta_2-R\beta_1
	}.
	\label{eq:u_in_terms_of_mu}
\end{equation}
Differentiating Eq.~\eqref{eq:u_in_terms_of_mu} then yields \(u'(z)\) in terms of \(\mu''(z)\), \(\mu'(z)\), \(\mu(z)\), and derivatives of the sources and coefficients. Substituting the result back into Eq.~\eqref{eq:eff_mode_eq1} produces a single second--order equation for \(\mu\). In full generality this equation has \(z\)--dependent coefficients and can be written schematically as
\begin{equation}
	-\,D_{\rm eff}(z)\,\mu''(z)
	+V_{\rm eff}(z)\,\mu'(z)
	+\Gamma_{\rm eff}(z)\,\mu(z)
	=
	S_{\rm eff}(z),
	\label{eq:general_second_order_mu}
\end{equation}
where \(D_{\rm eff}\), \(V_{\rm eff}\), \(\Gamma_{\rm eff}\), and \(S_{\rm eff}\) are determined by the original coefficients in Eqs.~\eqref{eq:eff_mode_eq1}--\eqref{eq:eff_mode_eq2}. Thus the familiar diffusion equation is the reduced form of the linear two--equation fluid system once the velocity perturbation is eliminated.

A particularly transparent limit is obtained when the coefficients vary slowly over the support of the source, or when one studies the asymptotic regions in which they are approximately constant. In that case Eq.~\eqref{eq:general_second_order_mu} reduces to the standard constant--coefficient form
\begin{equation}
	-\,D\,\mu''(z)+\gamma_w v_w\,\mu'(z)+\Gamma\,\mu(z)=S(z),
	\label{eq:single_effective_diffusion}
\end{equation}
where \(D\) is an effective diffusion constant, \(\Gamma\) an effective relaxation rate, and \(S\) an effective source. Equations of this form underlie many semi--analytic discussions of EWBG transport~\cite{Cohen:1994ss,Joyce:1994zt,Huet:1995sh,Cline:2000nw,Fromme:2006wx,Cline:2020jre,Kainulainen:2024qpm}. In the present derivation, however, one sees explicitly that the coefficient of the first derivative is the wall--frame drift term and therefore naturally appears as \(\gamma_w v_w\), rather than only in its low--velocity limit.

For constant \(D\) and \(\Gamma\), with boundary conditions \(\mu(\pm\infty)=0\), Eq.~\eqref{eq:single_effective_diffusion} is solved by a Green's function,
\begin{equation}
	\mu(z)=\int_{-\infty}^{+\infty}\!dz'\,G(z-z')\,S(z'),
	\label{eq:single_mode_convolution}
\end{equation}
with
\begin{equation}
	G(\Delta z)=
	\frac{1}{D(\lambda_+-\lambda_-)}
	\begin{cases}
		e^{\lambda_+ \Delta z}, & \Delta z<0,\\[4pt]
		e^{\lambda_- \Delta z}, & \Delta z>0,
	\end{cases}
	\label{eq:green_single_mode}
\end{equation}
and
\begin{equation}
	\lambda_\pm=
	\frac{\gamma_w v_w\pm\sqrt{\gamma_w^2 v_w^2+4D\Gamma}}{2D},
	\qquad
	\lambda_+>0,\quad \lambda_-<0.
	\label{eq:lambdapm_single_mode}
\end{equation}
The two characteristic lengths
\begin{equation}
	L_-\equiv \lambda_+^{-1},
	\qquad
	L_+\equiv |\lambda_-|^{-1},
\end{equation}
control the decay of the profile into the broken and symmetric phases, respectively. In particular, the symmetric--phase tail --- the one most relevant for baryon production through weak sphalerons --- is broader for small \(\Gamma\) and large \(D\), while increasing \(\gamma_w v_w\) enhances the directional drift and shortens the effective transport tail in front of the wall.

Equation~\eqref{eq:single_mode_convolution} makes the nonlocal nature of transport completely explicit. The profile at a given point is not determined by the local value of the source, but by a weighted average over the source in a neighbourhood whose size is set by \(L_\pm\). In the present context this means that \(\mu_{B_L}(z)\) should be interpreted as a smoothed, shifted response to the CP--violating source rather than as a local image of it.

\subsection{Asymptotic tails and the sign at infinity}

The Green--function solution also gives a precise answer to the question of how the profile approaches zero at large \(|z|\). For a source localised near the wall,
\begin{align}
	\mu(z)\;&\xrightarrow[z\to -\infty]{}\;
	A_-\,e^{\lambda_+ z},
	&
	A_-&=\frac{1}{D(\lambda_+-\lambda_-)}
	\int_{-\infty}^{+\infty}\!dz'\,e^{-\lambda_+ z'}\,S(z'),
	\label{eq:left_tail_sign}
	\\[4pt]
	\mu(z)\;&\xrightarrow[z\to +\infty]{}\;
	A_+\,e^{\lambda_- z},
	&
	A_+&=\frac{1}{D(\lambda_+-\lambda_-)}
	\int_{-\infty}^{+\infty}\!dz'\,e^{-\lambda_- z'}\,S(z').
	\label{eq:right_tail_sign}
\end{align}
Hence the profile goes to zero from above or from below according to the signs of the weighted moments \(A_\pm\). This is an important point: the asymptotic sign is not determined by the sign of the source at the edge of the wall, but by an exponentially weighted integral over the full source profile. In particular, even if the source changes sign inside the wall, the far tails can remain of definite sign; conversely, a source that is locally positive near its maximum need not imply a positive tail if the negative part is sufficiently enhanced by the transport weights.

Applied to the full network, the same statement remains true mode by mode. After diagonalisation, each transport eigenmode has asymptotic tails controlled by the analogue of Eqs.~\eqref{eq:left_tail_sign}--\eqref{eq:right_tail_sign}, and \(\mu_{B_L}(z)\) inherits its sign at infinity from the corresponding weighted sum over those modes. This is the precise sense in which one can discuss whether \(\mu_{B_L}(z)\) approaches zero from above or below.

\subsection{Controlled limits}

Several useful limits of Eq.~\eqref{eq:single_effective_diffusion} can be extracted immediately. It is convenient to characterise them in terms of the source width \(L_S\), keeping the wall--frame drift \(\gamma_w v_w\) explicit throughout.

In the reaction--dominated regime,
\begin{equation}
	\Gamma \gg \frac{\gamma_w v_w}{L_S},
	\qquad
	\Gamma \gg \frac{D}{L_S^2},
	\label{eq:reaction_dominated_cond}
\end{equation}
one may solve Eq.~\eqref{eq:single_effective_diffusion} iteratively in derivatives:
\begin{equation}
	\mu(z)=\frac{S(z)}{\Gamma}
	-\frac{\gamma_w v_w}{\Gamma^2}S'(z)
	+\frac{D}{\Gamma^2}S''(z)
	+\mathcal O(\partial_z^3).
	\label{eq:local_expansion}
\end{equation}
In this limit the chemical potential follows the source almost locally. Therefore the zeros of \(\mu\) are close to those of \(S\), up to derivative corrections. This is the regime in which the intuitive statement that the profile tracks the source is actually justified.

In the drift--dominated regime,
\begin{equation}
	\frac{\gamma_w v_w}{L_S}\gg \Gamma,
	\qquad
	\frac{\gamma_w v_w}{L_S}\gg \frac{D}{L_S^2},
	\label{eq:drift_dominated_cond}
\end{equation}
diffusion becomes subleading and Eq.~\eqref{eq:single_effective_diffusion} reduces approximately to
\begin{equation}
	\gamma_w v_w\,\mu'(z)+\Gamma\,\mu(z)\simeq S(z),
	\label{eq:drift_dominated_eq}
\end{equation}
with solution
\begin{equation}
	\mu(z)\simeq \frac{1}{\gamma_w v_w}\int_{-\infty}^{z}\!dz'\,
	\exp\!\left[-\frac{\Gamma}{\gamma_w v_w}(z-z')\right]\,S(z').
	\label{eq:advection_limit}
\end{equation}
The response is then manifestly one--sided: the profile at \(z\) depends mainly on points upstream along the plasma flow. This limit makes transparent why rapidly moving walls tend to suppress the transport tail and reduce the overlap with the weak--sphaleron region.

In the diffusion--dominated regime, by contrast,
\begin{equation}
	\frac{D}{L_S^2}\gg \Gamma,
	\qquad
	\frac{D}{L_S^2}\gg \frac{\gamma_w v_w}{L_S},
	\label{eq:diffusion_dominated_cond}
\end{equation}
the source is spread over a region much larger than its microscopic width. The profile broadens, its extrema are reduced in magnitude, and the tails are controlled by the diffusion length rather than by the source width. In this regime one should expect substantial smoothing: narrow sign changes in the source are inefficiently transmitted to the chemical potentials.\\

\subsection{Can one predict the number of zeros?}

The previous discussion already suggests the main conclusion: in general there is no theorem stating that the chemical potentials, or $\mu_{B_L}(z)$ in particular, must have the same number of zeros as the CP--violating source. The reason is structural. The transport solution is not a local algebraic function of the source, but a convolution of the source with positive or sign--definite kernels, preceded in the full network by a nontrivial projection onto transport eigenmodes. As a result, zeros of the solution are determined by cancellations among weighted integrals of the source, not by the local condition $S(z)=0$.

Some partial statements are nevertheless possible. If the source has a definite sign and the relevant transport kernel is positive, then the corresponding effective mode has a definite sign as well, so no zero is generated dynamically. If the source changes sign once, the solution may also change sign, but the zero is generally shifted relative to the source zero and can even disappear if one sign dominates after convolution. In the local limit of Eq.~\eqref{eq:local_expansion}, the zeros approximately coincide with those of the source. Away from that limit, however, the profile is smoothed and the relation is lost. In particular, for the full network entering Eq.~\eqref{eq:muBL-final}, the number of zeros of $\mu_{B_L}(z)$ is not fixed solely by the number of zeros of the microscopic CP--violating source.

For the purpose of baryogenesis, the practically relevant information is often weaker but sufficient: the sign of $\mu_{B_L}(z)$ in the symmetric--phase tail and the overall overlap with the weak--sphaleron kernel. The formulas above show that both are controlled by weighted integrals of the source rather than by pointwise properties. This is one reason why two source profiles with similar extrema but different sign structure can lead to very different values of $\eta_B$ once transport is solved.

The main lesson of this section is therefore simple. Even when the complete fluid network must be solved numerically, its analytic structure already tells us that the transport profiles are nonlocal, smoothed responses to the CP--violating source. In special limits the chemical potentials can closely track the source, but in general the mapping between source zeros and profile zeros is not one--to--one. This should be kept in mind when interpreting the numerical solutions presented below.

With this analytic understanding in place, we now turn to the numerical solution of the transport system and to the presentation of the \texttt{BARYONET} code.

\section{Numerical implementation in \texttt{BARYONET}}
\label{sec:Code}

\begin{figure}[t!]
\resizebox{\textwidth}{!}{%
\begin{tikzpicture}[
    >=latex,
    thick,
    node distance=2cm and 2cm,
    block/.style={draw, rounded corners, align=center,
                  minimum width=3cm, minimum height=1.2cm,
                  font=\large},
    layerlabel/.style={font=\large\bfseries},
    arrow/.style={->, thick}
]



\node[block] (solver) at (1,1)
  {\textbf{\color{red}{Transport} \color{blue}{solvers}} in \large \texttt{transport\_solver/} \\[1pt]
   FH04, FH06, CK, KV24 \\[1pt]
   mapping $u\!\in[-1,1] \leftrightarrow z$\\[1pt]
   system for $w_a(z)=(\mu_a(z) \ \ u_{a,\ell}(z))$};
   
\node[block, above left=0.4cm and -3cm of solver] (model)
  {\textbf{\color{red}{Models}} in \texttt{models/} \\[1pt]
   \large \texttt{xSM}, \texttt{xSMv2}, \texttt{Higgs6Model}, \texttt{TwoHDM}, ... \\[1pt]
   provide $h(z), \Theta(z)$, $m_a(z)$ and their derivative};

\node[block, above right=0.4cm and -3cm of solver] (interp)
  {\textbf{\color{red}{Transport functions}} in \\[1pt] \large \texttt{function\_interpolators/} \\[1pt]
   $D_\ell(x),\ K_\ell(x),\ Q_\ell(x, v_w),\ Q_\ell^{8o/9o}(x, v_w), \ \bar R(x, v_w), \ \dots$};

\node[block, right=0.5cm of solver] (symbolic)
  {\textbf{\color{blue}{Symbolic system}} in\\[1pt] \texttt{transport\_solver/Symbolic\_system/} \\[1pt]
symbolic ODE systems};

\node[block, left=0.5cm of solver] (source)
  {\textbf{\color{blue}{WKB Sources}} in\\[1pt] \texttt{transport\_solver/Sources.py} \\[1pt]
$\mathcal{S}_o\propto m'(z), \Theta'(z),~\dots$};


\node[block, below=0.4cm of solver] (bvp)
  {\textbf{\color{blue}{BVP solution}} \\[1pt]
   \large \texttt{scipy.solve\_bvp} + Lagrange multipliers \\[1pt]
   provide chemical potentials \& \\[1pt] velocity perturbations};

\node[block, below=0.4cm of bvp] (bau)
  {\textbf{\color{blue}{Baryon asymmetry}} in \\[1pt]
   \large \texttt{Baryon\_Asymmetry/baryon\_asymmetry.py} \\[1pt]
   compute $\mu_{B_L}(z)$, $\eta_B$, $\eta_B(z)$};

\node[block, left=1cm of bau] (plots)
  {\textbf{\color{forestgreen}{Plots}} \& \textbf{\color{forestgreen}{diagnostics}} \\[1pt]
   \large \texttt{utils/plotting\_complete.py} \\[1pt]
   wall profiles, $\mu_{a}(z)$, $\mu_{B_L}(z)$, $\eta_B(z)$};

\node[block, right=1cm of bau] (result)
  {\textbf{\color{forestgreen}{Final results}} $\eta_B$ vs.\ $\eta_B^{\mathrm{obs}}$, \\[1pt]
   parameter scans, model viability};

\node[block, below=0.5cm of bau] (literature)
  {\textbf{\color{forestgreen}{Literature validation}} in
  \texttt{literature/}\\[1pt]
  \texttt{hep-ph/0412366}, \texttt{hep-ph/0604159},\\[1pt] \texttt{arXiv:1110.2876}, \texttt{arXiv:2001.00568}, \\[1pt] \texttt{arXiv:2407.13639}, $\dots$};


\draw[arrow] (model) -| (solver);
\draw[arrow] (interp) -| (solver);
\draw[arrow] (solver) -- (bvp);
\draw[arrow] (solver)  -- (symbolic);
\draw[arrow] (solver)  -- (source);
\draw[arrow] (bvp)    -- (bau);
\draw[arrow] (bau)    -- (result);
\draw[arrow] (bau)  -- (literature);
\draw[arrow] (result)  |- (literature);

\draw[arrow] (bau)  -- (plots);
\draw[arrow] (plots)  |- (literature);

\end{tikzpicture}%
}
\caption{\textbf{Workflow overview with three conceptual layers.}
\textcolor{red}{\textbf{Physics:}} the only model-specific physics input enters through the definition of the particle-physics model (field profiles and masses) and, to a lesser extent, through the structure of the transport system via \emph{model-dependent interaction rates} that feed the source terms and collision operators.
\textcolor{blue}{\textbf{Numerics:}} given these inputs, the computation proceeds through a modular numerical pipeline: WKB sources and interpolated transport functions are assembled into transport solvers (optionally generated from a symbolic ODE system) and the resulting boundary-value problem is solved (e.g.\ with \texttt{scipy.solve\_bvp} and Lagrange multipliers) to obtain chemical potentials and velocity perturbations.
\textcolor{forestgreen}{\textbf{Outputs:}} the solutions are post-processed into diagnostics and wall-profile plots, the baryon asymmetry $\eta_B$, literature cross-checks, and final deliverables such as parameter scans and model viability summaries.}

\label{fig:structure_code}
\end{figure}

\begin{figure}[t]
  \centering
  \includegraphics[width=\linewidth]{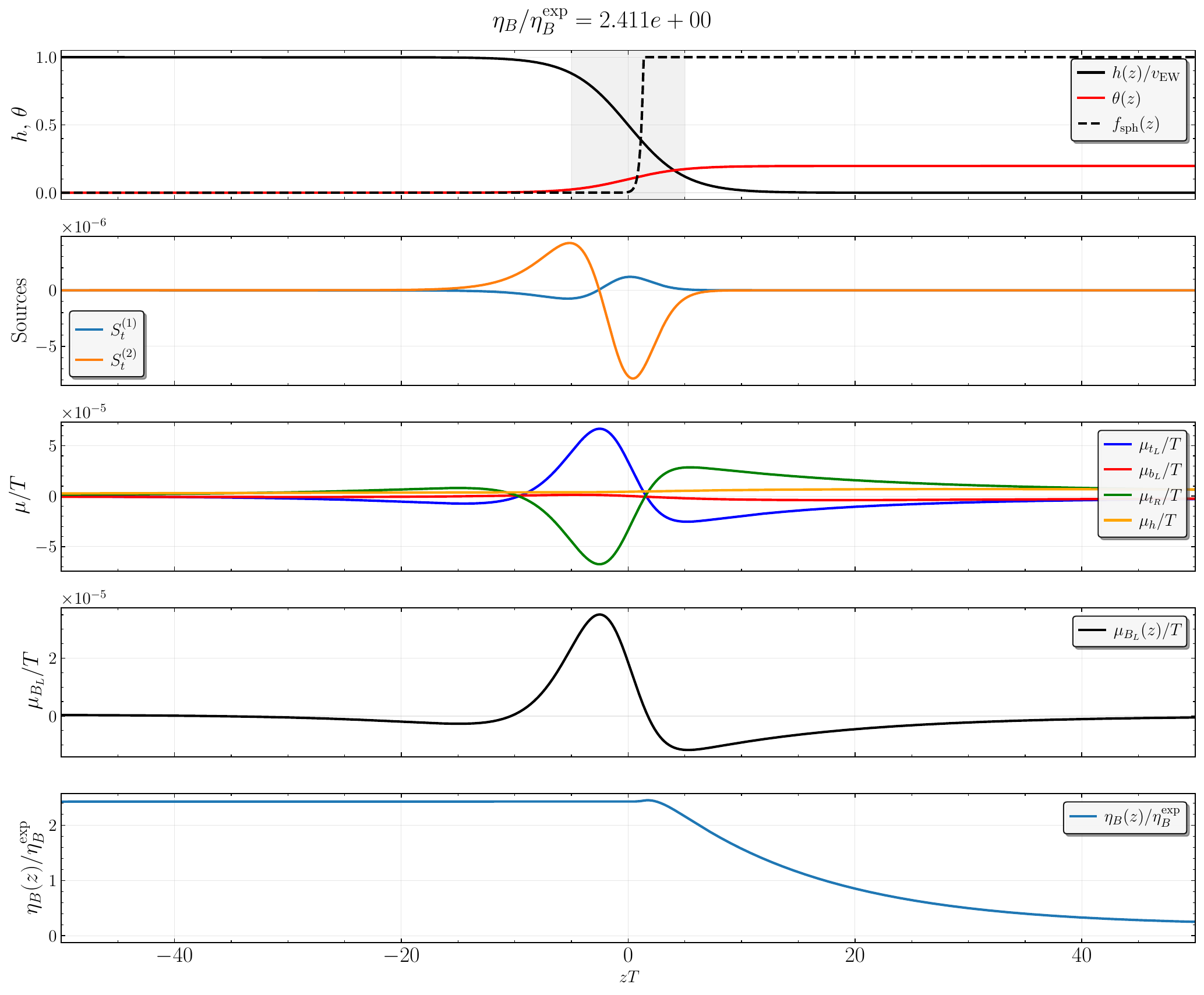}
  \caption{\textbf{From wall profiles to \(\eta_B\) with \texttt{BARYONET}.} \textit{Top row}: Higgs profile \(h(z)/v_{\rm EW}\), CP phase \(\theta(z)\), and weak–sphaleron suppression \(f_{\rm sph}(z)\). Gray shade region indicate $|z|<L_w$. \textit{Second row}: CP sources \(S_t^{(1)}(z)\) and \(S_t^{(2)}(z)\) obtained from the WKB/gradient expansion and thermally averaged with the fluid weights. \textit{Third row}: transport solution for \(\mu_{t_L}/T\), \(\mu_{b_L}/T\), \(\mu_{t_R}/T\), and \(\mu_h/T\). Fourth row: baryon–left chemical potential \(\mu_{B_L}(z)/T\) constructed from the heavy-sector potentials with thermal weights. \textit{Last row}: cumulative \(\eta_B(z)/\eta_B^{\rm exp}\) versus \(zT\); the plateau at negative \(zT\) is the frozen asymmetry (here \(\eta_B/\eta_B^{\rm exp}\simeq 2.4\)). Benchmark parameters as in \eqref{eq:BM1}.}
  \label{fig:comprehensive}
\end{figure}

{
Alongside this paper, we release \href{https://github.com/GiulioBarni/BARYONETv1}{\faGithub\ BARYONET}, a public Python package for computing the
baryon asymmetry generated by electroweak transport in the WKB framework. The code implements
the convention-fixed calculation derived in the previous sections. Its input is a stationary planar
bubble-wall background, specified by the wall parameters \(v_w,L_w\), the relevant mass profiles
\(m_i(z)\), CP-violating phases \(\theta_i(z)\), their derivatives, and the interaction rates entering
the transport network. Its output is the set of CP-odd transport profiles, the baryon-left chemical
potential \(\mu_{B_L}(z)\), the cumulative baryon asymmetry \(\eta_B(z)\), and the final frozen value
\(\eta_B\).

The structure of the package is shown in Fig.~\ref{fig:structure_code}. The implementation separates
model-dependent inputs from universal transport ingredients. Model classes provide wall profiles and
rates; thermal-kernel modules supply the precomputed momentum averages entering the WKB/Boltzmann
system; solver modules assemble and solve the framework-dependent linear boundary-value problem;
and the baryon-asymmetry module constructs \(\mu_{B_L}(z)\) and integrates the weak-sphaleron equation.
This modular organisation is a central part of the code design: the same wall background can be run
with different transport frameworks, and the same solver can be applied to different particle-physics
models once the required profiles are supplied.

The scope of \texttt{BARYONET} is deliberately restricted to the CP-odd transport problem. It does
not compute the finite-temperature effective potential, the nucleation action, the wall profile, or the
steady-state wall velocity from first principles. These quantities are treated as external physical input.
Given such an input, the code provides a reproducible implementation of the map
\begin{equation}
  \left\{
  m_i(z),\theta_i(z),\Gamma_a,v_w,L_w
  \right\}
  \longrightarrow
  \left\{
  S_i(z),\mu_i(z),u_i(z),\mu_{B_L}(z),\eta_B
  \right\},
  \label{eq:baryonet_code_map}
\end{equation}
with selectable transport prescriptions corresponding to the FH04, FH06, CK, and KV24 frameworks.

Figure~\ref{fig:comprehensive} illustrates a representative end-to-end run for a benchmark in the real
singlet extension, denoted here by \texttt{xSM},
\begin{equation}
\textbf{BM1}:\qquad
v_w=0.1,\qquad
L_w=5,\qquad
\xi=\frac{v_n}{T_n}=1.0,\qquad
y_t=0.7,\qquad
y_b=\lambda_h=0,\qquad
\frac{s_n}{\Lambda}=0.2 .
\label{eq:BM1}
\end{equation}
Starting from the wall background, \texttt{BARYONET} evaluates the CP-odd WKB sources, solves the
transport equations for the chemical potentials and velocity perturbations of the tracked species,
constructs \(\mu_{B_L}(z)\) with the appropriate thermal weights, and computes the cumulative profile
\(\eta_B(z)\). The plateau of \(\eta_B(z)\) in the broken phase gives the final baryon asymmetry. This
benchmark will be used below as a common reference point for displaying the main components of the
implementation and for validating the numerical pipeline.
}

\subsection{Code structure}
\label{subsec:baryonet-layers}

\texttt{BARYONET} is organised to let users run end-to-end electroweak baryogenesis calculations with minimal friction: pick a physics model, choose a transport framework, run an example or notebook, and inspect results. This is the structure inside the main folder:

\begin{itemize}
  \item \textbf{\texttt{models/}} Minimal, ready-to-use model definitions (e.g.\ SM extensions) that expose wall/VEV/mass profiles and parameters through a common interface.
  \item \textbf{\texttt{transport\_solver/}} Core solvers for the transport network (CK, FH04, FH06, KV24). Each solver reads a model, sources, and interpolated transport functions, then returns profiles and observables.
  \item \textbf{\texttt{function\_interpolators/}} Pre-tabulated transport coefficients and lightweight loaders; provides consistent interpolants over $(x\!=\!m/T,\,v_w)$.
  \item \textbf{\texttt{Baryon\_Asymmetry/}} Utilities to compute $\eta_B$ from solver outputs (two-step formulation).
  \item \textbf{\texttt{working\_examples/}} Small scripts showing canonical runs (choose model + solver, execute, write plots/tables). Ideal first contact for users.
  \item \textbf{\texttt{notebooks/}} Guided, executable notebooks that mirror the examples but with narrative, sanity checks, and optional diagnostic plots.
  \item \textbf{\texttt{results/}} Default output sink for profiles, figures, and text dumps.
      \item \textbf{\texttt{Estimates/}} Back–of–the–envelope notebooks (rates, diffusion constants, orders of magnitude) used for setup and sanity checks.
  \item \textbf{\texttt{CP\_sources\_derivation/}} Symbolic derivations of CP–violating sources (notebook + notes) underpinning the solver source terms.
\item \textbf{\texttt{literature/}} Reproductions and utilities tied to specific papers (\texttt{CK\_paper2020} \cite{Cline:2020jre}, \texttt{FH04\_paper2011} \cite{Espinosa:2011eu}, \texttt{FH06\_paper2006} \cite{Fromme:2006wx, Fromme:2006cm}, \texttt{KV24\_paper2024} \cite{Kainulainen:2024qpm}) with notebooks and scripts.
  \item \textbf{\texttt{utils/}} Shared helpers for plotting, logging, I/O, and configuration to keep user scripts concise and uniform.
  \item \textbf{\texttt{tests/}} Quick regression tests covering solvers, interpolators, and BAU assembly; useful to validate local installs.
\end{itemize}

In practice, a user selects a model from \texttt{models/}, pairs it with a solver in \texttt{transport\_solver/}, ensures the required tables are available via \texttt{function\_interpolators/}, and runs one of the \texttt{working\_examples/} or \texttt{notebooks/}. Outputs are written to \texttt{results/}, while \texttt{utils/} and \texttt{tests/} support reproducible, checkable runs without cluttering user code.

Inside the main folder, the user can find \texttt{BARYONET\_installation\_guide.pdf}, which provides installation instructions and a brief user manual for navigating \texttt{BARYONET}.

\subsection{Model definition and loading}
\label{subsec:baryonet-model-interface}

The first user-facing object in a \texttt{BARYONET} run is the model class. Its role is to provide the
wall background seen by the particles in the plasma. The transport solver does not need to know the
finite-temperature potential, the tunnelling solution, or the microscopic origin of the CP phase. It only
queries a fixed set of functions: the wall profiles, the mass profiles, their derivatives, the CP-violating
phase, and the wall parameters. This makes the model class the interface between the particle-physics
input and the transport engine.

For example, the benchmark used in Fig.~\ref{fig:comprehensive} is loaded by instantiating the
\texttt{xSM} model class and passing the wall parameters and couplings that determine the profiles:
\begin{lstlisting}[style=baryonetpython, caption={Instantiation of a benchmark model.}]
from models.xSM import xSM

model = xSM(
    vw=0.10,        # wall velocity
    Lw=5.0,         # wall width in units of T_n
    xi=1.0,         # v_n/T_n
    yt=0.7,         # top Yukawa coupling
    yb=0.0,         # bottom Yukawa coupling
    lambda_h=0.0,   # Higgs quartic used by the benchmark class
    g=0.65,         # SU(2) gauge coupling
    sn=200,         # singlet vev (GeV)
    Lambda=1000     # CP violating scale in dim. 5 op. (GeV)
)
\end{lstlisting}
The precise list of constructor arguments is model dependent. In the \texttt{xSM} example above, the
constructor stores the parameters that define the analytic Higgs and singlet wall profiles, and from
them it computes the top mass and CP phase. Other models can use a different constructor. What is
fixed by \texttt{BARYONET} is not the list of input parameters, but the interface exposed after the
model has been instantiated.

Concretely, the transport solver and the source builder only require that the model object can be
queried as a function of the wall coordinate \(z\). The standard way of passing \(z\)-dependent input to
\texttt{BARYONET} is therefore to define a model class. This class may contain analytic functions, as in
the benchmark models shipped with the code, or it may wrap numerical wall profiles obtained from an
external phase-transition calculation. In the latter case, the user can define interpolation functions
inside the model class and expose them through the same methods. Thus external wall profiles do not
need to be hard-coded in the repository: they only have to be packaged into an object with the expected
methods.

A minimal model interface has the following structure:
\begin{lstlisting}[style=baryonetpython, caption={Minimal structure of a model class.}, label={lst:minimal-model-interface}]
class MyModel:
    def __init__(self, vw, Lw, **params):
        self.vw = vw
        self.Lw = Lw
        # Store either analytic parameters or numerical/interpolated profiles.
        # For example:
        # self.h_profile = ...
        # self.theta_profile = ...
        # self.mt_profile = ...

    def gamma_w(self):
        return 1.0 / (1.0 - self.vw**2)**0.5

    # Wall or scalar background
    def f(self, z): ...
    def df_dz(self, z): ...
    def d2f_dz2(self, z): ...

    # Mass profiles and derivatives for transported species
    def mt(self, z): ...
    def mt2(self, z): ...
    def dmt2_dz(self, z): ...

    def mb(self, z): ...
    def mb2(self, z): ...
    def dmb2_dz(self, z): ...

    def mh(self, z): ...
    def mh2(self, z): ...
    def dmh2_dz(self, z): ...

    # CP-violating phase profile and derivatives
    def Theta(self, z): ...
    def dTheta_dz(self, z): ...
    def d2Theta_dz2(self, z): ...
\end{lstlisting}
The methods in Listing~\ref{lst:minimal-model-interface} can return closed-form expressions or
interpolated numerical functions. For example, \texttt{f(z)} may evaluate a kink profile, as in the
analytic benchmarks, or it may return the interpolation of a wall profile computed elsewhere. The
same applies to \texttt{Theta(z)}, the masses, and their derivatives. The solver does not use the
microscopic origin of these functions; it only requires that they are consistently defined on the
numerical grid used for the transport problem.

For additional transported species or model-specific sources, the same pattern is followed by adding
the corresponding mass profiles and derivatives to the model class. The solver therefore sees all
physical setups through the same functional interface, independently of whether the input is a
model-independent ansatz, a singlet extension, a Higgs--\(\phi^6\) model, a 2HDM, or a numerical wall
profile imported from another calculation.

Listing~\ref{lst:model-queries} shows a typical set of quantities queried by the source builder for the
top sector. This is an example of the interface used by the current model classes, not a restriction to
the particular analytic profiles used in the benchmark models.
\begin{lstlisting}[style=baryonetpython, caption={Typical quantities queried by the source builder.}, label={lst:model-queries}]
z = ...  # grid points in the wall frame

# Wall or scalar background
h       = model.f(z)
dh_dz   = model.df_dz(z)
d2h_dz2 = model.d2f_dz2(z)

# Mass profiles and derivatives (here for the top)
mt      = model.mt(z)
mt2     = model.mt2(z)
dmt2_dz = model.dmt2_dz(z)

# CP-violating phase and derivatives
theta       = model.Theta(z)
dtheta_dz   = model.dTheta_dz(z)
d2theta_dz2 = model.d2Theta_dz2(z)

# Wall kinematics
vw = model.vw
gw = model.gamma_w()
\end{lstlisting}

For additional transported species, the same naming pattern is used. For instance, a model with a
bottom source provides \texttt{mb}, \texttt{mb2}, and \texttt{dmb2\_dz}; a model with a Higgs mass
provides \texttt{mh}, \texttt{mh2}, and \texttt{dmh2\_dz}. The model object is therefore the standard
entry point through which \(z\)-dependent functions are passed to the solver. It does not need to be
one of the model files shipped in \texttt{models/}: a user can define a new class, in their own script or
notebook, whose methods return the required profiles. In this way, externally computed wall profiles
can be used by wrapping them in the same interface, without modifying the transport solver.

The solver therefore sees all physical setups through the same functional interface, independently of
whether the underlying input is a model-independent ansatz, a singlet extension, a Higgs--\(\phi^6\)
model, a 2HDM, or a numerical wall profile imported from another calculation.

The source builder then combines these functions into the two CP-odd wall structures appearing in
the WKB source,
\begin{equation}
  \big[m_i^2(z)\theta_i'(z)\big]',
  \qquad
  m_i^2(z)\big[m_i^2(z)\big]'\theta_i'(z).
  \label{eq:baryonet-source-combinations}
\end{equation}
Thus a new model can be added in two equivalent ways: either by providing \(m_i(z)\), \(m_i^2(z)\),
\([m_i^2]'(z)\), \(\theta_i(z)\), \(\theta_i'(z)\), and \(\theta_i''(z)\), or by directly implementing the
source combinations used by a specialised source routine. The first option is the default because it
keeps the physics input transparent and allows the same source code to be reused across models.

Once the model is instantiated, it is passed unchanged to the solver. The user then chooses a transport
framework and loads the corresponding thermal kernels:
\begin{lstlisting}[style=baryonetpython, caption={Loading a transport framework and its kernels.}]
from function_interpolators.interpolators import (
    load_interpolators,
    load_interpolators_2d
)
from transport_solver.CK import TransportSolverCK
from models.default_constants import Z_MAX_FACTOR

fun_1d = load_interpolators("CK")
fun_2d = load_interpolators_2d("CK")
interpolators = {**fun_1d, **fun_2d}

z_span = (-Z_MAX_FACTOR * model.Lw, Z_MAX_FACTOR * model.Lw)

solver = TransportSolverCK(
    model=model,
    interpolators=interpolators,
    z_span=z_span
)
solution = solver.solve()
\end{lstlisting}
At this stage, the model has supplied only the wall background. The choice of CK, FH04, FH06, or KV24
fixes the species basis, the moment truncation, and the collision closure. This separation is important:
the same model object can be run with different transport prescriptions, and differences in \(\eta_B\)
can then be traced to the transport approximation rather than to a different implementation of the
wall profile.

Finally, the solution is post-processed into the physical observable. The baryon-asymmetry module
constructs the framework-dependent \(\mu_{B_L}(z)\), solves the weak-sphaleron equation, and returns the
final value of \(\eta_B\):
\begin{lstlisting}[style=baryonetpython, caption={Post-processing the transport solution into \(\eta_B\).}]
from Baryon_Asymmetry.baryon_asymmetry import BaryonAsymmetry

bau = BaryonAsymmetry(
    solver=solver,
    sol=solution,
    model=model,
    set_name="CK"
)

eta_B = bau.compute_eta_B()
\end{lstlisting}

A new model can be added by implementing the same functional interface in a user-defined class, as illustrated by the \texttt{MyModel} skeleton in Listing~\ref{lst:minimal-model-interface}. The class does not need to reproduce the internal structure of the benchmark models shipped with the code. It only has to expose the methods queried by the source builder and the transport solver: the wall profile, the relevant mass profiles and mass-squared derivatives, the CP-violating phase and its derivatives, and the wall parameters. These methods may return analytic expressions, interpolated numerical profiles, or wrappers around wall profiles computed with an external phase-transition code.

This is the only model-specific input required for the transport calculation, apart from optional model-dependent choices in the collision sector. Once the model object provides the required functions of $z$, the rest of the pipeline---thermal-kernel interpolation, source construction, assembly and solution of the boundary-value problem, computation of $\mu_{B_L}(z)$, and integration of the weak-sphaleron equation---is shared by all models. This is the sense in which \texttt{BARYONET} factorises particle-physics input from the numerical implementation of the WKB transport problem.

\subsection{Numerical values of the parameters}
{A further original component of this work is the reassessment of the numerical inputs entering the
transport equations. To validate \texttt{BARYONET}, we start from the rates and diffusion constants
commonly used in the EWBG literature, but several of these values trace back to estimates that are now
decades old and are often quoted in mutually inconsistent conventions. We therefore recompute, update,
and consistently normalise the relevant diffusion constants, Yukawa and helicity-flip relaxation rates,
and strong- and weak-sphaleron rates. The resulting benchmark inputs, used throughout the validation
runs, are collected in Table~\ref{tab:rates_numerical}.}

\section{Results \& comparison with the literature}
\label{sec:Results}

{In this section, we present the numerical results obtained with \texttt{BARYONET} and benchmark them
against representative calculations in the electroweak baryogenesis literature. The primary goal is
\emph{validation}, not model-building novelty: we align conventions, match wall profiles and effective
rates, and reproduce published trends and normalisations within the stated frameworks
(FH04, FH06, CK/KV24). This cross-validation is an essential part of the construction of
\texttt{BARYONET}: it checks that the implementation contains no hidden convention or coding errors,
and it also provides an independent reproduction of technically involved calculations that, in several
cases, have not been systematically rechecked by other groups.

In this sense, the benchmark comparisons serve a dual purpose. First, they validate the code as a
reliable numerical implementation of the transport systems reviewed above. Second, they provide a
useful community cross-check of the existing EWBG literature, where small differences in signs,
normalisations, source definitions, rates, or wall-frame conventions can lead to sizable changes in
\(\eta_B\). Unless stated otherwise, we report the entropy-normalised asymmetry
\(\eta_B=n_B(-\infty)/s\) and often show the ratio \(\eta_B/\eta_B^{\rm obs}\), with
\(\eta_B^{\rm obs}\simeq 8.7\times 10^{-11}\), to make comparisons robust to unit conventions.

Each subsection follows a uniform template: \emph{(i)} a concise description of the physics setup and
inputs, including wall profiles, rates, and solver options; \emph{(ii)} the scan strategy and target
observables; and \emph{(iii)} a side-by-side comparison with the literature. For full reproducibility,
we provide the paths to the result files used in each figure, so the plots can be regenerated directly
from the repository. The subsection titles coincide with the model names used in the code, for example
the classes under \texttt{models/}, enabling a one-to-one mapping between text and implementation.}

\begin{table}
  \centering
  \begin{tabular}{|l|c|c|c|}
  \hline
  \multicolumn{4}{|c|}{\textbf{Interaction rates}}\\
    \hline
    \textbf{Interaction name} & \begin{minipage}{2.9cm}
    \centering\vspace{0.1cm}
        \textbf{Rates used in the literature}\\ \cite{Bodeker:2004ws, Fromme:2006wx,Konstandin:2013caa, Cline:2020jre} \vspace{0.1cm}
    \end{minipage} &\begin{minipage}{3.5cm}
    \centering\vspace{0.1cm}
        \textbf{Previous update}\\ \textbf{from} \cite{Cline:2021dkf} \vspace{0.1cm}
    \end{minipage} & \begin{minipage}{3cm}
    \centering\vspace{0.1cm}
        \textbf{Last update}\\ \textbf{(this work)} \vspace{0.1cm}
    \end{minipage}\\
    \hline
    Top Yukawa rate{\color{white}{$\dfrac{1}{1}$}} & $\widehat\Gamma_y^t \simeq 4.2 \times 10^{-3}T$ &$\widehat\Gamma_y^t \simeq 3 \times10^{-2}T$ &$\widehat\Gamma_y^t \simeq 10^{-2}T$\\
    Top Helicity flipping & $\widehat\Gamma_m^t \simeq \dfrac{m_t^2}{63 T}$ & $\widehat\Gamma_m^t \simeq \dfrac{0.79m_t^2}{T}$&$\widehat\Gamma_m^t \simeq \dfrac{0.26m_t^2}{T}$\\[8pt]
    Higgs Yukawa rate & $\widehat\Gamma_m^h \simeq \dfrac{m_W^2}{50 T}$ &$\widehat\Gamma_m^h \simeq \dfrac{m_W^2}{25 T}$ &$\widehat\Gamma_m^h \simeq \dfrac{1.5m_W^2}{T}$\\[8pt]
    W boson rate & $\widehat\Gamma_W \simeq \widehat\Gamma^h_{\rm tot}$ & —"— & —"— \\[6pt]
    \hline
    Strong sphaleron{\color{white}{$\dfrac{1}{1}$}}
      & $\Gamma_{ ss}\simeq 4.9\times10^{-4}\,T$ 
       & $\Gamma_{ss}\simeq8.7\times10^{-3}\,T$& $\Gamma_{ss}\simeq2.7\times10^{-3}\,T$ \\
    Weak sphaleron {\color{white}{$\dfrac{1}{1}$}}
      & ${\Gamma_{ws}}\simeq 10^{-6} T$ 
      & ${\Gamma_{ws}}\simeq 6.3\times10^{-6}T$& ${\Gamma_{ws}}\simeq 8\times10^{-7}T$ \\
    \hline
  \multicolumn{4}{|c|}{\textbf{Diffusion constants}}\\
  \hline
  Quarks{\color{white}{$\dfrac{1}{1}$}} & $D_q\simeq6/T$&$D_q\simeq{7.4}/{T}$ &$D_{q_L}\simeq{7.1}/{T}, \,D_{q_R}\simeq{7.6}/{T} $\\[8pt]
  \hline
  Higgs{\color{white}{$\dfrac{1}{1}$}} & $D_h\simeq20/T$& —"—&$D_{h}\simeq14/T$\\
  \hline
  Leptons{\color{white}{$\dfrac{1}{1}$}} & \begin{minipage}{2.9cm} \centering \vspace{0.1cm}
      \xmark \\(see \cite{Joyce:1994zn,Joyce:1994zt} and App. \ref{app:diffusion})\vspace{0.1cm}
  \end{minipage} & $D_{l_R}\simeq\dfrac{490}{T}, \, D_{l_L}\simeq\dfrac{90}{T}  $&$D_{l_R}\simeq\dfrac{412}{T}, \, D_{l_L}\simeq\dfrac{100}{T}  $\\
  \hline
  \multicolumn{4}{|c|}{\textbf{Couplings}}\\
  \hline
  Strong int. &$ \alpha_s \simeq 1/7$& \multicolumn{2}{|c|}{$\alpha_s(m_z)\simeq 0.1179$}\\
  \hline
  Weak int. &$ \alpha_w \simeq 1/30$& \multicolumn{2}{|c|}{$\alpha_w(m_z)\simeq 0.0338$}\\
  \hline
  Weinberg angle &$ \tan \theta_W \simeq 0.29$& \multicolumn{2}{|c|}{$\tan \theta_W \simeq 0.302$}\\
  \hline
  \end{tabular}
  \caption{Representative quantities used in solving the diffusion network, old and updated values. For details regarding diffusion constants, see App. \ref{app:diffusion}, for the Yukawa mediated relaxation rate see App. \ref{app:rate Gammay} and \ref{app:rate GammaM}, and for sphalerons rate see App. \ref{app:sphalerons}.}
  \label{tab:rates_numerical}
\end{table}

\subsection{\texttt{ModelIndependent}}

A clean way to expose the ingredients of electroweak baryogenesis without committing to a specific ultraviolet completion is to work with a \emph{model–independent} ansatz for the top quark mass across the wall, long used in the EWBG literature (see, e.g., \cite{Bodeker:2004ws,Konstandin:2013caa,Espinosa:2011eu}). We take
\begin{equation}
m_t(z)=y_t\,\phi(z)\,e^{i\Theta_t(z)}\,,
\end{equation}
where the real background $\phi(z)$ and the CP–violating phase $\Theta_t(z)$ are smooth kinks of width $L_w$ that interpolate between the symmetric and broken phases\footnote{Throughout this work we adopt the convention that the symmetric phase lies to the right of the wall ($z>0$) and the broken phase to the left ($z<0$). This orientation differs from that used in parts of the literature; physical results are unaffected as long as boundary conditions and sign conventions are implemented consistently.}
\begin{align} \phi(z) &= \frac{\phi_c}{2}\Big(1 - \tanh \frac{z}{L_w}\Big),\\ \Theta_t(z) &= \frac{\Delta\Theta_t}{2}\Big(1 - \tanh \frac{z}{L_w}\Big), \end{align} controlled by the wall width \(L_w\) and the total phase excursion \(\Delta\Theta_t\).
This choice isolates the two macroscopic sources that matter for charge transport—gradients of the mass modulus and of the CP phase—while keeping the rest of the computation agnostic about how $\Theta_t$ arises (two–doublet, singlet, higher–dimensional operators, \emph{etc}.). In practice one scans over the wall parameters $(v_w,L_w)$ and the thermodynamic strength, and for each point determines the \emph{minimal} phase excursion $\Delta\Theta_t$ that reproduces the observed baryon asymmetry. The transport sector delivers $\mu_{B_L}(z)$, the weak–sphaleron kernel converts it into $n_B(z)$, and the freeze–in value gives $\eta_B$; this pipeline is the same that will be used for all concrete models later on.

\begin{figure}
  \centering
  \includegraphics[width=0.78\linewidth]{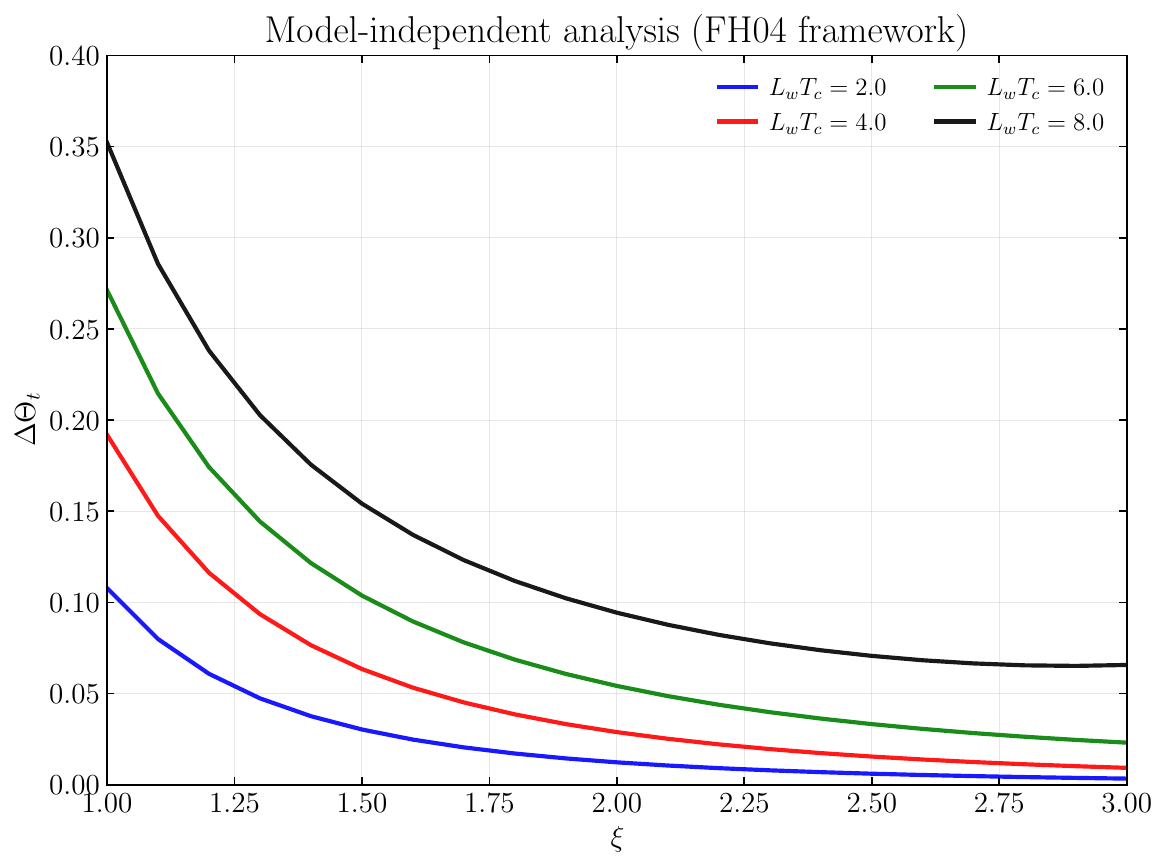}
  \caption{Model–independent benchmark in the FH04 framework: minimal CP–phase excursion $\Delta\Theta_t$ required to match the observed $\eta_B$ as a function of $\xi$, shown for several wall thicknesses $L_wT_c\in\{2,4,6,8\}$. Curves are generated by solving the transport equations for each $(\xi,L_wT_c)$, constructing $\mu_{B_L}(z)$, and using an overshoot method to find the phase reproducing the BAU's observed value. Benchmark parameters: \(v_w=0.01,\ y_t=0.7,\ y_b=0.0\). This reproduce Fig.~1 of \cite{Espinosa:2011eu} and Fig.~3 of \cite{Konstandin:2013caa}.}
  \label{fig:MI_FH04}
\end{figure}

Figure~\ref{fig:MI_FH04} illustrates a typical outcome in the FH04 closure: for several wall thicknesses (quoted as $L_w T_c$) one obtains the required $\Delta\Theta_t$ as a function of $\xi=\phi_c/T_c$. The plot is best read as a “response curve’’ of the charge–transport engine: thicker walls and weaker transitions demand larger CP–phase excursions, while thinner walls (or stronger transitions) reduce the burden on CP violation. The main virtue of this benchmark is pedagogical: it makes explicit which microscopic inputs the code uses (wall kinematics, scalar and phase profiles) and how they flow into the final observable $\eta_B$, independently of any detailed microphysics. This plot can be reproduced from the notebook \texttt{literature/FH04\_paper2011/FH04\_scans\_minimal.ipynb} .

\subsection{\texttt{xSM}}

In the real–singlet extension of the SM (xSM), as considered in \cite{Cline:2020jre}, a concrete realisation of a space–dependent complex top mass is provided by the dimension–five operator
\(i\,(s/\Lambda)\,\bar Q_3 H t_R\), which induces a phase wherever the singlet \(s\) acquires a wall profile. In the wall background the effective top mass term reads
\begin{equation}
  y_t\,h(z)\,\bar t_L\!\left(1+i\,\frac{s(z)}{\Lambda}\right)\!t_R+\text{h.c.},
\end{equation}
implying the magnitude–phase decomposition
\begin{align}
  m_t(z) &= y_t\,h(z)\,\sqrt{1+\frac{s^2(z)}{\Lambda^2}},\\
  \Theta_t(z) &= \arctan\!\Big(\frac{s(z)}{\Lambda}\Big).
\end{align}
We model the background with opposite–kink profiles for Higgs and singlet\footnote{In a genuine two–field background (e.g.\ Higgs $h$ and singlet $s$) the wall profiles need not share the same thickness: one should in principle allow for $L_h\neq L_s$, possible relative shifts in the wall centres, (see, e.g., \cite{DeCurtis:2023hil} for an explicit treatment). These differences modify the spatial gradients that source CP violation and can quantitatively affect transport. To simplify the analysis here, we adopt the assumption $L_h=L_s\equiv L_w$ and take the induced mass phase to follow a single smooth profile $\Theta_t(z)$.},
\begin{align}
  h(z) &= \frac{v_n}{2}\Big(1-\tanh\frac{z}{L_w}\Big),\\
  s(z) &= \frac{w_n}{2}\Big(1+\tanh\frac{z}{L_w}\Big),
\end{align}
so that \(\Theta_t(z)\) interpolates smoothly across the wall. Given \((v_w,L_w,v_n,w_n,\Lambda)\), the transport system yields \(\mu_{B_L}(z)\), which feeds the baryon equation for \(\eta_B\). 

\begin{figure}
  \centering
  \includegraphics[width=0.82\linewidth]{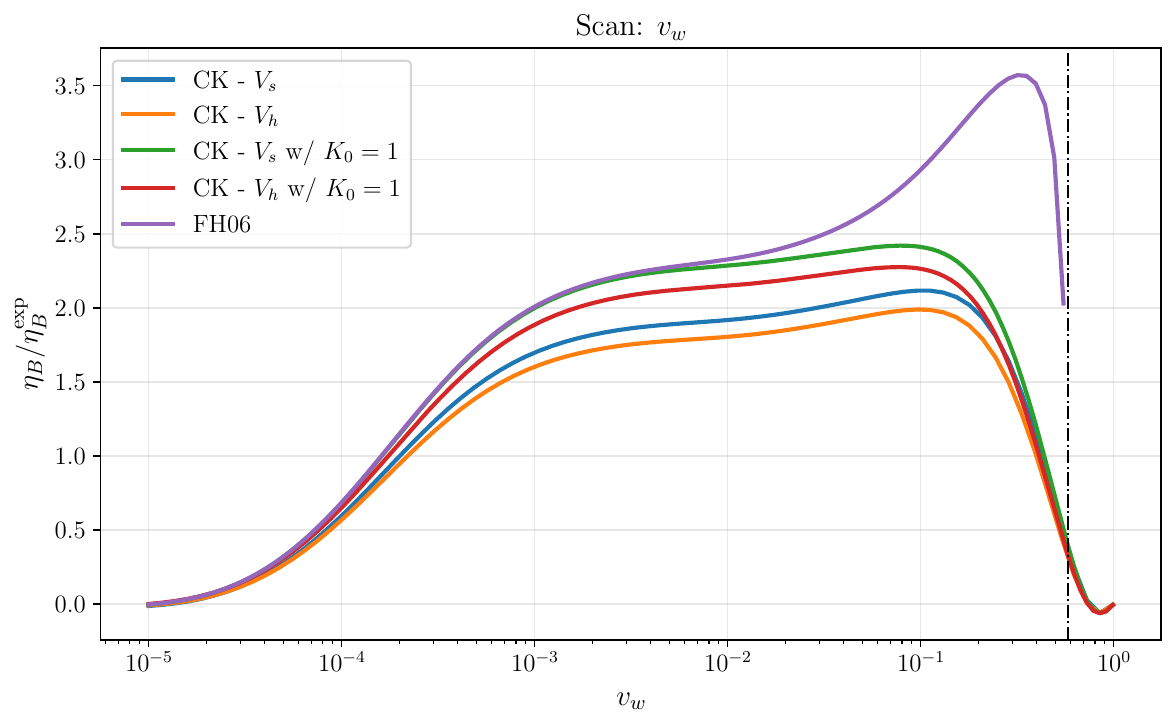}
  \caption{\textbf{xSM benchmark,} \(\eta_B/\eta_B^{\rm obs}\) versus \(v_w\).
  Blue: scheme with spin–\(s\) source (CK–\(V_s\));
  Orange: scheme with helicity–\(h\) source (CK–\(V_h\));
  Green: CK–\(V_s\) with \(K_0=1\);
  Red: CK–\(V_h\) with \(K_0=1\);
  Purple: FH06 with the spin–\(s\) source. 
  Curves coincide at small \(v_w\), while FH06 drops near \(c_s\) and the  scheme remains smooth; the \(s\)–vs–\(h\) difference is small. This reproduce Fig.~3 of \cite{Cline:2020jre}.}
  \label{fig:CKvsFH06}
\end{figure}

Following \cite{Cline:2020jre}, we use this setup to benchmark frameworks and sources. In Fig. \ref{fig:CKvsFH06}, we compare (i) CK fluid scheme versus the FH06 equations, and (ii) two closely related CPV sources: the \emph{spin}\textendash\(s\) source (identifying chirality with the spin eigenstate along \(\hat z\)) and the \emph{helicity}\textendash\(h\) source (identifying chirality with helicity), as you can see in Appendix \ref{app: kernel}. For small wall velocities, all curves agree—as expected from the common small–\(v_w\) limit. At larger \(v_w\), FH06 exhibits a strong suppression as \(v_w\) approaches the sound speed\footnote{In the FH06 setup, the transport is treated as a small–$v_w$ diffusion problem in a nearly static plasma. As the wall approaches the sound speed $c_s=1/\sqrt{3}$ from below (deflagration regime), hydrodynamics predicts the build–up of a compressed “sonic layer’’ (a sonic boom) in front of the wall. In that layer, the effective diffusion length ahead of the wall shrinks to zero, so the CP–odd left charge cannot outrun the interface and $\mu_{B_L}$ (hence the BAU) is strongly suppressed. Because FH06 linearises in $v_w$ and neglects the boosted background flow, this physical suppression manifests as a sharp drop near $v_w\simeq c_s$. The CK implementation uses boosted formulations to keep a smooth $v_w$ dependence and show a gradual decrease instead of a spurious cusp. For a hydrodynamic discussion of the sonic layer and its impact on charge transport, see \cite{Dorsch:2021nje}.} \(c_s=1/\sqrt{3}\), while the improved scheme degrades smoothly toward \(v_w\to1\). The difference between spin– and helicity–based sources is modest because the two bases coincide in the massless limit relevant ahead of the wall. We believe that in \cite{Cline:2020jre} the “\(h\)” curve appears to have been drawn with \(K_0\!\to\!1\); to disentangle these effects we show both source choices with the full \(K_0\) and with \(K_0=1\) for reference.

For a recent analysis, in the xSM, of how out--of--equilibrium corrections affect the bubble--wall dynamics—and thereby the resulting prediction for the baryon asymmetry—see~\cite{Branchina:2025adj}.

\subsection{\texttt{xSMv2}}

In this subsection, we introduce a refined version of the \textit{xSM} setup that makes the link between \emph{microscopic couplings} and \emph{macroscopic wall properties} explicit—most notably the wall width \(L_w\)--in the spirit of Ref.~\cite{Espinosa:2011eu}. Concretely, we consider the SM augmented by a real scalar singlet with a \(Z_2\)-symmetric potential, and we follow the finite–temperature dynamics (effective potential, phase structure, nucleation, and bubble–wall profiles) closely along the lines of Ref.~\cite{Espinosa:2011eu}. This construction provides a transparent mapping from the portal and singlet self–couplings to thermal parameters such as \(T_c\), \(\xi\!\equiv\!v(T)/T\), and, crucially for transport, the wall thickness \(L_w\) that enters the semiclassical sources and diffusion network. At mean–field level, and around the critical temperature, the scalar potential can be written as
\begin{equation}
V(h,s,T)=\frac{\lambda_h}{4}\!\left[h^2-v_c^2+\frac{v_c^2}{w_c^2}s^2\right]^2+\frac{\kappa}{4}\,s^2 h^2
+\frac{1}{2}\bigl(T^2-T_c^2\bigr)\bigl(c_h h^2+c_s s^2\bigr),
\end{equation}
with \(\kappa\equiv \lambda_m-2\lambda_h\,v_c^2/w_c^2\). For \(\kappa>0\) the potential exhibits a tree–level barrier and, at \(T=T_c\), degenerate minima at \((h,s)=(v_c,0)\) and \((0,\pm w_c)\). The thermal coefficients are
\begin{align}
c_h&=\frac{1}{48}\!\left(9g^2+3g'^2+12y_t^2+\lambda_h\bigl(24+4\,v_c^2/w_c^2\bigr)+2\kappa\right),\\
c_s&=\frac{1}{12}\!\left(\lambda_h\bigl(3(v_c^2/w_c^2)^2+4\,v_c^2/w_c^2\bigr)+2\kappa\right),
\end{align}
and electroweak breaking below \(T_c\) is ensured when
\begin{equation}
\frac{c_h}{c_s}>\frac{w_c^2}{v_c^2}.
\end{equation}
Standard relations then follow
\begin{equation}
T_c^2=\frac{\lambda_h}{c_h}\,(v_0^2-v_c^2),\qquad
m_h^2=2\lambda_h v_0^2,\qquad
m_s^2=\frac{1}{2}\kappa v_0^2+\lambda_h\,(v_0^2-v_c^2)\!\left(\frac{v_c^2}{w_c^2}-\frac{c_s}{c_h}\right).
\end{equation}
\begin{figure}[t]
  \centering
  \includegraphics[width=0.49\linewidth]{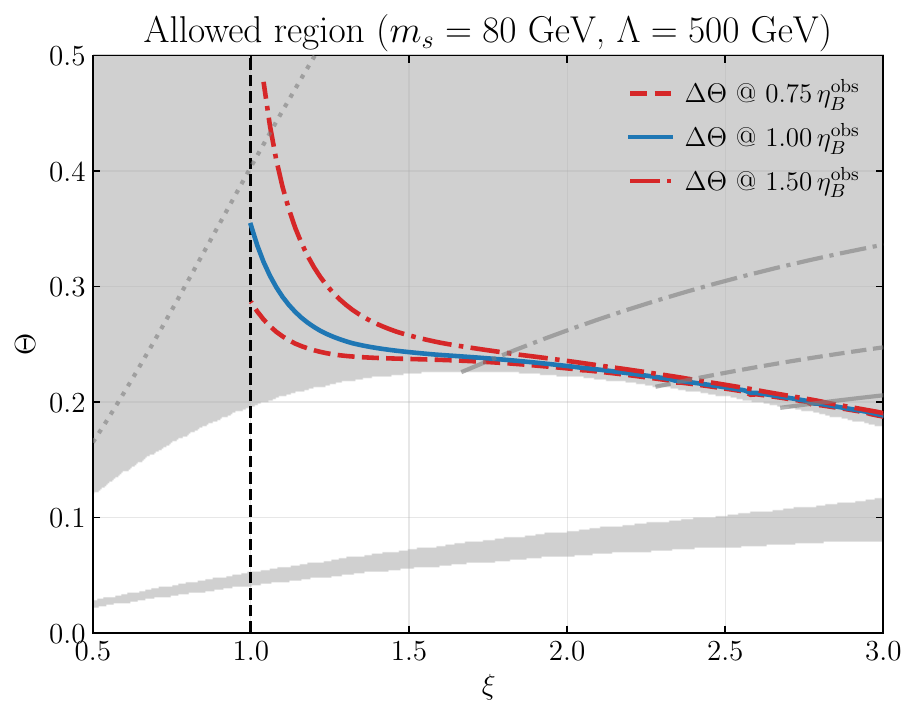}\hfill
  \includegraphics[width=0.49\linewidth]{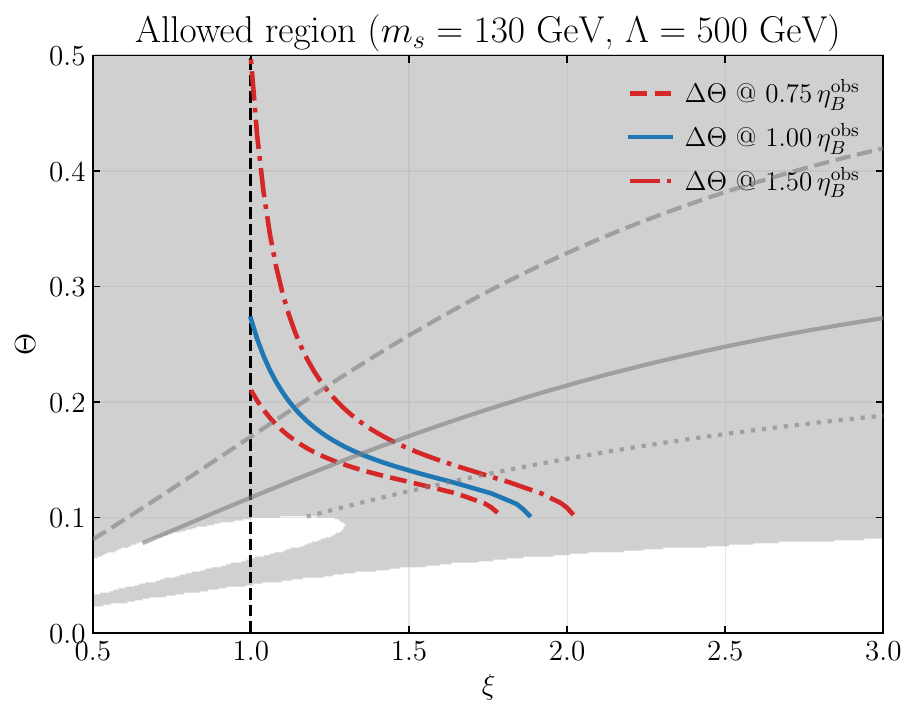}
  \caption{\textbf{xSMv2 benchmark scans} (\(m_h=120\ \mathrm{GeV}\), \(\Lambda=500\ \mathrm{GeV}\)). 
  Curves show the CP–phase excursion \(\Delta\Theta_t\) required to obtain \(0.75\eta_B^{\rm obs}\) (dash–dotted), \(\eta_B^{\rm obs}\) (solid), and \(1.5\eta_B^{\rm obs}\) (dashed) as a function of \(\xi=v_c/T_c\). 
  Left: \(m_s=80\ \mathrm{GeV}\). Right: \(m_s=130\ \mathrm{GeV}\). 
  The shaded band indicates the region required for the FOPT ($L_wT_c>0$ and $\lambda_{m}<2\pi)$; the vertical line marks \(\xi=1\), below which active sphalerons would erase the asymmetry. The grey lines (dotted above left, dot-dashed, dashed, solid, dotted below right) correspond to lines with fixed $\lambda_m =0.25, 0.5, 0.75, 1, 1.5$, respectively. These reproduce Fig.~2 from \cite{Espinosa:2011eu} and Fig.6 from \cite{Konstandin:2013caa}.}
  \label{fig:xSMv2}
\end{figure}
Across the wall, we use kink profiles of common thickness \(L_w\),
\begin{equation}
h(z)=\frac{v_c}{2}\!\left[1+\tanh\!\left(\frac{z}{L_w}\right)\right],\qquad
s(z)=\frac{w_c}{2}\!\left[1-\tanh\!\left(\frac{z}{L_w}\right)\right],
\end{equation}
interpolating between \((0,w_c)\) and \((v_c,0)\). A thin–wall estimate obtained by minimising the one–dimensional action gives\footnote{Using the kink profiles \(h(z),s(z)\)  modelled as 
\begin{align}
h(z) &= \frac{v_c}{\sqrt{2}} \sin\left( \frac{\pi}{4} \left( 1 + \tanh\left( \frac{z}{L_w} \right) \right) \right), \qquad s(z) = \frac{w_c}{\sqrt{2}} \cos\left( \frac{\pi}{4} \left( 1 + \tanh\left( \frac{z}{L_w} \right) \right) \right),
\end{align}
where, at the critical temperature, one can see that the wall tension splits into a kinetic piece and a potential piece and can be computed exactly
\[
S_E(L_w)=S_{E,{\rm kin}}(L_w)+S_{E,{\rm pot}}(L_w),
\qquad
S_{E,{\rm kin}}=\frac{\alpha}{L_w}\,(v_c^2+w_c^2),\quad
S_{E,{\rm pot}}=\beta\,\kappa\,v_c^2 w_c^2\,L_w,
\]
with
\[
\alpha=\frac{\pi^2}{48},\qquad
\beta=\frac{1}{32}\Big(\gamma_E-\mathrm{Ci}(2\pi)+\ln(2\pi)\Big),
\]
where \(\gamma_E\) is the Euler–Mascheroni constant and \(\mathrm{Ci}\) the cosine integral. Minimizing \(S_{E}\) gives the thin–wall estimate
\[
L_w^2=\frac{\alpha}{\beta}\,\frac{v_c^2+w_c^2}{\kappa\,v_c^2 w_c^2}
\;\;\simeq\;\;2.7\,\frac{v_c^2+w_c^2}{\kappa\,v_c^2 w_c^2},
\]
so the numerical prefactor \(\alpha/\beta\simeq 2.699\). Compared with the formula in the main text, the parametric dependence is not identical. For completeness and reproducibility, however, our analysis implements the wall–width expression exactly as written in~\cite{Espinosa:2011eu} when benchmarking their figures, even though our rederivation indicates a likely typo/inconsistency in its parametric scaling.
}
\begin{equation}
L_w^2\simeq \frac{2.7}{\kappa}\,\frac{v_c^2+w_c^2}{v_c^2 w_c^2}\left(1+\frac{\kappa\,w_c^2}{4\lambda_h v_c^2}\right).
\end{equation}
CP violation is communicated to the top sector by the dimension–five operator
\begin{equation}
\frac{s}{f}\,H\,\bar Q_3\,(a+i b\,\gamma_5)\,t+\text{h.c.},
\end{equation}
which induces a space–dependent complex top mass,
\begin{equation}
m_t(z)=|m_t(z)|\,e^{i\Theta_t(z)},\qquad
\Theta_t(z)\simeq \frac{\Delta\Theta_t}{2}\!\left[1+\tanh\!\left(\frac{z}{L_w}\right)\right],\qquad
\Delta\Theta_t\simeq \frac{b}{y_t}\,\frac{\Delta s}{f}\,,
\end{equation}
with \(\Delta s\!\approx\! w_c\) for the profiles above. Larger \(\Delta\Theta_t\) and stronger transitions (larger \(\xi\equiv v_c/T_c\)) enhance the CP–odd source, whereas thicker walls reduce gradients and suppress transport.

In \texttt{BARYONET} one fixes \((v_w,L_w,v_c,w_c,\lambda_h,\kappa)\) and the spurion parameters \((b/f)\), builds \(h(z),s(z),\Theta_t(z)\), evaluates the CP sources and collision blocks, solves the transport system to obtain \(\mu_{B_L}(z)\), and finally computes \(\eta_B\). We present scans in \(\Delta\Theta_t\), and \(\xi\) comparing \(\eta_B/\eta_B^{\rm obs}\) to the qualitative trends in Ref.~\cite{Espinosa:2011eu}: increasing \(\xi\) or \(m_s\) lowers the \(\Delta\Theta_t\) needed for successful baryogenesis (until washout reappears near \(\xi\simeq 1\)); thicker walls \(v_w\) demand larger phases.

\subsection{\texttt{Higgs6Model}}
\begin{figure}
  \centering
  \begin{minipage}{0.49\linewidth}
    \centering
    \includegraphics[width=\linewidth]{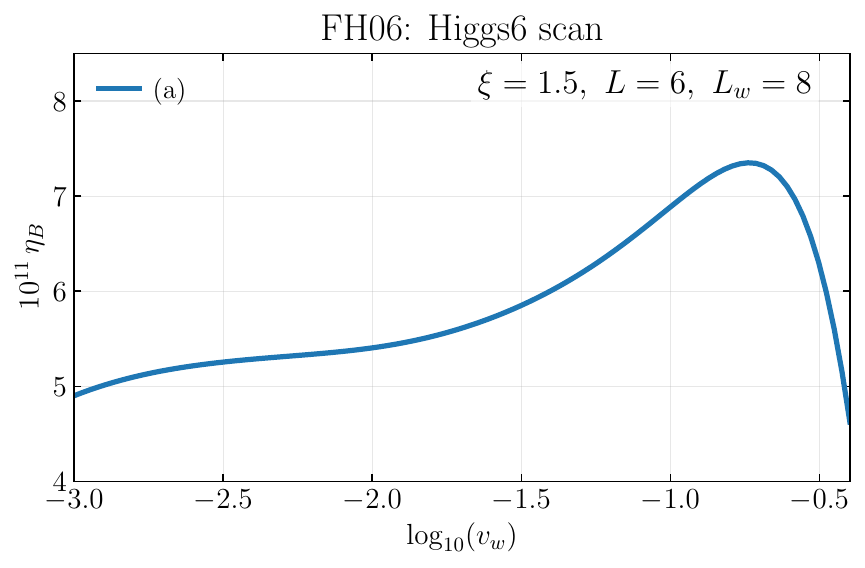}
  \end{minipage}\hfill
  \begin{minipage}{0.49\linewidth}
    \centering
    \includegraphics[width=\linewidth]{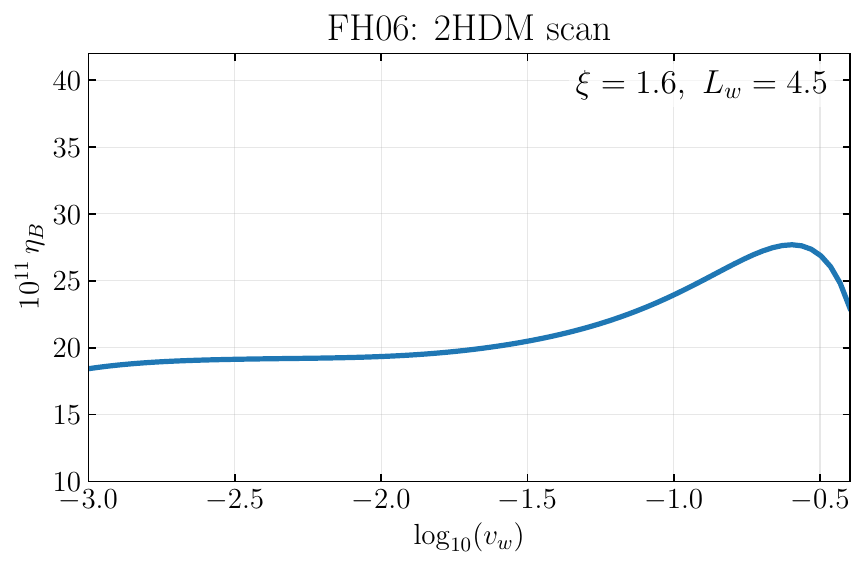}
  \end{minipage}
  \caption{\textbf{FH06 scans.} \emph{Left—Higgs6Model} (based on \cite{Fromme:2006wx}): baryon asymmetry $\eta_B$ versus wall velocity $v_w$ for representative $(\xi,L_w)$, obtained with the FH06 diffusion network and full $z$–dependent thermal averages; the curve corresponds to the “no approximations” case (line~(a) in Fig. 6 of \cite{Fromme:2006wx}). \emph{Right—2HDM} (next subsection, based on \cite{Fromme:2006cm}): analogous scan using the FH06 implementation and the two–doublet wall/phase profiles of Ref.~\cite{Fromme:2006cm}. Quantities in the top right corner are in units of $T_c$.}
  \label{fig:FH06_scans}
\end{figure}

As a minimal one–field benchmark, we follow Ref.~\cite{Fromme:2006wx} and augment the SM Higgs potential with a stabilising dimension–six operator. At finite temperature, the background is captured by
\begin{equation}
\label{eq:H6_potential}
V(\phi)=-\frac{\mu^2}{2}\phi^2+\frac{\lambda}{4}\phi^4+\frac{1}{8L^2}\phi^6\,,
\end{equation}
where the $\phi^6$ term allows $\lambda<0$ while keeping the potential bounded, yielding a tree–level barrier and a strong first–order transition over broad regions of parameter space. Typical walls are thick in the WKB sense, $3 \lesssim L_w\,T_c \lesssim 16$, and we model the profile by a kink,
\begin{equation}
\phi(z)=\frac{v_c}{2}\Big(1-\tanh\frac{z}{L_w}\Big)\!,
\end{equation}
interpolating from the symmetric to the broken phase. CP violation is introduced by dressing the top Yukawa through the lowest–dimensional operator,
\begin{equation}
\frac{x_t}{M^2}(\Phi^\dagger\Phi)\Phi\,\bar t_R Q_3+y_t\Phi\,\bar t_R Q_3+\text{h.c.},
\end{equation}
so that inside the wall, the top mass is complex,
\begin{equation}
m_t(z)=|m_t(z)|\,e^{i\Theta_t(z)},\qquad 
\tan\Theta_t(z)\simeq \sin\!\varphi_t\,\frac{\phi^2(z)}{2M^2}\Big|\frac{x_t}{y_t}\Big|\,,
\end{equation}
and the total phase excursion $\Delta\Theta_t$ controls the CP–odd source. For transport, we implement the diffusion network of Ref.~\cite{Fromme:2006wx} and keep the standard four species $\{t,b,t_c,h\}$ with the baryon–left potential built as in FH06, see Eq.~\eqref{eq:muBL_FH06}. We then solve the FH06 moment equations with the full $z$–dependence of the thermal averages and finite $W$ scatterings, and convert $\mu_{B_L}(z)$ into $\eta_B$ through the weak–sphaleron equation. The left panel of Fig.~\ref{fig:FH06_scans} reproduces the qualitative behaviour of line~(a) in Fig.~6 of Ref.~\cite{Fromme:2006wx}: for fixed $(\xi,L_w)$ the asymmetry grows from very small $v_w$, flattens for intermediate speeds, and eventually decreases as transport becomes inefficient.

\paragraph{Parameter choice.}
Following Ref.~\cite{Fromme:2006wx}, the parameters of the \texttt{Higgs6Model} potential are selected to ensure a strong first–order electroweak phase transition. The effective operator scale $L$ controls the $\phi^6$ deformation and the height of the tree–level barrier: smaller $L$ values strengthen the transition but eventually spoil perturbativity or EFT validity. For each $(m_h,L)$ pair, the critical temperature $T_c$ and the corresponding Higgs expectation value $v_c$ are computed, and the ratio $\xi\equiv v_c/T_c$ is required to exceed unity (typically $\xi\gtrsim1.1$) to avoid baryon washout in the broken phase. Viable regions are found for $5\lesssim L/T_c\lesssim 8.5$ with $\lambda<0$, where a stable, bounded potential and $\xi>1$ coexist.

\subsection{\texttt{TwoHDM}}

The two–Higgs–doublet model (2HDM) augments the SM with two $SU(2)_L$ doublets $H_1,H_2$ and allows for a soft, explicitly CP–violating mixing. Following Ref.~\cite{Fromme:2006cm}, we start from the neutral–field potential (soft $Z_2$ breaking by $\mu_3^2 e^{i\phi} H_1^\dagger H_2$) and include the one–loop Coleman–Weinberg and thermal pieces,
\begin{equation}
  V_{\rm eff}(h_1,h_2,\theta;T)=V_0+V_1+V_{\rm CT}+V^T_1,  
\end{equation}
where $H_1^0=h_1 e^{-i\theta_1}$, $H_2^0=h_2 e^{+i\theta_2}$ and $\theta\equiv\theta_1+\theta_2$. Minimizing $V_{\rm eff}$ at the critical temperature $T_c$ gives the broken–phase vevs $v_1\equiv\langle h_1\rangle_{T_c}$, $v_2\equiv\langle h_2\rangle_{T_c}$ and the CP phase $\theta_{\rm brk}$, degenerate with the symmetric minimum $(0,0,\theta_{\rm sym})$. The standard strength parameter and valley direction are
\begin{equation}
    \xi \equiv \frac{v_c}{T_c},\qquad v_c\equiv \sqrt{2\,(v_1^2+v_2^2)},\qquad \tan\beta_T \equiv \frac{v_2}{v_1}\Big|_{T_c}.
\end{equation}
Along the light “valley" direction $h$ the wall is well approximated by a kink of thickness $L_w$ determined by the barrier height $V_b$ at $T_c$
\begin{equation}
    h(z)\simeq \frac{v_c}{2}\Big[1-\tanh\!\Big(\frac{z}{L_w}\Big)\Big],\qquad L_w\simeq \sqrt{\frac{v_c^2}{8\,V_b}}.
\end{equation}
Projecting the vevs onto the two doublets along the valley gives $h_1(z)=h(z)\cos\beta_T$ and $h_2(z)=h(z)\sin\beta_T$. The top and bottom \emph{mass moduli} (the only quantities the transport tables need) follow directly
\begin{equation}
    m_t(z)=y_t\,h_2(z)=\frac{y_t}{\sqrt{2}}\,h(z)\sin\beta_T,\qquad
m_b(z)=y_b\,h_1(z)=\frac{y_b}{\sqrt{2}}\,h(z)\cos\beta_T,
\end{equation}
so $m_{t,b}^2(z)$ and their first/second $z$–derivatives are fixed once $(v_c,T_c,L_w,\beta_T)$ are known. The relative CP phase varies across the wall; we use the kink ansatz
\begin{equation}
    \theta(z)=\theta_{\rm brk}-\frac{\Delta\theta}{2}\Big[1+\tanh\!\Big(\frac{z}{L_w}\Big)\Big],\qquad \Delta\theta\equiv \theta_{\rm brk}-\theta_{\rm sym}.
\end{equation}
The phase that actually sources the transport is the \emph{top phase}
\begin{equation}
    \Theta(z)\equiv\theta_t(z)=\frac{\theta(z)}{1+\tan^2\beta_T}.
\end{equation}
For the diffusion network, we implement the moment equations of Ref.~\cite{Fromme:2006cm}; in the code, this option is labelled \texttt{FH06for2HDM}. A representative result is shown in the right panel of Fig.~\ref{fig:FH06_scans}: we reproduce the qualitative trends reported in Fig. 6 of \cite{Fromme:2006cm}—larger $\xi$ (stronger transitions) and moderate $v_w$ enhance the signal, while thicker walls reduce it.

\section{Higher–moment analysis}
\label{sec:higher-moments}

\begin{figure}[t]
    \centering
    \includegraphics[width=0.32\textwidth]{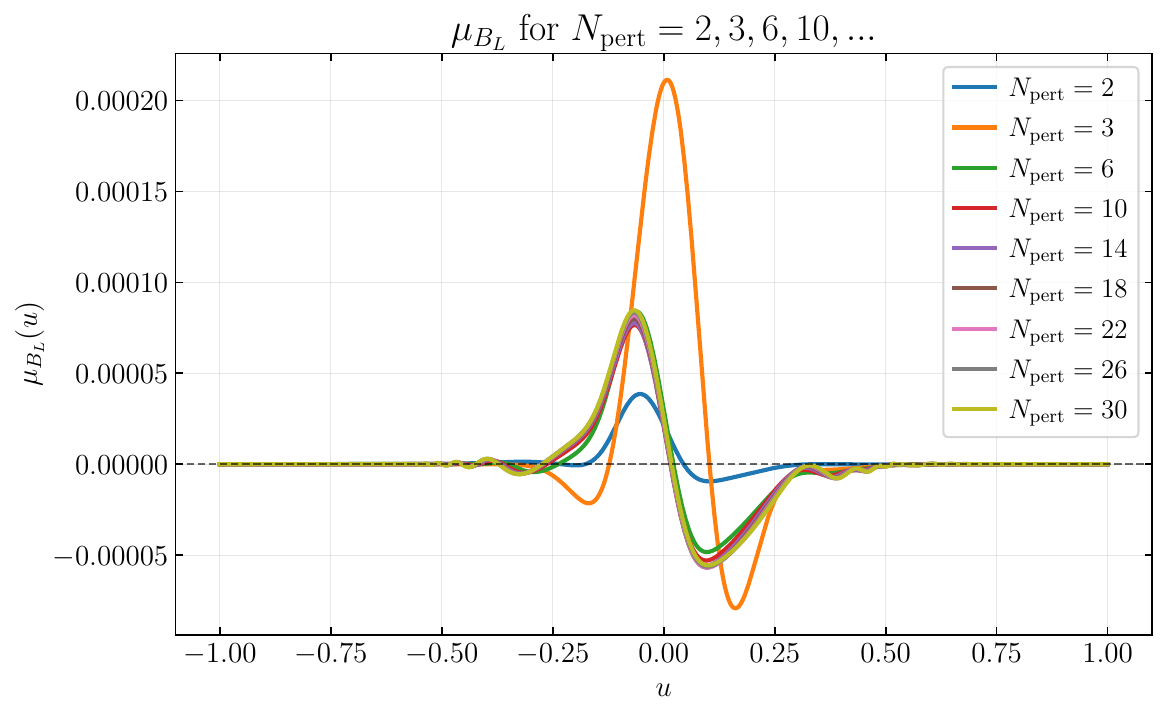}\hspace{4pt}
    \includegraphics[width=0.32\textwidth]{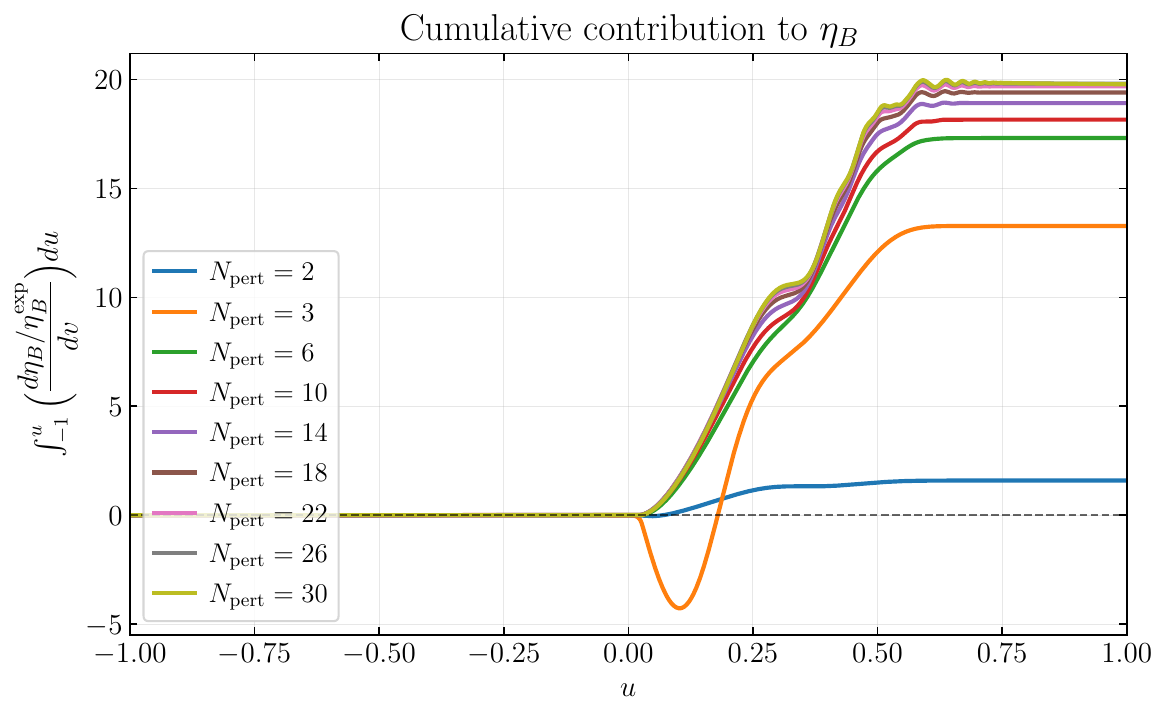}\hspace{4pt}
    \includegraphics[width=0.32\textwidth]{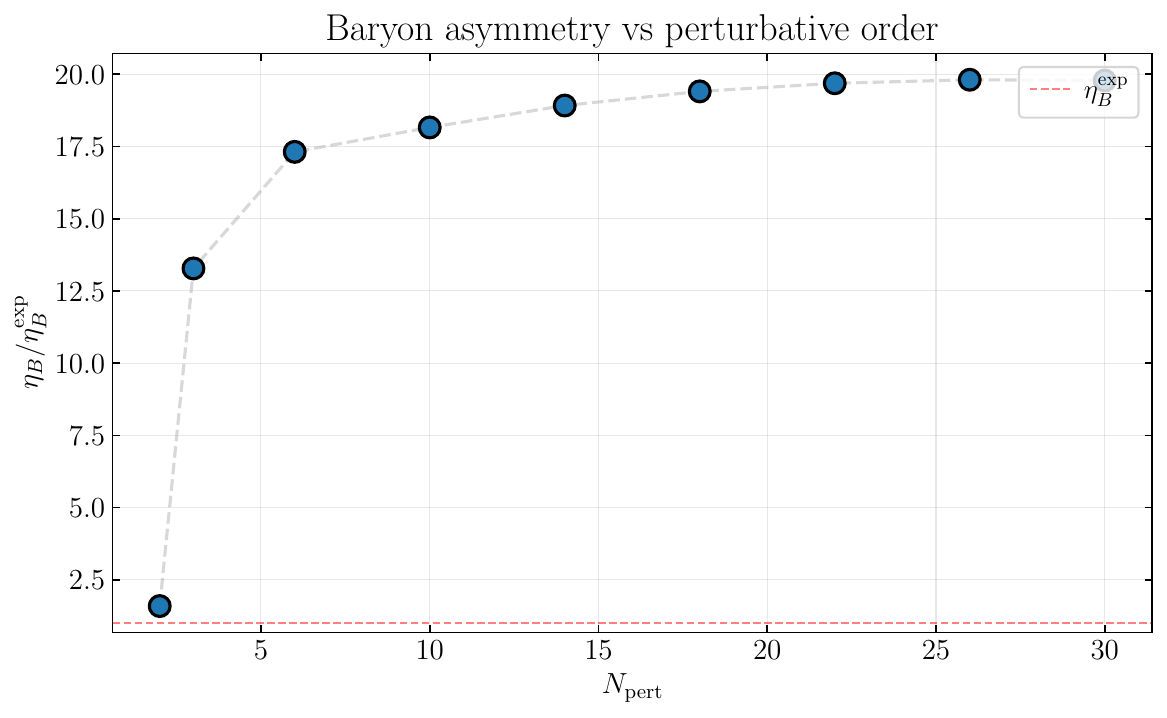}\\[6pt]
    \includegraphics[width=0.32\textwidth]{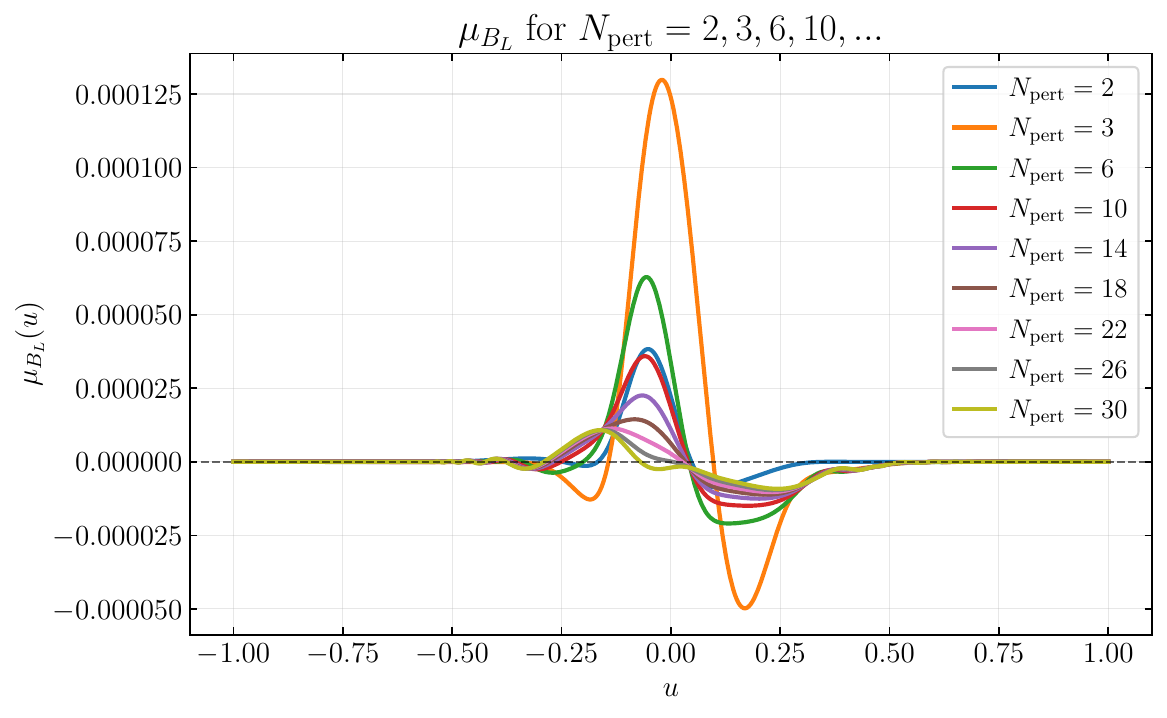}\hspace{4pt}
    \includegraphics[width=0.32\textwidth]{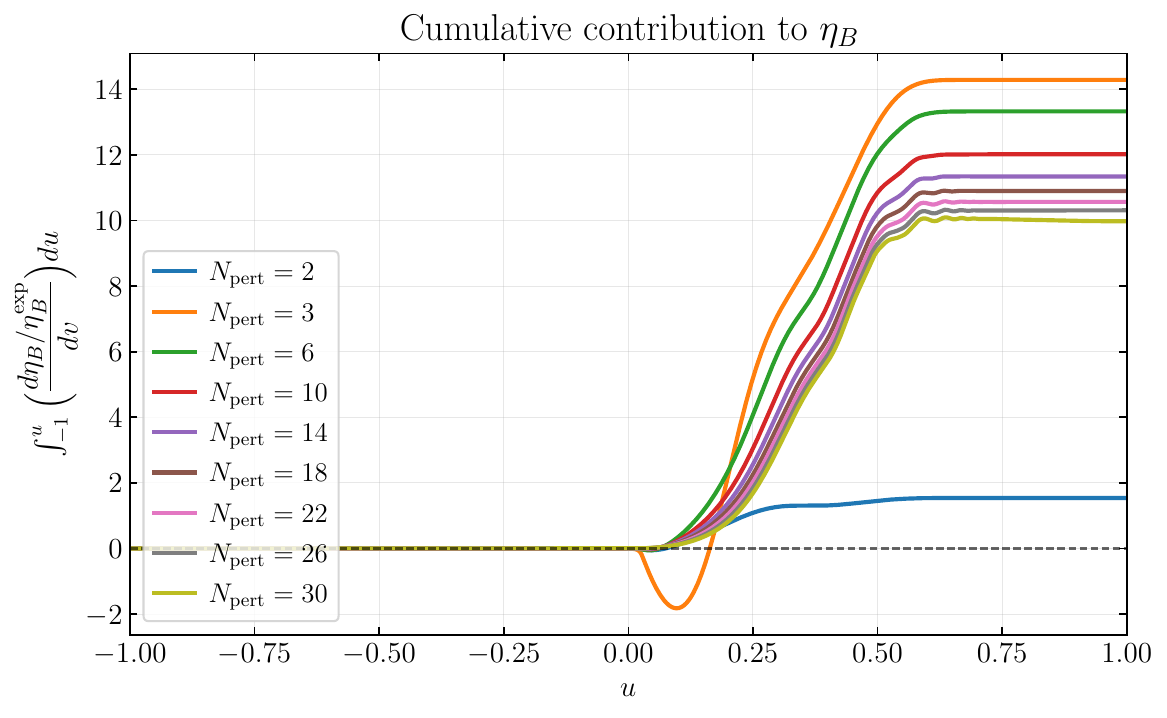}\hspace{4pt}
    \includegraphics[width=0.32\textwidth]{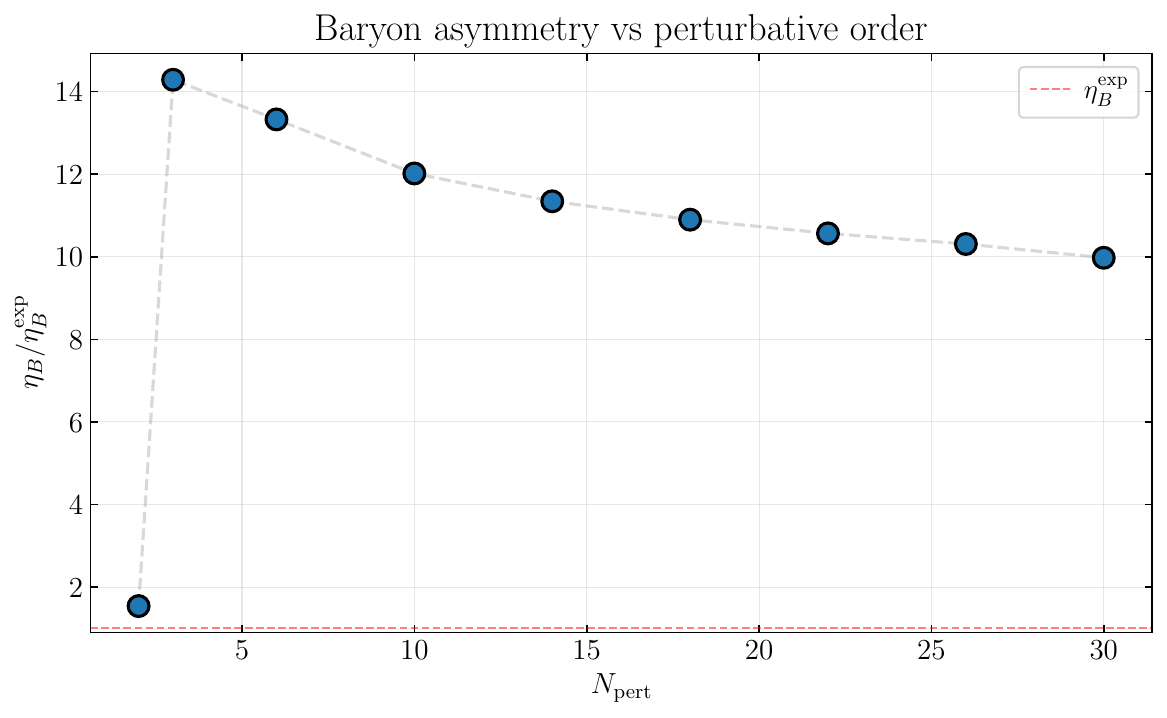}
    \caption{\textbf{Higher–moment convergence and truncation effects.} \emph{Top row} (constant–$R$ closure): 
    (left) $\mu_{B_L}(u)$ for $N_{\rm PERT}=2,6,\dots,30$; 
    (middle) cumulative BAU $\int_{-1}^u (d\eta_B/\eta_B^{\rm exp})\,dv$; 
    (right) $\eta_B/\eta_B^{\rm exp}$ versus $N_{\rm PERT}$. 
    \emph{Bottom row} (variance truncation): same panels, demonstrating earlier stabilisation and reduced spread. 
    The stronger drop when going from $N_{\rm PERT}=2$ to $N_{\rm PERT}\ge 6$ is evident in our data, while the overall convergence pattern and the stability of the $2\!+\!4k$ sequence agree with the findings of Ref.~\cite{Kainulainen:2024qpm}.}
    \label{fig:higher-moments}
\end{figure}

In this subsection, we streamline the higher–moment system into a compact, \emph{inverted} form that is convenient for both analysis and numerics. We {closely} follow the structure of Sec.~3.2 in Ref.~\cite{Kainulainen:2024qpm}, adapting the notation to our conventions and providing benchmark comparisons and convergence diagnostics. For each species, we assemble the CP-odd moment variables into
$
w(z)=\big(\mu,\,u_1,\,u_2,\,\dots,\,u_{\ell-1}\,\big)^{\!\top},
$. Stacking the \(\ell\) equations (one for \(\mu\) and \(\ell-1\) for moments), the system takes the form as in Eq.~\eqref{eq: diffusion system matrix}
with $A$ and $B$ now \(\ell\!\times\!\ell\) matrices of the following form\footnote{We believe that the $1/2$ in the definition of the $B$ matrix in \cite{Kainulainen:2024qpm} is a typo and should not be there.}
\begin{equation}
\label{eq:Ahat}
{A}\;=\;
\begin{pmatrix}
 -D_1 & 1 & 0 & \cdots & 0 \\
 -D_2 & 0 & 1 & \cdots & 0 \\
 \vdots & \vdots & \ddots & \ddots & \vdots \\
 -D_{\ell-1} & 0 & \cdots & 0 & 1 \\
 -D_{\ell} & R_1 & \cdots & R_{\ell-2} & R_{\ell-1}
\end{pmatrix}\!,
\qquad
B=(\partial_z m^2)\,
\begin{pmatrix}
Q_1\,v_w\gamma_w & 0 & 0 & \cdots & 0 \\[3pt]
Q_2\,v_w\gamma_w & \bar R & 0 & \cdots & 0 \\[3pt]
Q_3\,v_w\gamma_w & 0 & 2\,\bar R & \cdots & 0 \\[3pt]
\vdots & \vdots & \ddots & \ddots & \vdots \\[3pt]
Q_{\ell}\,v_w\gamma_w & 0 & 0 & \cdots & (\ell-1)\,\bar R
\end{pmatrix}\!,
\end{equation}
and vector sources/collisions \(\mathcal {S}_\ell,\,\delta \mathcal{C}_\ell\) given component–wise by the CP-odd semiclassical source (built from \(Q^{8\!o}_{\ell},Q^{9\!o}_{\ell}\)) and by the moment–reduced collision integrals. 

It is useful, especially numerically, to write explicitly the inverse matrix of the matrix $A$, which reads
\begin{equation}
\label{eq:Ainv}
{A}^{-1}\;=\;\frac{1}{D_{\ell}}
\begin{pmatrix}
 R_1 & R_2 & \cdots & R_{\ell-1} & -1 \\
 R_1D_1 & R_2D_1 & \cdots & R_{\ell-1}D_1 & -D_1 \\
 \vdots & \vdots & \ddots & \vdots & \vdots \\
 R_1D_{\ell-1} & R_2D_{\ell-1} & \cdots & R_{\ell-1}D_{\ell-1} & -D_{\ell-1}
\end{pmatrix}
\;+\;
\begin{pmatrix}
 0 & 0 & \cdots & 0 \\
 1 & 0 & \cdots & 0 \\
 \vdots & \ddots & \ddots & \vdots \\
 0 & \cdots & 1 & 0
\end{pmatrix}\!,
\end{equation}
where we defined 
\begin{equation}
\label{eq:Ddet}
D_{\ell}\;\equiv\;(-1)^{\ell}\det{A}\;=\;D_{\ell}\;-\;\sum_{i=1}^{\ell-1} R_i\,D_i \ .
\end{equation}
\noindent
Therefore, we can write the full diffusion system in the following compact form
\begin{equation}
w' = {A}^{-1}\!\left(\mathcal{S} \;+\; \delta \mathcal{C} -{B}w\right) .
\label{eq: compact form}
\end{equation}
Following the strategy of Ref.~\cite{Kainulainen:2024qpm} (their Sec.~8), we performed a systematic moment expansion to assess both convergence and truncation–scheme systematics of our transport network. Concretely, we solved the longitudinal hierarchy up to $N_{\rm PERT}\!=\!\ell_{\max}\!=\!30$ using the same two closures considered in \cite{Kainulainen:2024qpm}—a constant–$R$ closure and the variance truncation—and we monitored (i) the baryon–left chemical potential $\mu_{B_L}(z)$, (ii) the cumulative build–up of the BAU as a function of the compactified coordinate $u\!\in\![-1,1]$, and (iii) the final $\eta_B$ versus $N_{\rm PERT}$. Our qualitative trends mirror those reported in \cite{Kainulainen:2024qpm}: increasing $N_{\rm PERT}$ reduces $\mu_{B_L}$ and stabilises $\eta_B$, with the variance truncation systematically more robust. Quantitatively, however, the transition from the canonical two–moment system to multi–moment truncations is \emph{more pronounced} in our setup, leading to a sharper suppression of $\mu_{B_L}$ as $N_{\rm PERT}$ increases. This is visible in the variance–closure profiles of $\mu_{B_L}(u)$ (top–right panel of Fig.~\ref{fig:higher-moments}), where the $N_{\rm PERT}=2$ curve lies noticeably above the $N_{\rm PERT}\ge 6$ family and the separation persists across the wall region.  The same pattern appears in the cumulative integral for $\eta_B(u)$ and in $\eta_B$ versus $N_{\rm PERT}$, for both closures, with the variance scheme exhibiting earlier saturation and smaller spread (top row and bottom–middle/right panels of Fig.~\ref{fig:higher-moments}).\!\!

\begin{figure}
  \centering
  \includegraphics[width=\linewidth]{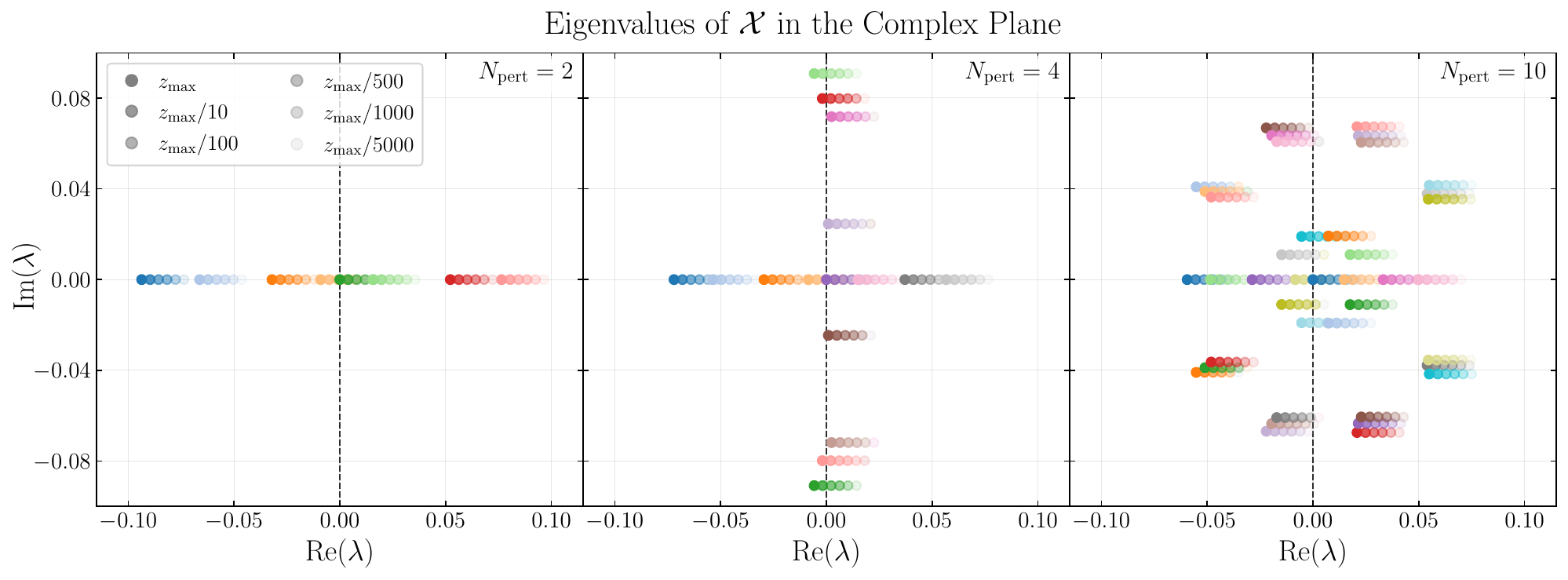}
  \caption{Eigenvalues \(\lambda\) of the far–from–wall propagation matrix \(\mathcal X\) in the complex plane for three truncations: \(\ell_{\max}=2\) (left), \(4\) (middle), and \(10\) (right). The \(\ell_{\max}=4\) spectrum exhibits conjugate pairs with \(|\Re\lambda|\ll |\Im \lambda|\), which produce oscillatory spatial behaviour, while \(\ell_{\max}=2\) and \(10\) are dominated by real, more strongly damped modes. Opacity illustrates changes in the eigenvalues across the wall. This aligns with Fig.~6 of \cite{Kainulainen:2024qpm}. }
  \label{fig:eigs}
\end{figure}

A second point of agreement with Ref.~\cite{Kainulainen:2024qpm} concerns the spectrum of the \emph{asymptotic evolution matrix} $\mathcal X$. Recall that, from \eqref{eq: compact form}, the far–from–wall limit (constant background, vanishing sources) yields the autonomous system
\[
w'(z)=\mathcal X\,w(z),\qquad \mathcal X=\mathcal X(x\!\equiv\!m/T,\,v_w;\,\text{closure}),
\]
with $\mathcal X$ a sparse matrix fixed by the chosen truncation/closure. The large–$|z|$ behaviour is $w(z)\propto e^{\lambda z}$ with $\lambda\in\text{spec}(\mathcal X)$, and physical boundary conditions select the decaying eigenmodes (the sign of $\Re\lambda$ required for decay is set by the wall orientation). We find the most stable truncation sequence to be $N_{\rm PERT}=2+4k$ ($k\in\mathbb{N}$): along $2,6,10,\ldots$ the eigenvalues cluster away from the imaginary axis, as can be seen from Fig.~\ref{fig:eigs}, so $\Re\lambda$ remains safely negative and oscillatory tails are strongly damped. Numerically, this manifests as monotonic profiles and smooth convergence of the BAU. In the complementary sequences, one encounters more conjugate pairs with $|\Re\lambda|\ll|\Im\lambda|$. This makes the computation numerically unstable, and every prediction does not make much sense, so we are reporting the results only from the stable sequence.

Despite the positive convergence diagnostics, we encountered practical difficulties when integrating the full diffusion network at large $N_{\rm PERT}$. In our implementation, the boundary conditions at $z=\pm N L_w$ are enforced by introducing \emph{constant-in-$z$} Lagrange multiplier fields, $\lambda_{\mu_i}$ and $\lambda_{v_i}$, obeying $\partial_{\tilde z}\lambda_{\mu_i}=\partial_{\tilde z}\lambda_{v_i}=0$ and entering the flow equations as linear “boundary sources,” e.g.
\[
\partial_{\tilde z} \tilde\mu_i=\dots+\lambda_{\mu_i}(\tilde z+N\tilde L_w),\qquad
\partial_{\tilde z} u_i=\dots+\lambda_{u_i}(\tilde z-N\tilde L_w),
\]
with $\tilde L_w=L_w/T$ and the sign choice in the $v_i$–equation made for stability, see Appendix~\ref{app: lagrange multiplier} for more details. This augmentation effectively doubles the state dimension and inserts blocks with vanishing characteristic speeds, which amplifies stiffness and degrades the condition number of the discretised operator as $N_{\rm PERT}$ grows. In practice, sporadic instabilities can be traced to the interplay between this doubling and nearly defective sub-blocks associated with eigenmodes having small $|\Re\lambda|$ in the asymptotic evolution matrix, so that the multiplier–coupled rows/columns render the overall system close to singular. A robust cure is to eliminate the multipliers analytically, or to replace them with constraint imposition strategies that do not enlarge the dynamical state. Implementing these improvements is feasible but beyond the scope of the present release. A refined convergence study and a cross–validation of reliable bands against the moment methodology of Ref.~\cite{Kainulainen:2024qpm} will be presented in the updated version \texttt{BARYONETv2}.

\section{Conclusions and outlook}
\label{sec:Conclusions}

In this work we have delivered a \emph{self-contained} and pedagogical review of electroweak baryogenesis in the semiclassical WKB framework and turned it into a reproducible, public pipeline. Starting from first principles, we derived the gauge-invariant WKB kinematics (shifted dispersion relations, group velocities, and semiclassical forces) for fields with space-dependent complex masses across a planar bubble wall, and built the stationary Boltzmann equation in the wall frame. We then performed the CP projection and a controlled moment reduction to obtain a compact fluid network for chemical potentials and velocity perturbations, expressed entirely in terms of \emph{universal thermal kernels}. A key outcome is a uniform notation and set of conventions that reconcile and organise three decades of approaches (FH04, FH06, CK, KV24) within a single language, including a precise operational definition of the final baryon asymmetry $\eta_B$ with a transparent weak–sphaleron profiling.

On the phenomenology side, we implemented these results in the open-source code \texttt{BARYONET}, providing an \emph{end-to-end}, automated pipeline from wall profiles (masses, phases, derivatives) to $\eta_B$. The collision sector is closed in a model-independent way via thermally averaged rates, enabling consistent applications across a broad class of electroweak-scale extensions. We validated the framework by reproducing representative results from the literature and by supplying benchmark overlays for widely studied scenarios—singlet extensions of the SM (including variants), two-Higgs-doublet models, and Higgs–$\phi^6$ setups—thereby establishing both correctness and portability of the implementation.

A central technical contribution of this work lies in the appendices. There we clarify and, where necessary, rederive the present literature, the ingredients that have long underpinned EWBG computations—\emph{diffusion constants}, \emph{Yukawa-mediated relaxation rates} and \emph{helicity-flip rates}, as well as \emph{strong} and \emph{weak sphaleron} rates—clarifying conventions, reconciling formulae scattered across the literature, and, importantly, \emph{updating all numerical inputs to current parameter values}. These updated, ready-to-use quantities are provided with clear normalisation choices and are wired into \texttt{BARYONET} so that phenomenological studies can rely on a consistent and modern baseline.

Taken together, the self-contained derivation, the unification of the approaches, the public and reproducible code, the explicit validation against prior results, and the systematic update of transport inputs advance EWBG studies from a collection of heterogeneous formalisms to a coherent, documented standard. We hope \texttt{BARYONET} will serve as a reliable reference implementation and a platform for extensions, thereby tightening the connection between microscopic model building and cosmological predictions.

\paragraph{Outlook.}  
Several directions can further improve both the numerical robustness and the physical reach of the present framework:
\begin{enumerate}
    \item \textbf{Numerical methods.} The current enforcement of boundary conditions via constant Lagrange multipliers becomes unstable at large perturbation order. Replacing this scheme with direct constraint elimination or implicit boundary enforcement methods would substantially improve stability and scalability.
    \item \textbf{Flavour-source mixing.} The inclusion of models featuring CP-violating flavour mixing, see \cite{Li:2024mts}, is a natural and essential extension, as such scenarios can enhance the generated baryon asymmetry and probe qualitatively new physical effects.
    \item \textbf{EDMs constraints.} An important future direction will be the inclusion of electric dipole moment (EDM) constraints in our analysis. EDM measurements provide extremely sensitive tests of CP violation, directly connected to the sources responsible for baryogenesis, see \cite{vandeVis:2025efm} for a review. Implementing these bounds will be essential to identify regions of parameter space that are simultaneously compatible with successful electroweak baryogenesis and current experimental limits.
    \item \textbf{Beyond the fluid network.} A more ambitious direction is to rethink the approach entirely by abandoning the moment-truncated Boltzmann hierarchy and solving directly for the full out-of-equilibrium distribution (or its perturbation). This can follow the strategy already successful in the CP-even sector—e.g. spectral discretisations in $z$ and momentum using Chebyshev polynomials (see \cite{Laurent:2022jrs}) or finding the eigenfunctions of the kinetic operator (see~\cite{DeCurtis:2022hlx, DeCurtis:2023hil,DeCurtis:2024hvh}). Such methods would bypass closure ambiguities, capture nonlocal angular structure, and offer systematically improvable accuracy with well-understood convergence diagnostics.
    \item \textbf{Spatial dependence of $\eta_B$.} Computing the full spatial profile of the baryon asymmetry $\eta_B(z)$ across and ahead of the wall would provide valuable insight into the inhomogeneities imprinted by bubble propagation, with potential implications for baryon-number diffusion, baryogenesis efficiency, and subsequent cosmological observables.
\end{enumerate}

    Taken together, these improvements would bring \texttt{BARYONET} from a state-of-the-art WKB solver to a comprehensive numerical laboratory for baryogenesis and transport phenomena in the early Universe. Future releases will incorporate these developments, providing a benchmark for cross-framework validation and a bridge between microscopic particle physics and cosmological phenomenology.

\paragraph{Acknowledgment}
This project grew out of a long journey through the literature, during which I benefited from the insight and generosity of many colleagues. I am profoundly and gratefully indebted to \textbf{Carlo Branchina}, \textbf{Luigi Delle Rose}, and \textbf{Salvador Rosauro-Alcaraz}. Their sharp understanding, perseverance in grasping every aspect, and patience in discussing them with me were not merely helpful but truly decisive—shaping my comprehension of the subject and resolving pivotal technical issues without which this work (and the code that implements it) would simply not exist.

I thank {Thomas Biekötter}, {Francesco Costa}, and {Miguel Vanlasselaer} for thoughtful comments on the draft and for their help during the code’s beta testing phase. I am indebted to {Enrique Fernández-Martínez}, {Toshihiko Ota}, and {Salvador Rosauro-Alcaraz} for illuminating discussions that helped me clarify several foundational aspects of the theory and for pointing me to parts of the literature I had missed. For early advice at the outset of this project, I warmly thank {Benoît Laurent} and {Mateusz Zych}. Finally, I would like to acknowledge {José Miguel No} and {Javi Serra}: this work originated from a joint project with them, without which it would not have been possible.
I also thank Kimmo Kainulainen for clarifications on the historical development of the semiclassical formalism and for constructive comments on the manuscript.

The author acknowledges support from the Spanish Research Agency (Agencia Estatal de Investigación, MCIN/AEI/10.13039/501100011033) via the IFT Severo Ochoa Center of Excellence grant CEX2020-001007-S. In addition, the author is supported by the grant CNS2023-145069 funded by MICIU/AEI/10.13039/501100011033 and by the European Union NextGenerationEU/PRTR.

\newpage 
\appendix
\section{Universal kernel}
In this section, we present a master formula for computing all the thermally averaged functions appearing in the main text, following closely the notation introduced in \cite{Cline:2020jre, Kainulainen:2024qpm}.
All coefficient functions entering the moment equations are Lorentz–invariant
integrals over the equilibrium distribution in the \emph{wall frame}. They depend only on
the local ratio \(x\equiv m/T\) and on the wall velocity \(v_w\).
We use dimensionless plasma–frame variables \(w\equiv E/T\),
\(\tilde p_w =\sqrt{w^2-x^2}\), and \(y\equiv \cos\theta=p_z/|\mathbf p|\in[-1,1]\).
A boost with speed \(v_w\) along \(\hat z\) gives
\begin{equation}
\tilde E=\gamma_w\,(w-v_w\,y\,\tilde p_w),\qquad
\tilde p_z=\gamma_w\,(y\,\tilde p_w-v_w\,w),\qquad
\tilde E_{0z}=\sqrt{\tilde p_z^{\,2}+x^2}.
\label{eq:boosted-vars-A}
\end{equation}
The boosted equilibrium is \(f_{0w}(\tilde E)\), with derivatives
\(f'_{0w}\equiv \partial_{\tilde E}f_{0w}\) and \(f''_{0w}\equiv \partial_{\tilde E}^2 f_{0w}\).
Angle brackets \(\langle\cdot\rangle\) denote the moment averages used in the main text,
normalised by \(N_1\).

\subsection{Master integral and choices of \(V\)}
All coefficient functions can then be written as a two-dimensional integral of a generic form\footnote{In \cite{Cline:2020jre, Kainulainen:2024qpm} they consider $f_{0w}^{(k)}$ as the $k-$th derivative with respect to its argument, but not considering the temperature. This brings down the power of the temperature for each derivative.}
\begin{equation}
    \left \langle {p_z^n \over E^m} V \mathcal{F}_{0w}^{(k)}\right \rangle=T^{n-m-k-1} K(\mathcal F_0;V;n,m)
\end{equation}
where $k=0,1,2$ for $\mathcal{F}_{0w}^{(k)}=f_{0w}^{(k)}$, and the dimensionless master integral
\begin{equation}
K(\mathcal F_0^{(k)};V;n,m)
=
-\frac{3}{\pi^2\,\gamma_w}
\int_{x}^{\infty}\!dw\int_{-1}^{1}\!dy\;
\frac{\tilde p_w\tilde p_z^{\,n}}{E^{m-1}}\;V(w,y;v_w,x)\;\mathcal F_0^{(k)}(\tilde E).
\label{eq:K-master-A}
\end{equation}
For the CP–-odd source one may use the spin–\(s\) or helicity basis, defined by
\begin{equation}
V_s=\frac{|\tilde p_z|}{E_{0z}}
=\frac{|\tilde p_z|}{\sqrt{\tilde p_z^{\,2}+x^2}},
\label{eq:Vs-A}
\end{equation}
\begin{equation}
V_h=V_s^{\,2}\left(1-\frac{x^2}{w^2}\right)^{-1/2}.
\label{eq:Vh-A}
\end{equation}
When no extra structure is needed, one simply takes \(V=1\).
The other functions defined in the main text can be expressed in term of the master integral as follows
\begin{align}
D_\ell(x,v_w)&=K\!\left(f'_{0w};\,1;\,\ell,\ell\right),
\label{eq:Dl-A}\\
T^2Q_\ell(x,v_w)&=\dfrac12 K\!\left(f''_{0w};\,1;\,\ell-1,\ell\right),
\label{eq:Ql-A}\\
T^2Q_\ell^e(x,v_w)&=\dfrac12 K\!\left(f'_{0w};\,1;\,\ell-1,\ell\right),
\label{eq:Ql-e}\\
T^2Q^{8o}_\ell(x,v_w)&=\frac{1}{2}\,K\!\left(f'_{0w};\,V;\,\ell-2,\ell\right),
\label{eq:Q8-A}\\
T^4Q^{9o}_\ell(x,v_w)&=\frac{1}{4}\left[
K\!\left(f'_{0w};\,V;\,\ell-2,\ell+2\right)
-\gamma_w\,K\!\left(f''_{0w};\,V;\,\ell-2,\ell+1\right)
\right],
\label{eq:Q9-A}
\end{align}
where \(V\) is chosen as in \eqref{eq:Vs-A} or \eqref{eq:Vh-A}. For fixed \((x,v_w)\) the kernels
\eqref{eq:Dl-A}-\eqref{eq:Q9-A} are universal and can be tabulated once on a grid.

\begin{figure}
    \centering
    \includegraphics[width=\linewidth]{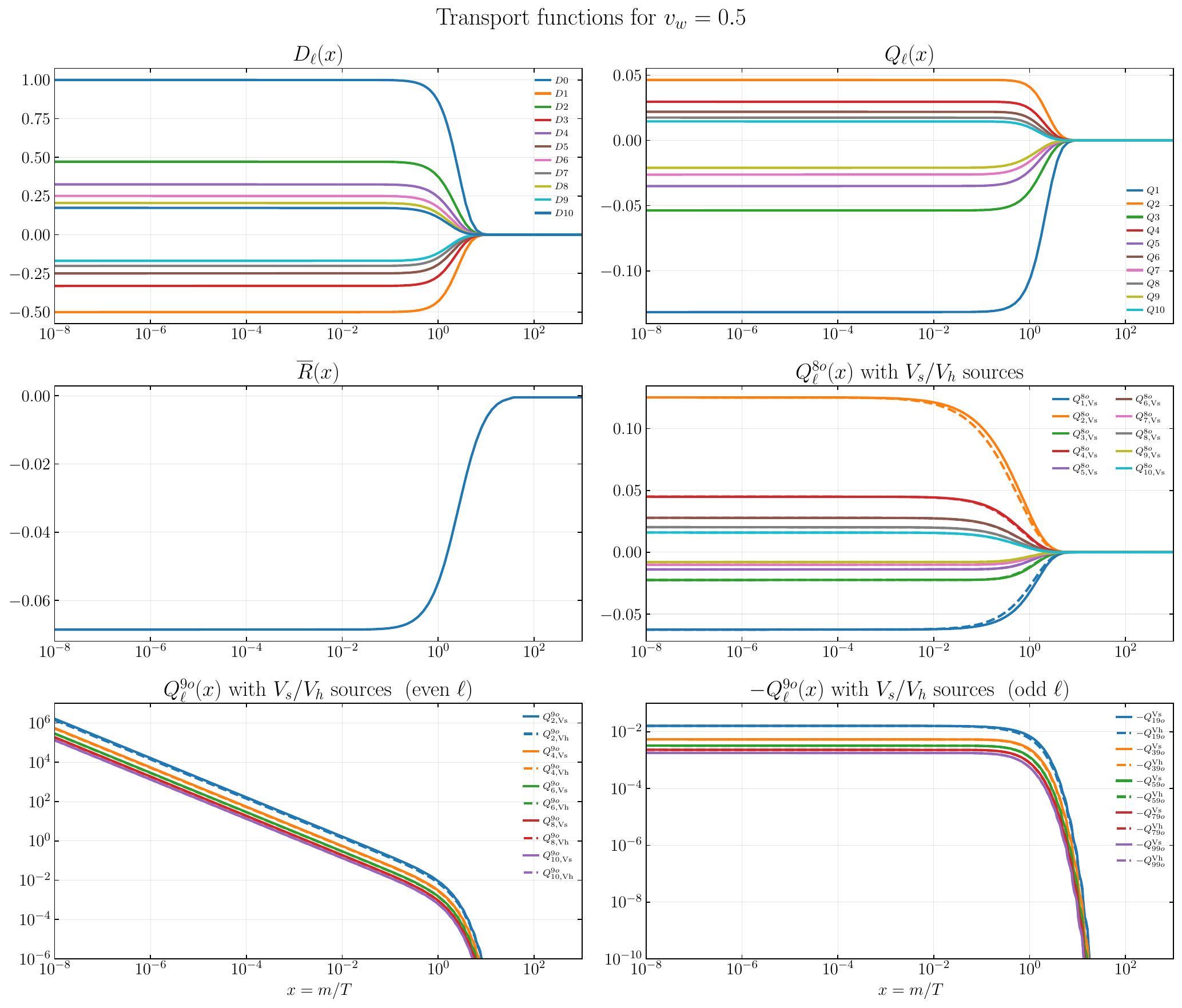}
    \caption{Universal transport kernels at fixed wall velocity \(v_w=0.5\) as functions of \(x=m/T\). 
    \textit{Top left}: \(D_\ell\) for \(\ell=0,\dots,10\). 
    \textit{Top right}: \(Q_\ell\) for \(\ell=1,\dots,10\). 
    \textit{Middle left}: \(\bar R(x)\). \textit{Middle right}: CP–odd source kernels \(Q_\ell^{\,8o}\) for spin/helicity weights \(V_s,V_h\);
    \textit{Bottom (two panels)}: CP–odd source kernels \(Q_\ell^{\,9o}\) for spin/helicity weights \(V_s,V_h\); functions with even/odd \(\ell\) are shown separately to highlight the alternating sign pattern and the large difference in value. 
    These curves are model–independent inputs that depend only on \(x\) and \(v_w\) and are reused across all scenarios.}
    \label{fig:Table_of_function}
\end{figure}

Figure~\ref{fig:Table_of_function} collects the dimensionless functions that enter the two–moment fluid equations: the equilibrium kernels \(D_\ell(x,v_w)\) and \(Q_\ell(x,v_w)\), the function \(\bar R(x,v_w)\), and the source kernels \(Q_\ell^{\,8o}(x,v_w;V)\) and \(Q_\ell^{\,9o}(x,v_w;V)\) for the two standard weights \(V\in\{V_s,V_h\}\) (spin and helicity choices). The example shown fixes \(v_w=0.5\) and scans the mass–to–temperature ratio \(x=m/T\) over many decades; changing \(v_w\) smoothly deforms the curves without altering their qualitative behaviour. These kernels are \emph{universal} (model–independent) and are computed once, tabulated, and then interpolated along the wall. In \texttt{BARYONET}, they can be generated with 
\texttt{function\_interpolators/function\_high\_vw/ ComputeTransportFunctions.cpp}, which produces the tables used by all solvers.

\label{app: kernel}

\section{Linearised collision operators}
\label{app: collision operators}

In this appendix, we derive, step by step and in a model-independent way, the linearised collision terms that enter the transport equations. We closely follow the approximations and derivations of Refs.~\cite{Cline:2000nw,Cline:2001rk,Cline:2020jre,Kainulainen:2024qpm}. 

We consider a generic $2\to 2$ process,
$
i_1\, i_2 \to f_1\, f_2 \,,
$
in order to be explicit, and we track the evolution of the distribution of species $i_1$. The corresponding collision operator can be written as
\begin{equation}
    {C}[f_j]=-\frac{1}{2E_{i_1}}\int_{\{p_{i_2},\,p_{f_{1}},\,p_{f_{2}}\}}|\mathcal{M}|^2(2\pi)^4\delta^{(4)}(p_{i_1}+p_{i_2}-p_{f_1}-p_{f_2})\mathcal{P}[f_{i_{1,2}},f_{f_{1,2}}]\,,
\end{equation}
where $f_j$ denotes the distribution function of species $j$, where $\int_{\{p\}}=\int {d^3p \over 2p_0(2\pi)^3}$, and the statistical factor is
\begin{equation}
    \mathcal{P}[f_{i_{1,2}},f_{f_{1,2}}]\equiv f_{i_1}f_{i_2}(1\mp f_{f_1})(1\mp f_{f_2})-f_{f_1}f_{f_2}(1\mp f_{i_1})(1\mp f_{i_2})\,.
    \label{eq:statistical_factor}
\end{equation}
The upper (lower) signs correspond to fermions (bosons). Equation~\eqref{eq:statistical_factor} simply encodes detailed balance with quantum-statistical blocking (for fermions) or enhancement (for bosons).

To perform a systematic gradient expansion, we split each distribution into an equilibrium piece—allowing for a small chemical potential in the wall frame—and a perturbation
\begin{equation}
    f_i\equiv f_i(\mu_i)+\delta f_i={1 \over \mathrm{exp}\left[\gamma_w(E_i+v_wp_{j,z})/T-\mu_i/T\right]}+\delta f_i\,.
\end{equation}

For a generic $2\!\to\!2$ process with incoming legs $\mathrm{In}=\{i_1,i_2\}$ and
outgoing legs $\mathrm{Out}=\{f_1,f_2\}$, we linearize around equilibrium
$f=f(\mu)+\delta f$ and write
\begin{equation}
\label{eq:P_linear_ck_general}
\mathcal{P}[f_{i_{1,2}},f_{f_{1,2}}]
\simeq
\mathcal{P}[f_{i_{1,2}}(\mu_{i_{1,2}}),f_{f_{1,2}}(\mu_{f_{1,2}})]
+\sum_{k\in\{i_1,i_2,f_1,f_2\}} c_k\,\delta f_k,
\qquad
c_k \equiv \left.\frac{\partial \mathcal{P}}{\partial f_k}\right|_{f=f^0},
\end{equation}
where $f^0=f(\mu=0)$. Each coefficient $c_k$ is the sum of a \emph{loss} term and a \emph{gain} term and,
keeping the upper/lower sign for Fermi--Dirac/Bose--Einstein statistics, is given by
\begin{equation}
\label{eq:ck_general}
c_k \;=\;
\begin{cases}
\displaystyle
\underbrace{\prod_{\,\ell\in\mathrm{In}\setminus\{k\}} f_\ell^0
\prod_{\,\ell\in\mathrm{Out}} \big(1\mp f_\ell^0\big)}_{\text{loss}}
\;\;\pm\;\;
\underbrace{\prod_{\,\ell\in\mathrm{Out}} f_\ell^0
\prod_{\,\ell\in\mathrm{In}\setminus\{k\}} \big(1\mp f_\ell^0\big)}_{\text{gain}},
& k\in\mathrm{In}, \\[2.2ex]
\displaystyle
-\underbrace{\prod_{\,\ell\in\mathrm{Out}\setminus\{k\}} f_\ell^0
\prod_{\,\ell\in\mathrm{In}} \big(1\mp f_\ell^0\big)}_{\text{loss}}
\;\;\mp\;\;
\underbrace{\prod_{\,\ell\in\mathrm{In}} f_\ell^0
\prod_{\,\ell\in\mathrm{Out}\setminus\{k\}} \big(1\mp f_\ell^0\big)}_{\text{gain}},
& k\in\mathrm{Out}.
\end{cases}
\end{equation}
For $k=i_1$ and $k=i_2$, Eq.~\eqref{eq:ck_general} gives, for example
\begin{align}
\label{eq:ck_i1_example}
c_{i_1} &= f^0_{i_2}\,\big(1\mp f^0_{f_1}\big)\big(1\mp f^0_{f_2}\big)
\;\;\pm\;\; f^0_{f_1} f^0_{f_2}\,\big(1\mp f^0_{i_2}\big),\\
\label{eq:ck_i2_example}
c_{i_2} &= f^0_{i_1}\,\big(1\mp f^0_{f_1}\big)\big(1\mp f^0_{f_2}\big)
\;\;\pm\;\; f^0_{f_1} f^0_{f_2}\,\big(1\mp f^0_{i_1}\big).
\end{align}

\noindent To make the structure more explicit, we use the statistical identity
$1\mp f \;=\; f\,\exp\!\big[\gamma_w(E+v_w p_z)/T\big]
$, together with
energy--momentum conservation,
$
E_{i_1}+E_{i_2}=E_{f_1}+E_{f_2},
~
\mathbf{p}_{i_1}+\mathbf{p}_{i_2}=\mathbf{p}_{f_1}+\mathbf{p}_{f_2}.
$
After a straightforward rearrangement, the result can be cast as
\begin{equation}
\label{eq:P_first_order_rearranged}
\begin{split}
\mathcal{P}[f_{i_{1,2}},f_{f_{1,2}}]
\simeq\;&
\mathcal{P}[f_{i_{1,2}}(\mu_{i_{1,2}}),f_{f_{1,2}}(\mu_{f_{1,2}})]
+{f^0_{i_1} f^0_{i_2}}
\left(
\frac{\delta f_{i_1}}{f^0_{i_1}} + \frac{\delta f_{i_2}}{f^0_{i_2}}
- \frac{\delta f_{f_1}}{f^0_{f_1}} - \frac{\delta f_{f_2}}{f^0_{f_2}}
\right),
\end{split}
\end{equation}
where we used the fact that in the Maxwell--Boltzmann (MB) limit and using
energy conservation, one finds
$
f_{f_1}^0 f_{f_2}^0\,
\exp\!\Big[\gamma_w\big(E_{i_1}+E_{i_2}+v_w(p_{i_1,z}+p_{i_2,z})\big)\Big]=1,
$ and exponentially Boltzmann suppressed
terms have been dropped, as they are subleading at this order.

\subsection{Linearised out-of-equilibrium contributions}

To describe departures from equilibrium, we expand the collision operator to first
order in the perturbations $\delta f_j$.  Starting from
Eq.~\eqref{eq:statistical_factor}, only terms containing a single $\delta f$ are
retained.  These contributions naturally split into two parts: one proportional to
the deviation of the species under consideration, $\delta f_{i_1}$, and another
proportional to the perturbations of its interaction partners, weighted by the
equilibrium distribution of $i_1$.  The collision term can then be expressed as
\begin{equation}
\label{eq:coll_delta}
\begin{split}
C[f_i,f_j] \;\simeq\;&\;
-\,\frac{\delta f_{i_1}}{2E_{i_1}}
\int_{\{p_{i_2},\,p_{f_1},\,p_{f_2}\}}
|\mathcal{M}|^2\,
f^0_{i_2}\,
(2\pi)^4\,
\delta^{(4)}\!\!\left(\sum p_k\right)
\\[3pt]
&+\;
f^0_{i_1}
\Bigg[
\int_{\{p_{i_2}\}}\!
\delta f_{i_2}\,\mathcal{A}(p_{i_1},p_{i_2})
-
\sum_{n}
\int_{\{p_{f_n}\}}\!
\delta f_{f_n}\,\mathcal{G}_{f_n}(p_{i_1},p_{f_n})
\Bigg]
\\[3pt]
\equiv\;&\;
-\,\Gamma(p_{i_1})\,\delta f_{i_1}
+\,\Delta \mathcal F\,,
\end{split}
\end{equation}
where the first term represents the direct relaxation of $i_1$ toward equilibrium,
and $\Delta F$ collects the influence of the perturbations of the other
species involved in the reaction.  The functions
$\mathcal{A}(p_{i_1},p_{i_2})$ and $\mathcal{G}_{f_n}(p_{i_1},p_{f_n})$ are
phase-space kernels built from the squared matrix element $|\mathcal{M}|^2$ and
equilibrium distributions.  They are obtained by taking the derivative of the
statistical factor $\mathcal{P}[f_j]$ with respect to $f_{i_2}$ or $f_{f_n}$ and
then evaluating at equilibrium.  The kernel functions appearing in Eq.~\eqref{eq:coll_delta} take the standard
forms
\begin{equation}
\label{eq:A_G_standard}
\begin{aligned}
\mathcal{A}(p_{i_1},p_{i_2})
&=
\frac{1}{2E_{i_1}}
\int_{\{p_{f_1},\,p_{f_2}\}}
(2\pi)^4
\delta^{(4)}\!\!\left(\sum p_k\right)
|\mathcal{M}|^2\,
(1\mp f^0_{f_1})(1\mp f^0_{f_2}),
\\[4pt]
\mathcal{G}_{f_n}(p_{i_1},p_{f_n})
&=
\frac{1}{2E_{i_1}}
\int_{\{p_{i_2},\,p_{f_m\neq f_n}\}}
(2\pi)^4
\delta^{(4)}\!\!\left(\sum p_k\right)
|\mathcal{M}|^2\,
f^0_{i_2}\,
(1\mp f^0_{f_m}).
\end{aligned}
\end{equation}
The first kernel, $\mathcal{A}$, encodes the effect of the incoming partner $i_2$ on
the relaxation of $i_1$, while $\mathcal{G}_{f_n}$ describes the feedback from the
final-state particles $f_n$. In the following, we neglect the kernel functions
$\mathcal{A}$ and $\mathcal{G}$ altogether; only the effective
interaction rates $\Gamma_{ij}$ are retained in the subsequent analysis.

\medskip
\noindent
The relaxation rate $\Gamma(p_{i_1})$ appearing in Eq.~\eqref{eq:coll_delta}
corresponds to the familiar thermal scattering rate of the process
$i_1 i_2 \!\to\! f_1 f_2$:
\begin{equation}
\label{eq:Gamma_new}
\begin{split}
\Gamma_{i_1,i_2\to f_1,f_2}
&=\frac{1}{2E_{i_1}}
\int_{\{p_{i_2}\}} f^0_{i_2}\,\langle\sigma v\rangle
\\[3pt]
&=\frac{1}{2E_{i_1}}
\int_{\{p_{i_2}\}} f^0_{i_2}
\int_{\{p_{f_1},\,p_{f_2}\}}
(2\pi)^4\,\delta^{(4)}\!\!\left(\sum p_k\right)\!
|\mathcal{M}|^2,
\end{split}
\end{equation}
where $\langle\sigma v\rangle$ is the standard thermal average of the cross section
times relative velocity.  As usual, $|\mathcal{M}|^2$ is summed or averaged over
internal degrees of freedom.  The same reasoning applies to processes with any
number of external particles.

\paragraph{Moment of the linearised collision operator.}
When computing the velocity moments of the collision operator (see Sec.~\ref{subsec:collisions}), a standard and consistent procedure is to replace the momentum–dependent $\Gamma(p)$ that appear in the linearised operator with their \emph{thermally averaged, per–particle} counterparts $\widehat{\Gamma}$—the so–called relaxation–time approximation (RTA). Invoking detailed balance for the equilibrium pieces (so that the collision term vanishes on equilibrium), for the first two moments, one obtains
\begin{equation}
    \big\langle \delta {C}\big\rangle \simeq 0\,,\qquad \quad
    \left\langle \dfrac{p_z}{E_{i_1}}\,\delta {C}\right\rangle \simeq -\,\left\langle \dfrac{p_z}{E_{i_1}}\,\delta f_{i_1}\right\rangle\,\kappa^X \,\widehat\Gamma
    \;=\; -\,u_1\kappa^X \,\widehat\Gamma\,,
\label{eq:averages_CLine}
\end{equation}
where $\langle \cdot \rangle$ denotes the thermal average as in the main text and, in the last equality, we used the definition of the velocity perturbation $u_1 \equiv \langle (p_z/E_{0})\,\delta f\rangle$. For the definition of the coefficient $\kappa^X$ see Sec~\ref{subsec:collisions}. The first relation reflects particle-number conservation in elastic $2\!\to\!2$ scatterings (no source in the zeroth moment at linear order), while the second encodes the relaxation of the momentum perturbation at a rate set by $\widehat \Gamma$.

\subsection{Linearised chemical potential contributions}

We now turn to the contribution proportional to $\mathcal{P}[f_{i_{1,2}}(\mu_{i_{1,2}}),f_{f_{1,2}}(\mu_{f_{1,2}})]$ in Eq.~\eqref{eq:statistical_factor}. As in the previous subsection, we expand the distributions to first order in gradients. Writing $\xi_i \equiv \mu_i/T$, one finds
\begin{equation}
    \begin{split}
        f_i(\mu_i)\simeq f_i^0\left[1+(1\mp f_i^0) \xi_i\right]\,,
    \end{split}
\end{equation}
where the factor $(1\mp f_i^0)$ encodes the fermionic blocking or bosonic enhancement. Using again Maxwell--Boltzmann statistics for the equilibrium pieces (so that $1\mp f_i^0 \to 1$ in the linearised expansion) and energy conservation, the statistical factor reduces to
\begin{equation}
    \mathcal{P}[f_{i_{1,2}}(\mu_{i_{1,2}}),f_{f_{1,2}}(\mu_{f_{1,2}})]\simeq f_{i_1}^0 f_{i_2}^0\left[\xi_{i_1}+\xi_{i_2}-\xi_{f_1}-\xi_{f_2}\right]\,,
\end{equation}
which is the familiar driving combination of chemical potentials dictated by detailed balance. Inserting this into the collision integral and using the same steps as for the $\delta f_i$ piece yields
\begin{equation}
    \delta {C}[\xi_{i_{1,2}},\xi_{f_{1,2}}]\simeq - f^0_{i_1}\sum_{i,j}\Gamma_{i \to j}  s_{ij} \xi_j\,,
\end{equation}
with $s_{ij}=\pm 1$ for initial (final) state particles, respectively. This term provides the number-changing (source/sink) component of the collision operator induced by departures from chemical equilibrium.

\subsection{Thermal averaging}

Collecting the contributions proportional to $\delta f_i$ and to the chemical potentials $\mu_i$, the collision operator for species $a$—including both number–changing channels and elastic scatterings—can be organized as in Eqs.~\eqref{eq:integrand-raw} and~\eqref{eq:integrand-raw 1}, where the sum runs over all relevant processes and $\Gamma^{\rm wf}$ denotes the wall–frame rate for $1\!\to\!2$ or $2\!\to\!2$ reactions. Following Ref.~\cite{Cline:2000nw}, we drop the $\Delta \mathcal F$ term from Eq.~\eqref{eq:coll_delta}: in the linearised moment expansion and under our working assumptions, it does not contribute to the first two moments.

We replace the momentum–dependent rates by their thermally averaged, per–particle counterparts $\widehat\Gamma$ (see Sec.~\ref{subsec:collisions} for the averaging conventions and normalisation $\kappa^X$). Projecting onto the first two velocity moments then gives
\begin{equation}
  \begin{aligned}
    \delta \mathcal{C}_1 \;=\; \big\langle \delta C \big\rangle 
      &\simeq - \langle f_{0w}\rangle \,\kappa^X \sum_{i,j}\widehat \Gamma_{i \to j}\, s_{ij}\,\xi_j
       \;=\; -K^{(0)}\,\kappa^X \sum_{i,j}\widehat \Gamma_{i \to j}\, s_{ij}\,\xi_j \;,\\[4pt]
    \delta \mathcal{C}_2 \;=\; \Big\langle \frac{p_z}{E_0}\,\delta C \Big\rangle 
      &\simeq -\Big\langle\frac{p_z}{E_0} f_{0w}\Big\rangle \kappa^X \sum_{i,j}\widehat \Gamma_{i \to j}\, s_{ij}\,\xi_j
          \;-\; \kappa^X\,\widehat \Gamma_{\rm tot}\,\Big\langle\frac{p_z}{E_0}\,\delta f\Big\rangle \\[2pt]
      &= -K^{(1)} \kappa^X \sum_{i,j}\widehat \Gamma_{i \to j}\, s_{ij}\,\xi_j
         \;-\; \kappa^X\,\widehat \Gamma_{\rm tot}\,u_1 \;,
  \end{aligned}
\end{equation}
and similarly for higher moments. Here $K^{(0)}\!\equiv\!\langle f_{0w}\rangle$ and $K^{(1)}\!\equiv\!\big\langle (p_z/E_0)\,f_{0w}\big\rangle$ are the universal projection weights; $u_1\!\equiv\!\big\langle (p_z/E_0)\,\delta f\big\rangle$ is the first velocity moment; $s_{ij}$ keeps track of the sign/stoichiometry of species $j$ in channel $i$; and $\widehat\Gamma^{\rm tot}_a\!=\!\sum_{i\in a}\big(\widehat\Gamma_i^{\rm el}+\widehat\Gamma_i^{\rm inel}\big)$ is the total (per–particle) scattering rate for species $a$. Inelastic processes relax the appropriate linear combinations of $\xi_j$ in both moments, while the elastic (and elastic part of inelastic) contribution appears as a flow–damping term in the second moment through $\widehat\Gamma^{\rm tot}_a\,u_{1,a}$.

\medskip
\noindent\textit{Remark on conventions.}
The compact form above can look different from the presentation in Ref.~\cite{Kainulainen:2024qpm}. Apparent discrepancies typically trace back to (i) whether wall– or rest–frame rates are inserted before averaging, (ii) the precise normalisation ($\kappa^X$ vs.\ $12/T^3$ or $1/N_1$) and averaging measure, (iii) the sign and assignment of $s_{ij}$, (iv) whether the $\delta F$ term is retained, and (v) MB versus quantum (FD/BE) weights in the linearisation. Accounting for these choices reconciles the formulas without affecting the underlying physics.

\medskip\noindent We are grateful to Salvador Rosauro-Alcaraz for incisive clarifications on the derivation of the linearised collision integrals and for guidance on aligning our conventions and averaging prescriptions with the existing literature.

\section{Diffusion constants}
\label{app:diffusion}
In this appendix, we revisit in detail the calculation of diffusion constants for relativistic particles in a hot plasma. 
These quantities were computed systematically for the first time in 
\cite{Joyce:1994zn} and \cite{Huet:1995sh}, and here we provide a pedagogical rederivation. 
We follow a linear-response approach starting from the Boltzmann equation, 
clarifying how the transport rate is defined, how scattering amplitudes 
enter the calculation, and how thermal masses regulate the infrared divergences 
arising from soft gauge-boson exchange.

We consider the relation between a particle–number current and a small spatial gradient of its density. 
Let \(n(\mathbf{x},t)\) be the (particle minus antiparticle) number density of the species, and let \(\mathbf{J}(\mathbf{x},t)\) be its spatial number current density. 
Decomposing the current into a convective part carried by the bulk flow \(\mathbf{u}\) and a diffusive part, \(\mathbf{J}=n\,\mathbf{u}+\mathbf{J}_d\), the diffusion constant \(D\) is defined by Fick’s law
\begin{equation}
    \mathbf{J}_d \;=\; -\,D\,\boldsymbol{\nabla} n \, .
\end{equation}
In kinetic theory, the evolution of the distribution function $f(x,p)$ is 
governed by the Boltzmann equation
\begin{equation}
    \frac{p^\mu}{p^0}\,\partial_\mu f(x,p) = - C[f](x,p) ,
\end{equation}
where $C[f]$ is the collision integral\footnote{In this subsection only we adopt the sign convention used in much of the literature: we factor the overall minus sign out of the collision integral and define the (linearised) collision operator with the opposite sign. This is purely notational—the minus sign simply migrates from the explicit definition of the diffusion coefficient \(D\) to the total relaxation rate \(\Gamma_t\)—so all physical results are unchanged.
}. 
We assume a slowly varying local chemical potential $\xi(x)=\beta\mu(x)$, with $\beta=1/T$, so that the equilibrium distribution takes the form
\begin{equation}
    f_0(p,x) = \frac{1}{e^{\beta p^0 + \xi(x)} + 1} \qquad 
    \text{for fermions}.
\end{equation}
We now linearise around the equilibrium
\begin{equation}
    f(p,x) = f_0(p,x) + \delta f(p,x), \qquad |\delta f|\ll f_0 .
\end{equation}
Because the perturbation is induced by a spatial gradient in $\xi(x)$ 
(say, in the $x$-direction), the most general ansatz consistent with rotational symmetry in the transverse plane is
\begin{equation}
    \delta f_{\mathbf{p}} = g(p)\, \frac{p_x}{p^0},
\end{equation}
where $g(p)$ is an unknown function of the momentum magnitude.
Inserting this ansatz and linearising the Boltzmann equation gives
\begin{equation}
    f_0'(p^0)\,\partial_x \xi \;\simeq\; -\, g(p)\,\Gamma_t(p),
\end{equation}
which defines the \emph{transport rate} $\Gamma_t(p)$. 
It characterises the relaxation of the anisotropic part of the distribution.

\subsection{Expression for the diffusion constants}

The number density is related to the chemical potential by
\begin{equation}
    n \simeq n_0 - \frac{T^3}{12}\,\xi \qquad 
    \text{(relativistic fermions)}.
\end{equation}
The associated current in the $z$-direction is
\begin{equation}
    J_z = \int \frac{d^3p}{(2\pi)^3}\, \delta f(p)\, \frac{p_z}{p^0}.
\end{equation}
Combining these expressions yields the master formula
\begin{equation}
    D = \frac{12}{T^3} \int \frac{d^3p}{(2\pi)^3}\,
    \frac{-f_0'(p^0)}{\Gamma_t(p)}\left(\frac{p_z}{p^0}\right)^2.
    \label{eq:D-master}
\end{equation}
Equation \eqref{eq:D-master} shows that $D$ is essentially the 
momentum-space average of the inverse transport rate, weighted by the thermal distribution.

For a generic $2\to2$ scattering process with incoming states $\{p,k\}$ 
and outgoing states $\{p',k'\}$, the collision integral is
\begin{equation}
    C[f] = \frac{1}{2p_0}\int_{\{p',k,k'\}}\;
    \delta^{(4)}(p+k-p'-k')\,|{\cal M}|^2\, {\cal P}[f],
\end{equation}
where $\int_{\{p\}}=\int {d^3p \over 2p_0(2\pi)^3}$, and where the statistical factor has been defined in Eq. \eqref{eq:statistical_factor}. Linearising and projecting onto the perturbation ansatz leads to
\begin{equation}
    \Gamma_t(p) = \frac{1}{2p_0^3}
    \int_{\{p',k,k'\}}\,(2\pi)^4\delta^{(4)}\Big(\sum p_i\Big)\,
    |{\cal M}|^2\,(-f_0')\,(p\cdot p').
\end{equation}
We now evaluate this expression in the centre–of–mass (CM) frame, where the kinematics simplify. Throughout, the subscript \(t\) indicates that we are considering the \(t\)-channel contribution.

The dominant contribution arises from $t$-channel exchange of gauge bosons. 
After performing spin and colour sums, one obtains, for massless external fermions,
\begin{align}
    |{\cal M}_B|^2 &= \frac{80\,Y^2\,g_1^4\,s^2}{(t - M_B^2)^2},\\
    |{\cal M}_W|^2 &= \frac{36\,g_2^4\,s^2}{(t - M_W^2)^2},\\
    |{\cal M}_g|^2 &= \frac{32\,g_3^4\,(s^2+u^2)}{(t - M_g^2)^2}.
\end{align}
Here $s=(p+k)^2$, $t=(p-p')^2$, $u=(p-k')^2$, 
$Y$ is the hypercharge of the external fermion, 
and $g_{1,2,3}$ are the gauge couplings for $U(1)_Y$, $SU(2)_L$, and $SU(3)_c$.
The denominators contain the \emph{Debye masses} $M_{B,W,g}$ which regulate the 
infrared divergence at $t=0$.

\begin{WideBox}[label={box:Debye masses}]{Debye screening masses in the SM-symmetric phase}
The Debye (electric screening) mass is defined from the longitudinal self-energy in the static, soft limit
\begin{equation}
m_D^2 \equiv -\Pi_{00}(\omega=0,\mathbf{k}\to 0).
\end{equation}
At leading order (hard-thermal-loop) for a gauge coupling \(g\),
\begin{equation}
m_D^2 = g^2 T^2\Big[\frac{C_A}{3} + \frac{1}{12}\sum_{\text{Weyl}}T(R_f) + \frac{1}{3}\sum_{\text{complex scalars}}T(R_s)\Big],
\end{equation}
with \(C_A\) the adjoint Casimir (\(C_A=N\) for \(SU(N)\)) and \(T(R)\) the index of representation (\(T(F)=1/2\) for the fundamental). 
Counting shortcut: 1 Dirac \(=\) 2 Weyl so \(\tfrac{1}{6}\sum_{\text{Dirac}}T(R_f)\) is equivalent; 1 complex scalar \(=\) 2 real so \(\tfrac{1}{6}\sum_{\text{real}}T(R_s)\) is equivalent. 
For \(U(1)_Y\), replace \(T(R)\to Y^2\) in the chosen normalisation.

\paragraph{(1) Masses for \(SU(2)_L\).} Gauge: \(C_A/3=2/3\). Fermions: only LH doublets; per generation \(6\) Weyl quarks \(+\) \(2\) Weyl leptons \(\Rightarrow\) \(24\) Weyl over 3 generations, each with \(T(F)=1/2\), giving \(\tfrac{1}{12}\sum T=1\). Scalars: one complex Higgs doublet, \(\tfrac{1}{3}\sum T=1/6\). Therefore
\begin{equation}
m_{D,2}^2 = g_2^2 T^2\Big(\frac{2}{3}+1+\frac{1}{6}\Big) = \frac{11}{6}g_2^2 T^2 = \frac{22\pi}{3}\alpha_w T^2.
\end{equation}
The author's choice in \cite{Joyce:1994zn} (used as IR regulator inside logs) drops the scalar \(+1/6\) piece:
\begin{equation}
M_W^2 = \frac{5}{3}g_2^2 T^2 = \frac{20\pi}{3}\alpha_w T^2.
\end{equation}

\paragraph{(2) Masses for \(SU(3)_c\).} Gauge: \(C_A/3=1\). Fermions: \(N_f=6\) Dirac flavours in \(F\), so \(\tfrac{1}{6}\sum_{\text{Dirac}}T(F)=\tfrac{1}{6}\cdot 6 \cdot \tfrac{1}{2}=1/2\). No coloured scalars in the SM:
\begin{equation}
m_{D,3}^2 = 2 g_3^2 T^2 = 8\pi\alpha_s T^2 \quad(N_f=6).
\end{equation}
At EW temperatures all six quark flavours are active, matching the paper’s $M_g^2 = 8\pi\alpha_s T^2$.
\paragraph{(3) Masses for \(U(1)_Y\).} Full HTL sum in the SM symmetric phase yields
\begin{equation}
m_{D,Y}^2 = \frac{11}{6}g^{\prime 2} T^2.
\end{equation}
The Ref. \cite{Joyce:1994zn} instead uses the simpler effective regulator
$
M_B^2 = \frac{1}{3}g^{\prime 2} T^2 = \frac{4\pi}{3}\alpha_w \tan^2\theta_W T^2,
$
which differs from the full HTL by a factor \((1/3)/(11/6)=2/11\).

\paragraph{SM values adopted.}
With \(\alpha_w=g_2^2/(4\pi)\), \(\alpha_s=g_3^2/(4\pi)\), \(\tan\theta_W=g_1/g_2\),
\begin{equation}
M_B^2 = \frac{4\pi}{3}\alpha_w\tan^2\theta_W T^2,\quad
M_W^2 = \frac{20\pi}{3}\alpha_w T^2,\quad
M_g^2 = 8\pi\alpha_s T^2.
\end{equation}

\footnotesize\emph{Note on magnetic screening.} Only longitudinal (electric) modes are Debye screened at one loop. Transverse (magnetic) modes lack perturbative screening; a magnetic mass \(m_M\sim g^2 T\) arises nonperturbatively and is not included in leading-log transport estimates.
\end{WideBox}

In computing the transport rate, it is convenient to work in the centre-of-mass (CM) frame of the $2\to2$ process, since the integrand of the $k'$ and $p'$ phase-space integrals is Lorentz invariant. In the CM frame, one has
\[
\vec p+\vec k=0=\vec p\, '+\vec k',\qquad p^0=k^0\equiv E,\qquad p'^0=k'^{0}\equiv E,
\]
and the linearised collision operator projected onto the transport harmonic yields
\begin{equation}
\Gamma_t(E)=\frac{1}{2E^3}\int_{\vec k}\frac{(-f_0'(k^0))}{2k^0}\cdot\frac{1}{32\pi^2}\int d\Omega\,|\mathcal{M}|^2\,(p\!\cdot\! p'),
\label{eq:Gammat-start}
\end{equation}
where $f_0$ is the equilibrium distribution and $f_0'(x)\equiv \partial f_0/\partial x$.
Approximating all external legs as massless wherever permitted (masses only enter as IR regulators), one finds
\begin{equation}
p\!\cdot\! p' = p^0 p'^0 - \vec p\!\cdot\!\vec p'\simeq E^2-|\vec p|^2\cos\theta\simeq p^2(1-\cos\theta),
\end{equation}
with $p\equiv|\vec p|=E$. In the CM frame $s\simeq 4p^2$ and for small-angle scattering
\begin{equation}
t\simeq -2p^2(1-\cos\theta)\equiv -2p^2 y,\qquad y\equiv 1-\cos\theta\in[0,2].
\end{equation}
For $t$-channel gauge exchange the squared matrix element has the universal near-forward structure
\begin{equation}
|\mathcal{M}|^2 \propto \frac{\text{(couplings)}\times\text{(kinematics)}}{(t-M^2)^2}\ \ \longrightarrow\ \ \frac{Ag^4}{(t-M^2)^2},
\end{equation}
where $M^2$ is the appropriate Debye mass and $A$ collects group and spinor factors.
Using $d\Omega=2\pi\,d(\cos\theta)= -2\pi\,dy$ and $p\!\cdot\!p'=p^2 y$, the angular integral in \eqref{eq:Gammat-start} reduces to
\begin{equation}
\int d\Omega\,\frac{p^2 y}{(t-M^2)^2}
= 2\pi \int_{0}^{2} dy\,\frac{p^2 y}{\big(-2p^2 y - M^2\big)^2}
= \frac{\pi}{2p^2}\int_{0}^{2} dy\,\frac{y}{(y+a)^2},
\end{equation}
where we defined the dimensionless IR regulator
$
a\equiv \frac{M^2}{2pk}\,.
$
The remaining elementary integral can be done in closed form\footnote{%
Screening by a Debye mass $M\equiv m_D$ is implemented by excluding $|t|<m_D^2$. In the CM parametrisation $t=-(s/2)\,y$ this corresponds to
$y_{\min}=a\simeq M^2/(2pk)$. Evaluating
\(
\mathcal I(a)
\)
from $y_{\rm min}$ to 2, gives
\(
\mathcal I(a)=\ln(1/a)+\mathcal O(1)
\)
for $a\ll1$. Thus the Debye cutoff replaces $\ln(2/a)\to\ln(1/a)$, i.e.\ reduces the result by a constant $\ln 2$. Below, we follow the standard leading–log convention.
}

\begin{equation}
\begin{aligned}
    {\cal I}(a)=\int_{0}^{2}\!dy\,\frac{y}{(y+a)^2} 
= \Big[\ln(y+a)+\frac{a}{y+a}\Big]_{0}^{2}
&=\ln\!\Big(\frac{2+a}{a}\Big)+\frac{a}{2+a}-1
\\&=\ln\!\Big(\frac{2}{a}\Big)+\mathcal O(1)\quad (a\ll 1).
\label{eq:D:ang-I}
\end{aligned}
\end{equation}
Keeping only the leading logarithm and restoring the overall prefactors one obtains the characteristic CM-frame kernel
\begin{equation}
\Gamma_t(E)\ \propto\ \int_{\vec k}\frac{-f_0'(k^0)}{2k^0}\ \frac{1}{p^2}\ \ln\!\Big(\frac{2}{a}\Big)
=\int_{\vec k}\frac{-f_0'(k^0)}{2k^0}\ \frac{1}{p^2}\ \ln\!\Big(\frac{4pk}{M^2}\Big).
\end{equation}
Thermal averaging over the hard momentum $k^0\sim T$ and using the standard “slow–log” approximation
\begin{equation}
\int dx\,x^nf_0'(x)\ln(1/x)^{-1} \simeq \ln(1/x_n)^{-1}\! \int dx\,x^nf_0'(x),\qquad x_n\simeq n,
\label{eq: approx int 1}
\end{equation}
leads to the compact leading-log estimate
\begin{equation}
\Gamma_t \approx {A \over 32} g^4\frac{T^3}{6\pi\,p^2}\ln\Big(\frac{4pTx_2}{M^2}\Big),
\end{equation}
where $x_2\simeq 2.399$ encodes the precise thermal averaging convention and $M^2$ is the Debye mass of the exchanged gauge boson, and $A$ distinguishes between the different vector bosons exchange.
\paragraph{From $\Gamma_t$ to $D^{-1}$.}
Inserting this result in the master formula \eqref{eq:D-master}
and using $\langle (p_z/p)^2\rangle=1/3$ together with the exact Fermi–Dirac moment
\(
\int \frac{d^3p}{(2\pi)^3}(-f_0'(p))\,p^2=\frac{7\pi^2}{60}T^4
\),
we arrive at the compact leading–log expression
\begin{equation}
D^{-1}=\frac{5}{448\pi^3}Ag^4T\,
\ln\!\Big(\frac{2x_2 x_4T^2}{M^2}\Big),
\label{eq:Dgeneric}
\end{equation}
which makes all phase–space and thermal factors explicit, and where $x_4\simeq 4.131$.

Writing $g^4=(4\pi\alpha)^2$ and inserting the group-theory weights
\(
A_{B,W,g}=\{80\,Y^2,\;36,\;32\times 2\}
\),
Eq.~\eqref{eq:Dgeneric} gives, channel by channel,
\begin{equation}
\begin{aligned}
D_B^{-1}&\simeq \frac{100}{7\pi}\,\alpha_w^2\tan^4\!\theta_W\,Y^2\,T\,
\ln\!\Big(\frac{2\,x_2 x_4\,T^2}{M_B^2}\Big),\\[3pt]
D_W^{-1}&\simeq \frac{45}{7\pi}\,\alpha_w^2\,T\,
\ln\!\Big(\frac{2\,x_2 x_4\,T^2}{M_W^2}\Big),\\[3pt]
D_G^{-1}&\simeq \frac{80}{7\pi}\,\alpha_s^2\,T\,
\ln\!\Big(\frac{2\,x_2 x_4\,T^2}{M_g^2}\Big).
\end{aligned}
\label{eq:DWDBDG}
\end{equation}
The authors in \cite{Joyce:1994zn} absorb the product $2x_2x_4$ into the reference choice
\(
\ln(32T^2/M^2)
\),
yielding the familiar forms quoted in their paper. The total diffusion inverse for a given species is the sum over the channels it couples to\footnote{The exact derivation of these combinations is presented in the Box~\ref{box:channel combination}.},
\begin{equation}
\begin{aligned}
D^{-1}_{\ell_R}&=D_B^{-1},\\
D^{-1}_{\ell_L}&=D_W^{-1}+Y_{\ell_{L}}^{2}D_B^{-1},\\
D^{-1}_{q_{L,R}}&=Y_{q_{L,R}}^{2}\,D_B^{-1}+\epsilon_{L,R}\,D_W^{-1}+D_G^{-1},
\end{aligned}
\qquad \epsilon_L=1,\ \epsilon_R=0.
\end{equation}
Adopting the benchmark couplings
\(
\alpha_w\simeq 1/30,\ \alpha_s\simeq 1/7,\ \alpha_w\tan^2\!\theta_W\simeq 1/103
\)
and the \emph{paper’s} Debye regulators, Eqs.~\eqref{eq:Dgeneric}-\eqref{eq:DWDBDG} with
\(4x_2x_4=32\)
numerically give
\[
D_B\simeq \frac{350}{T\,Y^2},\qquad
D_W\simeq \frac{115}{T},\qquad
D_G\simeq \frac{6}{T},
\]
and, combining channels as above,
\[
D_{\ell_R}\simeq \frac{350}{T},\quad
D_{\ell_L}\simeq \frac{106}{T},\quad
D_{q_L}\simeq \frac{5.8}{T},\quad
D_{q_R}\simeq \frac{6.1}{T}.
\]
If instead one uses the \emph{full HTL} screening masses for the SM (e.g.\ $M_W^2=\tfrac{22\pi}{3}\alpha_w T^2$ and $M_B^2=\tfrac{11}{6}g'^2T^2=\tfrac{11\pi}{3}\alpha_w\tan^2\!\theta_W T^2$), the updated values $\alpha_s(m_Z)\simeq 0.1179, \ \alpha_w(m_z)\simeq 0.0338, \ \alpha_w (\tan \theta_w)^2\simeq 0.0102$, and the true values for $x_{2,4}$, then the logs become smaller and the corresponding numbers shift mildly
\[
D_B\simeq \frac{410}{T\,Y^2},\quad
D_W\simeq \frac{110}{T},\quad
D_G\simeq \frac{7.6}{T},\quad
D_{\ell_L}\simeq \frac{100}{T},\quad
D_{q_L}\simeq \frac{7.2}{T},\quad
D_{q_R}\simeq \frac{7.6}{T}.
\]
These \emph{slight differences} relative to the compact literature values trace back to two deliberate choices: (i) the precise IR regulator used inside the leading logarithm, and (ii) keeping the explicit $2x_2x_4$ from CM kinematics and thermal averaging rather than setting it to $32$. Both effects only enter the argument of the log and therefore produce $\mathcal O(10\%)$ variations without changing the characteristic LL scaling or the channel hierarchy.

\begin{WideBox}[label={box:channel combination}]{How the channel combination is built.}
In the linearised Boltzmann equation, the momentum–relaxation (transport) rate is additive over
independent interaction channels:
\(\Gamma_t=\sum_{X\in\{B,W,g\}}\Gamma_t^{(X)}\).
Since \(D\) follows from Eq.~\eqref{eq:D-master} as a thermal average of \(1/\Gamma_t\),
the leading–log (LL) evaluation—where each \(\Gamma_t^{(X)}\) is dominated by the same near–forward
region and differs only by its coupling/group factor—allows one to write the inverse diffusion
constant as a \emph{sum} of channel contributions,
\[
D^{-1}(\psi)\simeq D_B^{-1}(\psi)+D_W^{-1}(\psi)+D_G^{-1}(\psi).
\]
For hypercharge exchange, the squared matrix element carries \(g_1^4\,Y_\psi^2\,Y_{\rm bath}^2\);
after summing/averaging over the bath, the probe’s factor \(Y_\psi^2\) remains explicit. It is thus
convenient to define a reference \(B\)-channel kernel \(D_B^{-1}\) and then scale it by the
\emph{probe} hypercharge,
\[
D_B^{-1}(\psi)=Y_\psi^2\,D_{B,\,Y^2=1}^{-1},
\]
so that, e.g., for leptons \(Y_{\ell_R}^2=1\) and \(Y_{\ell_L}^2=1/4\), implying the familiar quarter
relation for the \(B\) piece of \(\ell_L\) compared to \(\ell_R\).
The \(W\)-channel contributes only for \(SU(2)_L\) doublets, captured by
\(\epsilon_\psi=1\) (LH doublets) or \(0\) (RH singlets):
\[
D_W^{-1}(\psi)=\epsilon_\psi\,D_W^{-1},\qquad
\epsilon_{\ell_L,q_L}=1,\ \ \epsilon_{\ell_R,u_R,d_R}=0.
\]
The gluon piece is universal for colored probes:
\(D_G^{-1}(\psi)=D_G^{-1}\) for all quarks, and vanishes for leptons.
Altogether,
\[
D^{-1}(\psi)=Y_\psi^2\,D_{B,\,Y^2=1}^{-1}+\epsilon_\psi\,D_W^{-1}+\delta_\psi\,D_G^{-1},
\]
with \(\delta_\psi=1\) for quarks and \(0\) for leptons. In the compact notation used in the text, this is equivalently written as
\[
D^{-1}_{\ell_R}=D_B^{-1},\quad
D^{-1}_{\ell_L}=D_W^{-1}+\tfrac14 D_B^{-1},\quad
D^{-1}_{q_{L,R}}=Y_{q_{L,R}}^2 D_B^{-1}+\epsilon_{L,R}D_W^{-1}+D_G^{-1}.\]
\end{WideBox}

\subsection{Diffusion constant of the Higgs field: \texorpdfstring{$D_h$}{Dh}}

The leading–log derivation of the diffusion constant for the Higgs, proceeds as in the fermionic case
(linearised Boltzmann equation, near–forward $t$–channel exchange, regulated by
Debye screening slow–log thermal averaging), but three inputs are \emph{different}
for a relativistic \emph{scalar}:
The rate uses the Bose–Einstein derivative
  \[
    -f_B'(E)\;=\;\frac{e^{E/T}}{(e^{E/T}-1)^2}\,,
  \]
  instead of the Fermi–Dirac one. Consequently all thermal moments entering the
  collision integral are \emph{bosonic},
  \begin{equation}
    I_m^{(B)} \;\equiv\; \int_0^\infty dx\,x^m\big(-f_B'(x)\big)
    \;=\;\Gamma(m{+}1)\,\zeta(m)\qquad (m>1),
  \end{equation}
  while for fermions
  \(I_m^{(F)}=\Gamma(m{+}1)\big(1-2^{\,1-m}\big)\,\zeta(m)\).
  In particular, \(I_2^{(B)}/I_2^{(F)}=2\) and \(I_4^{(B)}/I_4^{(F)}=8/7\), so the bosonic prefactor is larger by the factor
  \((I_2^{(B)}I_4^{(B)})/(I_2^{(F)}I_4^{(F)}) = 16/7\).

For a complex scalar in representation $R$
  the gauge–scalar–scalar vertex is \(ig_V(p{+}p')^\mu T^a_R\).
  The squared matrix element therefore carries the kinematic numerator \((s-u)^2\),
  in contrast to the \(s^2\) coming from the spinor current \(\bar u\gamma^\mu u\).
  
  When factoring out the slowly varying
  logarithm, the effective points are the \emph{bosonic} ones (denoted
  \( x^{(B)}_m\)), not the fermionic ones; see below.

\paragraph{Master formula.}
Starting from the linearised Boltzmann equation and the standard anisotropy ansatz for
a scalar perturbation, one obtains
\begin{equation}
  D_h \;=\; \frac{12}{T^3}\int\!\frac{d^3p}{(2\pi)^3}\,
  \frac{-f_B'(p)}{\Gamma_t^{(h)}(p)}\left(\frac{p_z}{p}\right)^2,
  \label{eq:Dh-master-new}
\end{equation}
where \(\Gamma_t^{(h)}\) is the transport (momentum–relaxation) rate of the Higgs mode,
to be computed at leading log from small–angle $t$–channel exchange.

For \(2\!\to\!2\) scattering mediated by a gauge boson \(V=B,W\) between the Higgs
\(H\) and a plasma constituent \(X\), the scalar–gauge vertices yield
\begin{align}
  |{\cal M}_B(HX)|^2 &=
  16\,g'^4\,Y_H^{2}\,Y_X^{2}\;
  \frac{(s-u)^2}{(t-M_B^2)^2}, \label{eq:MB-scalar}\\
  |{\cal M}_W(HX)|^2 &=
  16\,g_2^4\,C_2(H)\,C_2(X)\;
  \frac{(s-u)^2}{(t-M_W^2)^2}, \label{eq:MW-scalar}
\end{align}
with \(Y\) the hypercharge and \(C_2(R)\) the quadratic Casimir in representation \(R\)
(\(C_2(H)=3/4\) for the SM Higgs doublet). The overall factor \(16\) is the familiar
kinematic factor from two scalar vertices \((p{+}p')^\mu\) and their contraction.
In the near–forward (small–angle) region relevant for the leading logarithm,
\begin{equation}
  (s-u)^2 \xrightarrow[\text{forward}]{} 4\,s^2,
  \label{eq:forward-limit}
\end{equation}
which changes the transport weight relative to a pure \(s^2\) numerator, prior to the
angular average.

\paragraph{Leading–log expression and bosonic slow–log points.}
Summing over all scattering partners \(X\) present in the plasma (with degeneracies
\(d_X\): spin/polarisations, colour, isospin, particle/antiparticle) and performing the
same transport–weight angular average as in the fermionic derivation, the small–angle
contribution gives
\begin{equation}
  D_h^{-1}\simeq\sum_{V=B,W}\;
  \frac{5\mathcal{A}_V^{(h)}}{49\pi^3}g_V^{4}T
  \ln\!\Big(\frac{2\bar x^{(B)}_2 x^{(B)}_4T^2}{M_V^2}\Big),
  \qquad g_B\equiv g',\ \ g_W\equiv g_2,
  \label{eq:Dh-LL-new}
\end{equation}
where \(M_V^2\) are the longitudinal Debye screening masses that regulate the IR.
Here the prefactor \(1/(49\pi^3)\) is the \emph{bosonic} one, obtained from the
fermionic \(1/(448\pi^3)\) by multiplying by the moment ratio \(16/7\) and by 4 that is coming from the forward–limit enhancement (\ref{eq:forward-limit}), and \(\bar x^{(B)}_{2}\), \( x^{(B)}_{4}\) are the
bosonic “slow–log” evaluation points determined by the replacements
\begin{equation}
  \int_0^\infty\!dx\,x^m f_B'(x)\frac{1}{\ln(1/x)}
  \approx \frac{1}{\ln\!\big(1/ x_m^{(B)}\big)}
  \int_0^\infty\!dx\,x^m f_B'(x)\,,
  \qquad x_m^{(B)}\simeq m\;.
\end{equation}
Strictly speaking, this approximation is trustworthy only for \(m>2\), and it becomes
accurate for \(m\gtrsim 4\). In practice, it is convenient to adopt effective values
\(\bar x_m^{(B)}\), for $m<2$, that reproduce the full numerical integral, and the one from the above expression for $m\geq 4$. We will use
\(\bar x_{2}^{(B)} \simeq 1.593\) and \( x_{4}^{(B)} \simeq 3.83\), which retain
the same functional form as in the fermionic case while matching the bosonic integrals.\footnote{%
  The “slow–log” replacement assumes that the weight
  \(w_m(x)=x^m\big(-f_0'(x)\big)\) is sharply peaked at some scale \(x_\star\), so a
  slowly varying factor \(g(x)\) can be evaluated at \(x_\star\) and factored out. For
  fermions, \(w_m\) is well peaked already at small \(m\), but for bosons the IR
  behaviour \(-f_B'(x)\sim x^{-2}\) pushes weight toward small \(x\). As a result,
  the integrand has \emph{no} interior maximum for \(m=2\), and for
  \(m=3,4\) the peak is broad. Only for \(m\gtrsim 5\) does the approximation become
  parametrically good. In practice, one may (i) introduce a physical IR regulator
  (thermal mass) or (ii) define \(\bar x_m^{(B)}\) by \emph{matching} the exact integral
  to the slow–log ansatz; we follow (ii) here just for the case of $\bar x_2^{(B)}$.}

The other difference in the bosonic case is encoded in the following coefficients
\begin{equation}
  \mathcal A_B^{(h)}=16\,Y_H^2\sum_{X} d_X\,Y_X^2,
  \qquad
  \mathcal A_W^{(h)}=16\,C_2(H)\sum_{X\in SU(2)} d_X\,C_2(X)\; .
  \label{eq:A-coeffs-boson}
\end{equation}
For Higgs–Higgs scattering one simply inserts \(X=H\), so the hypercharge factor
becomes \(Y_H^4\), with \(Y_H=1/2\) and \(d_H=4\) (two isospin components and
particle/antiparticle), and for \(SU(2)\) one has \(C_2(H)^2=(3/4)^2\).
If \(X\) is a fermion, replace \(Y_X^2\) or \(C_2(X)\) accordingly and include its
spin and colour degeneracies in \(d_X\).

As a concrete benchmark, keeping only \(X=H\) (and \(H^\dagger\)) in the plasma sums above yields the simple estimate
\begin{equation}
  D_h \;\simeq\; \frac{14}{T}\,,
\end{equation}
obtained by inserting the Higgs quantum numbers and degeneracy into
Eqs.~(\ref{eq:Dh-LL-new})–(\ref{eq:A-coeffs-boson}) together with the bosonic
slow–log points quoted above. Adding the abundant light fermions (and gauge bosons) in the thermal bath increase the total transport rate \(\Gamma_t^{(h)}\) and correspondingly reduces \(D_h\).

\section{Yukawa–mediated relaxation rates: $\widehat \Gamma_y$}
\label{app:rate Gammay}

In this appendix, we derive, from first principles and in a self–contained way, the leading–log expressions for the \emph{per–particle} relaxation rates induced by Yukawa interactions in a hot plasma. We keep all phase–space and group/Yukawa factors explicit and regulate the forward ($t$–channel) singularity with thermal masses. The notation and normalisations (distribution functions, measures, and the overall $12/T^3$ factor converting volume rates to per–particle rates) follow the conventions already fixed in the previous section; we do not repeat them here.

The rate per particle per unit time can be defined as
\begin{equation}
    \widehat \Gamma= {12 \over T^3}\int_{\{p, p', k, k'\}} f_p \tilde f_k (1+\tilde f_{p'})(1-f_{k'}) \delta^4 (p+k-p'-k') |\mathcal{M}|^2\ ,
\end{equation}
where $f_p$ and $\tilde f_k$ and the distribution functions for the ingoing fermion and boson, while $(1+\tilde f_{p'})$ and $(1-f_{k'})$ are the spin blocking for the outgoing particles.

The chirality–changing (or flavour–relaxing) $2\!\to\!2$ reactions driven by a Yukawa coupling $y_\psi$ are dominated, at weak coupling, by near–forward $t$–channel exchange of a gauge boson $X\in\{B,W,g\}$ in the thermal bath:
\[
\psi + X \ \longleftrightarrow\  \psi' + \phi,
\]
where $\phi$ is the Higgs doublet and $\psi,\psi'$ the appropriate chiral components. After spin/colour sums, the squared matrix element has the universal small–angle form
\begin{equation}
\;
|{\cal M}_X|^2 \;=\; \frac{A_X\,y_\psi^2\,s\,(-t)}{(t-M^2)^2}\,,
\;
\qquad
A_X \equiv c_X\,(4\pi\alpha_X),
\label{eq:D:M2-def}
\end{equation}
with $s=(p+k)^2$, $t=(p-p')^2$, $\alpha_X$ the gauge fine–structure constant, and $M$ an \emph{asymptotic} thermal mass acting as an IR regulator listed earlier. The channel–dependent coefficients $c_X$ encode group theory and bath–species sums; a convenient and widely–used choice is\footnote{%
At weak coupling, the chirality–changing $2\!\to\!2$ process is dominated by near-forward $t$–channel exchange of a bath gauge boson $X\in\{B,W,g\}$:
$\psi(p)+X(k)\to\psi'(p')+\phi(q)$. The tree amplitude with one Yukawa and one gauge vertex can be written schematically as
\[
\mathcal M_X \;\sim\; y_\psi\,g_X\;
\frac{1}{t-M^2}\;
\big[\bar u(p')\,\Gamma_Y\,u(p)\big]\;
\big[\varepsilon_\mu^{\,a}(k)\,J^\mu_a\big],
\]
where: $y_\psi$ is the Yukawa coupling; $g_X$ is the gauge coupling ($g_Y\equiv g'$ for $B$, $g$ for $W$, $g_s$ for $g$); 
$t\equiv(p-p')^2$ is the momentum transfer; $M$ is the (asymptotic/screening) thermal mass of the exchanged gauge boson regulating the IR; 
$u,\bar u$ are Dirac spinors; $\Gamma_Y$ denotes the chiral Yukawa vertex (projector $P_{L/R}$ as appropriate); 
$\varepsilon_\mu^{\,a}(k)$ is the polarisation vector of the external bath gauge boson with adjoint index $a$; and 
$J^\mu_a \equiv \bar u(p)\,\gamma^\mu T^a_R\,u(p)$ is the gauge current of the fermion in representation $R$ ($T^a_R$ are the generators, with 
$\mathrm{tr}\,T^a_R T^b_R=\tfrac12\,\delta^{ab}$). After summing/averaging over spins and polarisations and using the small-angle limit $|t|\ll s$ with $s\equiv(p+k)^2$, one obtains the universal kernel
\[
\sum_{\text{spins, pols}}|\mathcal M_X|^2 \;=\; y_\psi^2\,g_X^2\;
\frac{s(-t)}{(t-M^2)^2}\;
\underbrace{\Big[8\,C_2(R)\Big]}_{\text{spin/pol kernel}\times\text{group}},
\]
with $C_2(R)$ the quadratic Casimir of $\psi$ under the exchanged gauge group ($C_2(F)_{\mathrm{SU(2)}}=\tfrac34$, $C_2(F)_{\mathrm{SU(3)}}=\tfrac43$; for $\mathrm{U}(1)_Y$, replace $T^a\!\to\!Y_\psi$ so $C_2\!\to\!Y_\psi^2$). 
For the Abelian case, there is an extra factor $1/2$ from the initial-state polarisation average not compensated by adjoint sums, yielding
\[
c_W=8\,C_2(F)_{\mathrm{SU(2)}}=8\times\tfrac34=6,\qquad
c_g=8\,C_2(F)_{\mathrm{SU(3)}}=8\times\tfrac43=\tfrac{32}{3},\qquad
c_B=\frac{8}{2}\,Y_\psi^2=4\,Y_\psi^2.
\]
}%
\begin{equation}
c_B=4\,Y_\psi^2,\qquad c_W=6,\qquad c_g=\frac{32}{3}\,,
\label{eq:D:cX}
\end{equation}
so that $A_B=16\pi\alpha_w\tan^2\theta_W\,Y_\psi^{2}$, $A_W=24\pi\alpha_w$, and $A_g=\tfrac{128}{3}\pi\alpha_s$. Now, ignoring the spin blocking terms, where this amounts to overestimating the rate by a factor $1-4$, we arrive at
\begin{align}
        \widehat\Gamma&={12 \over T^3} \int_{\{p, k\}} f_p \tilde f_k \mathcal{J}, \\ \mathcal{J}&=\int_{\{p', k'\}} \delta^4(p+k-p'-k') A_Xy_\psi^2 {-ts \over (t-M^2)^2}= {A_X y_\psi^2 \over 8\pi}\mathcal{I}\left(M^2 \over 4 p k\right) \ ,
\end{align}
where $\mathcal{I}(a)$ has already been defined in Eq. \eqref{eq:D:ang-I}. In order to perform the integral over $p$ and $k$ we use the following relations
\begin{align}
    \eta_m(n)&=\int dx x^m e^{-nx} \ln x ={m! \over n^{m+1}}\left[ H_m-\gamma_E-\ln n\right], \\ 
    &~~~~~\int dx x^m f_x \ln {1 \over x}\approx \ln {1 \over x_{m+1}}\int dx x^m f_x\ ,
\end{align}
where $H_m=\sum_{\ell=1}^{m}{1 \over \ell}$ is the harmonic $m-$number. The integral expressed in the full form is
\begin{equation}
    \widehat\Gamma= {3Ay_\psi^2T\over 32\pi^5} \int _0^\infty dx {x \over e^x+1}{x \over E_p/T}\int _0^\infty dy {y \over e^y+1} {y \over E_k/T}\left[ \ln \left( 1+\lambda x y\right)-1+{1 \over 1+\lambda x y}\right]\ ,
\end{equation}
where $\lambda = 4T^2/M^2$, $x=p/T$ and $y=k/T$. Looking for LL contributions, we are considering the limit $\lambda \gg1$. We explicitly wrote the energy appearing in the normalisation of the integral so that for massless particles $E=p$ and a different power of $x$ or $y$ appears. In order to reproduce the result in \cite{Joyce:1994zn} (their App. D), one needs to consider the fermion to be massless and the boson to be massive\footnote{In the leading-log approximation, the dominant infrared divergence is regulated by the mass of the exchanged boson, not the fermion. Thus, treating the fermion as massless while keeping the boson massive is sufficient and captures the main effect. Other approaches, see \cite{Huet:1995sh} include small fermion masses, but this only adds subleading corrections.}. In order to fully perform the integral, it is useful to introduce the following quantities
\begin{align}
    J^{(F)}_m &= \int_0^\infty dx \, {x^m \over e^x+1}=\left( 1-2^{-m}\right) \Gamma(m+1)\zeta(m+1), \quad J^{(B)}_m = \int_0^\infty dx \, {x^m \over e^x-1}= \Gamma(m+1)\zeta(m+1), \\
    L^{(F)}_m &= \int_0^\infty dx \, {x^m \ln x \over e^x+1}= \sum_{n=1}^\infty (-1)^{m+1}\eta_m(n), \quad ~~~~~~~~~~~~~L^{(B)}_m = \int_0^\infty dx \, {x^m \ln x \over e^x-1}= \sum_{n=1}^\infty \eta_m(n)
\end{align}
So in the end writing $\ln(1+\lambda x y)\approx \ln \lambda x y$ and forgetting about the the last term in the $\mathcal{I}(a)$ function the final result is
\begin{align}
    \widehat\Gamma&\approx {3Ay_\psi^2T\over 32\pi^5} \left[ \left(\ln (\lambda x_2)-1 \right)J_2^{(B)}J_1^{(F)}+J_1^{(F)}L_2^{(B)}\right]\ ,\\
    &={\zeta(3) A y_\psi^2\over 64 \pi^3}T \left[ \ln\left(4 x_2 T^2 \over M^2\right)-1+{1\over 2 \zeta(3)}\sum_{n=1}^\infty \eta_2 (n)\right].
\end{align}
In order to see if the LL approximation makes sense, we need to compare the last two terms in the bracket with the LL containing the thermal masses for the fermions appearing in the diagrams.
For a relativistic chiral fermion \(\psi\) in a hot gauge plasma, the one-loop
HTL self-energy gives the asymptotic (large-momentum) thermal mass
\begin{equation}
m_\psi^2 \;=\; \frac{T^2}{8}\left[\,
\sum_{a\in\{SU(3)_c,\,SU(2)_L\}} g_a^2\,C_2^{(a)}(\psi)
\;+\; g^{\prime\,2}\,Y_\psi^2
\right].
\end{equation}
Here \(C_2^{(a)}(\psi)\) is the quadratic Casimir of the representation of \(\psi\) under
the gauge group \(a\):
\[
C_F(SU(3))=\frac{4}{3},\qquad C_F(SU(2))=\frac{3}{4},
\]
and \(Y\) is the hypercharge in the convention \(Q=T_3+Y\). Writing \(g_a^2=4\pi\alpha_a\) and \(g^{\prime\,2}=4\pi\,\alpha_w\tan^2\theta_W\),
one finds for each SM chiral multiplet the chirality-resolved HTL (asymptotic) thermal masses are
\begin{equation}
\begin{aligned}
M_{q_L}^2 &= \frac{2}{3}\pi\,\alpha_s\,T^2 + \frac{3}{8}\pi\,\alpha_w\,T^2 + \frac{1}{72}\pi\,\alpha_w\,\tan^2\theta_W\,T^2,\\
M_{u_R}^2 &= \frac{2}{3}\pi\,\alpha_s\,T^2 + \frac{2}{9}\pi\,\alpha_w\,\tan^2\theta_W\,T^2,\\
M_{d_R}^2 &= \frac{2}{3}\pi\,\alpha_s\,T^2 + \frac{1}{18}\pi\,\alpha_w\,\tan^2\theta_W\,T^2,\\
M_{\ell_L}^2 &= \frac{3}{8}\pi\,\alpha_w\,T^2 + \frac{1}{8}\pi\,\alpha_w\,\tan^2\theta_W\,T^2,\\
M_{\ell_R}^2 &= \frac{1}{2}\pi\,\alpha_w\,\tan^2\theta_W\,T^2.
\end{aligned}
\end{equation}
Here \(q_L\) denotes any left-handed quark doublet, \(u_R\) and \(d_R\) the right-handed up- and down-type singlets, \(\ell_L\) a left-handed lepton doublet, and \(\ell_R\) a right-handed charged-lepton singlet. These masses enter dispersion relations and the regulation of soft regions in rates where appropriate; in the diffusion constants at leading log, the dominant regulators are the Debye masses of the exchanged gauge bosons above.
What we get (using the updated value of the couplings) is
\begin{equation}
    M_{q_L}\simeq 0.58 T, \quad M_{u_R}\simeq0.28T , \quad M_{d_R}\simeq0.27 T, \quad M_{\ell_L}^2\simeq0.21T , \quad M_{\ell_R}\simeq 0.12T,
\end{equation}
and so the LL are
\begin{equation}
    \text{LL}_{q_L} \simeq 3.3\ , \quad 
    \text{LL}_{u_R}\simeq 4.8\ , \quad 
    \text{LL}_{d_R}\simeq 4.9\ , \quad 
    \text{LL}_{\ell_L} \simeq 5.4\ , \quad 
   \text{LL}_{\ell_R}\simeq 6.4\ .
\end{equation}
where LL$_{i}=\ln \left(4 x_2 T^2 \over M_{i}^2 \right)$ and these has to be compared with ${1\over 2 \zeta(3)}\sum_{n=1}^\infty \eta_2 (n)-1\simeq -0.24$. The final answer is

\begin{equation}
\begin{aligned}
\widehat\Gamma_{\ell_R} &= 2\,\widehat\Gamma_{l_L} \;=\;
\frac{3\zeta(3)}{8\pi^2}\,\alpha_w\,y_\ell^2\,T
\Bigg[
\Big(1+\tfrac{1}{6}\tan^2\theta_W\Big)
\text{LL}_{\ell_L}
+ \tfrac{2}{3}\tan^2\theta_W\,
\text{LL}_{\ell_R}
\Bigg], \\
&\simeq 0.32 \alpha_w y_\ell^2 T \ ,\\
\widehat\Gamma_{q_L} &= \frac{2\zeta(3)}{3\pi^2}\,\alpha_s\,y_{q_L}^2\,T\,
\text{LL}_{q_L} \simeq 0.27 \alpha_s y_{q_L}^2 T\ ,\\
\widehat\Gamma_{q_R} &= \frac{2\zeta(3)}{3\pi^2}\,\alpha_s\,y_{q_R}^2\,T\,
\text{LL}_{q_R} \simeq 0.4 \alpha_s y_{q_R}^2 T \ .
\end{aligned}
\label{eq:final rates}
\end{equation}
\paragraph{Colour normalisation.}
The per–particle rates computed in this subsection are \emph{colour–summed}: they include the factor $N_c$ and thus correspond to a rate “per particle \emph{and per number of colours}.” In the diffusion network we track one fluid per flavour with colour already averaged, so the rate that must enter the transport equations in \eqref{eq:pheno collision 1}-\eqref{eq:pheno collision 4} is the colour–averaged one,
\begin{equation}
\widehat\Gamma_y=\frac{1}{N_c}\,\widehat{\Gamma},\qquad N_c=3.
\label{eq: colour normalisation}
\end{equation}
This is why, in the final line, see \eqref{eq:pheno collision 4}, where the Higgs is considered, a factor \(3\) appears: it comes from explicitly summing over the three quark colours in the microscopic calculation. Consistently dividing by \(N_c\) removes this degeneracy when inserting \(\widehat \Gamma_y\) into the fluid system.

\subsection{Reconciliation with the literature for \texorpdfstring{$\widehat\Gamma_y$}{Gamma\_y}}
There has been long–standing confusion in the literature regarding the Yukawa (mixing) relaxation rate $\widehat\Gamma_y$. 
Below, we summarise the logic of the main estimates and place our result in that context, making all assumptions explicit.

\medskip

\noindent\textbf{Early estimates.}
The first quantitative estimates were given in Refs.~\cite{Joyce:1994zn,Joyce:1994zt}. 
In those works, the fermion thermal mass appears only as an \emph{infrared regulator} of the forward $t$-channel singularity in gluon exchange; 
equivalently, the kinematics are treated with massless external lines, and the collinear region is cut off by introducing the mass in the internal propagator. 
Reproducing that derivation in our conventions (see Eq.~\eqref{eq:final rates}), we emphasised that this choice underlies the numerical size of $\Gamma_y$ and the logarithmic sensitivity to the regulator.

A subsequent estimate was quoted in \cite{Huet:1995sh} (their Eq.~(3.18)) with minimal derivational detail. 
The logic appears to follow the same reasoning as \cite{Joyce:1994zn,Joyce:1994zt}, but this time \emph{both} the fermion and the boson are treated as massive throughout the computation. 
Repeating the calculation in that setup, we obtain
\begin{align}
\widehat\Gamma
&= \frac{9\,A_g\,y_q^2}{32\pi}\!\left(\frac{\zeta(3)}{\pi^2}\right)^{\!2}\! T
\left[
\ln\!\left(\frac{4 x_2 T^2}{M_q^2}\right)
-1
+\frac{1}{3\,\zeta(3)^2}\sum_{m=1}^\infty(-1)^{m+1}\eta_2(m)\sum_{n=1}^\infty \eta_2(n)
\right],
\label{eq:GammaY_exact}
\\[4pt]
&\simeq
12\,\alpha_s\,y_q^2\!\left(\frac{\zeta(3)}{\pi^2}\right)^{\!2}\! T\,
\ln\!\left(\frac{4 x_2 T^2}{M_q^2}\right),
\label{eq:GammaY_log}
\end{align}
where $A_g$ is the standard colour factor as before, $y_q$ is the Yukawa coupling, $M_q$ denotes the mass scale regulating the forward singularity, and $x_2$ is a numerical constant from the phase–space integral.
Matching normalisations to the transport equations used in EWBG, the prefactors imply
\begin{equation}
\widehat\Gamma_{q_L}\;\simeq\;0.6\,\alpha_s\,y_{q_L}^2\,T,
\qquad
\widehat\Gamma_{q_R}\;\simeq\;0.87\,\alpha_s\,y_{q_R}^2\,T,
\label{eq: gamma y numerical}
\end{equation}
up to the logarithmic factor and mild scheme dependence encoded in $x_2$ and $M_q$.

\medskip

\noindent\textbf{Numerical values used in the literature.}
The value $\widehat\Gamma_y = 4.2\times 10^{-3}\,T$ was first quoted in \cite{Bodeker:2004ws} and subsequently adopted by several phenomenological studies \cite{Fromme:2006cm,Fromme:2006wx,Espinosa:2011eu,Konstandin:2013caa,Cline:2020jre,Kainulainen:2024qpm}. 
This number was later revised: \cite{Cline:2021dkf} reported $\widehat\Gamma_y \simeq 3\times 10^{-2}\,T$, tracing the discrepancy to an improved treatment and to the separation of channels advocated in \cite{deVries:2017ncy}.

\medskip

\noindent\textbf{Modern viewpoint and channel decomposition.}
Following \cite{deVries:2017ncy}, it is useful to decompose
\begin{equation}
\widehat\Gamma_y =\widehat \Gamma_y^{(3)} +\widehat \Gamma_y^{(4)},
\end{equation}
where $\widehat\Gamma_y^{(3)}$ arises from the trilinear interaction $h\,\bar t_L t_R$, while $\widehat\Gamma_y^{(4)}$ is induced by the gluon–assisted vertex $hg\,\bar t_L t_R$. 
In the kinematic regime relevant for electroweak baryogenesis, the \emph{trilinear} contribution $\widehat\Gamma_y^{(3)}$ is strongly suppressed, so that the dominant piece is the \emph{gluon–assisted} channel $\widehat\Gamma_y^{(4)}$.
This perspective is consistent with the historical derivations \cite{Joyce:1994zn,Joyce:1994zt} and with the finite–temperature transport treatments of \cite{Cirigliano:2006wh}, as emphasised in \cite{deVries:2017ncy}.

\medskip

\noindent\textbf{Take–home message.}
Apparent discrepancies in $\widehat\Gamma_y$ across the literature can be traced to: 
(i) different infrared regulators (and whether it enters only the propagator or the full kinematics); 
(ii) inclusion or omission of the gluon–assisted $4$-point channel and/or the $3$-point channel ; 
and (iii) distinct normalisation conventions in the transport equations. 
Equations \eqref{eq:GammaY_exact}-\eqref{eq: gamma y numerical} make these dependencies explicit and reconcile older and newer results once the same assumptions are imposed.

\medskip\noindent
Using Eq.~\eqref{eq:final rates} together with the colour normalisation in Eq.~\eqref{eq: colour normalisation}, and inserting the updated numerical inputs, the colour–averaged per–particle $t$–channel Yukawa rate entering the diffusion network is
\begin{equation}
  \boxed{\;\widehat{\Gamma}_y^{t} = \frac{1}{N_c}\,\widehat{\Gamma}_{q_L}^{t} \;\simeq\; 10^{-2}\,T\,,\qquad N_c=3.\;}
\end{equation}
If one were to consider the approximations in \eqref{eq:GammaY_log} the result is $\widehat\Gamma_y^t\simeq 2.5 \times 10^{-2}T$.

\begin{WideBox}[label={box:Thermal vs Debye}]{Thermal mass vs.\ Debye mass}
Two distinct thermal scales are often called ``thermal masses'':

\begin{itemize}
  \item \textbf{Debye mass} $m_D$: static electric screening of longitudinal gauge fields, defined from the longitudinal self–energy in the static soft limit $(k_0=0,|\vec{k}|\!\to\!0)$. For $SU(N)$,
  \[
  m_D^2=\frac{g^2T^2}{3}\Big(C_A+\tfrac{N_f}{2}\Big).
  \]
  \emph{Use:} as the IR regulator in $t$–channel exchange when computing leading–log transport/diffusion (e.g.\ $D^{-1}$) and small–angle scattering kernels; it appears only inside the logarithm.

  \item \textbf{Asymptotic mass} $m_\infty$: HTL correction to the dispersion of \emph{propagating} hard modes ($k\!\sim\!T$), entering the pole of retarded propagators. For gauge bosons
  \[
  m_\infty^2=\frac{g^2C_A}{6}T^2+\frac{g^2N_f}{12}T^2,
  \]
  and for fermions $m_{\infty,f}^2$ follows from their gauge Casimirs (plus Yukawa terms).
  \emph{Use:} in quasiparticle energies $\omega=\sqrt{k^2+m_\infty^2}$ within CTP/kinetic kernels (e.g.\ $\widehat\Gamma_M$ for fermions and Higgs), together with thermal widths $\Gamma$; see next section.
\end{itemize}

\noindent In short: $m_D$ regulates \emph{static, longitudinal} screening in exchanged propagators (inside logs); $m_\infty$ controls \emph{dynamic, transverse/fermionic} propagation in the medium and is the relevant mass in relaxation rates used in EWBG.
\end{WideBox}

\section{Yukawa helicity flipping rate: $\widehat \Gamma_M$}
\label{app:rate GammaM}
\subsection{Fermionic case for $\widehat\Gamma_m^q$}

The purpose of this section is to follow the steps to derive, in a self–contained way, the expression for the mixing (Yukawa) relaxation rate $\widehat\Gamma_M$ that appears in electroweak transport. The derivation follows the real–time Schwinger–Keldysh (CTP) framework and the treatment of Refs.~\cite{deVries:2017ncy,Cirigliano:2006wh}, with thermal fermions described by hard–thermal–loop (HTL) quasiparticles. Consider a chiral fermion doublet $(\psi_L,\psi_R)$ coupled to a slowly varying Higgs background, $\mathcal{L}\supset y_f\,\phi_b(z)\,\bar\psi_L\psi_R+\mathrm{h.c.}$. The object that measures the loss of $L$–$R$ coherence is the off–diagonal two–point function $S_{LR}$, whose equation of motion is a Kadanoff–Baym (KB) equation with a collision kernel determined by the imaginary parts of retarded self–energies. After Wigner transformation $S^{<,>}(X,k)$ and gradient expansion around the wall background (``VEV–insertion approximation''), the fermionic quasiparticle poles lie at complex energies $\varepsilon_i=\omega_i-i\Gamma_i$, with $\omega_i(k)=\sqrt{k^2+m_{i,\text{th}}^2}$ the HTL asymptotic dispersion and $\Gamma_i$ the single–particle damping rate. Linear response in the off–diagonal density amounts to keeping terms quadratic in the Yukawa insertion and evaluating the collision term with thermal propagators. The collision integral controlling the decay of $S_{LR}$ can be written in CTP notation as $\,\mathcal{C}_{LR}(X,k)=\tfrac{i}{2}\big[\Sigma^>_{LL}\,S^<_{LR}-\Sigma^<_{LL}\,S^>_{LR}-(L\leftrightarrow R)\big]$, and using KMS relations together with the spectral decomposition $S^{<,>}= \mp n_F(\pm k_0)\,A$ turns all statistical factors into derivatives of the Fermi function. Specifically, one encounters the combination $h(k_0)\equiv e^{k_0/T}/(e^{k_0/T}+1)^2=-\partial n_F/\partial (k_0/T)=n_F(1-n_F)$, which plays the role of the thermal weight for fluctuations. 

Carrying out the $k_0$ integration by closing the contour in the lower or upper half–plane selects the retarded/advanced poles at $k_0=\varepsilon_{L}$ and $k_0=\varepsilon_{R}$ (and their conjugates), and the KB commutator structure produces two distinct denominators: a difference $\big(\varepsilon_R^*-\varepsilon_L\big)^{-1}$ originating from the retarded–advanced subtraction and an energy sum $\big(\varepsilon_L+\varepsilon_R\big)^{-1}$ corresponding to particle–hole mixing. The Dirac trace over chiral projectors yields numerators proportional to $p\!\cdot\!p'$ with opposite signs in the two channels; in the isotropic plasma frame and after angular integration, this becomes the pair $(\varepsilon_L\varepsilon_R^*-k^2)$ and $(\varepsilon_L\varepsilon_R+k^2)$. The three–momentum phase space $\int d^3k/(2\pi)^3$ reduces to $(2\pi^2)^{-1}\!\int_0^\infty k^2\,dk$, and summing over the colour of the fermion introduces $N_c$. Finally, when one matches the kinetic equation to the transport equations written for number densities or chemical potentials, the standard susceptibility of a relativistic fermion gas, $n_f=\chi_f\,\mu_f$ with $\chi_f=T^2/6$ per chiral degree of freedom, pulls out a universal conversion factor $6/T^2$ in front of the momentum integral. Putting all elements together, one finds
\begin{equation}
\widehat\Gamma_M^{\pm}
=\frac{6}{T^2}\,\frac{N_c}{2\pi^2 T}\,|f|^2
\int_0^\infty \frac{k^2\,dk}{\omega_L\omega_R}\;
\Im\!\left[
-\frac{h(\varepsilon_L)\mp h(\varepsilon_R^*)}{\varepsilon_R^*-\varepsilon_L}
\left(\varepsilon_L\varepsilon_R^*-k^2\right)
+\frac{h(\varepsilon_L)\mp h(\varepsilon_R)}{\varepsilon_L+\varepsilon_R}
\left(\varepsilon_L\varepsilon_R+k^2\right)
\right],
\label{eq:GammaM_master_derivation}
\end{equation}
with $\varepsilon_{i}=\omega_i-i\Gamma_i$ and $\omega_i=\sqrt{k^2+m_{i,\text{th}}^2}$ the HTL quasiparticle energies for $i=L,R$, and $h(x)=e^x/(1+e^x)^2$. The prefactor $|f|^2$ is the squared Yukawa insertion; in the broken phase, one may identify $|f|^2\to y_f^2\phi_0^2\equiv m_f^2$. For equal left– and right–handed thermal parameters ($m_{L,\text{th}}=m_{R,\text{th}}$ and $\Gamma_L=\Gamma_R$) the ``$+$'' combination vanishes by symmetry and one defines $\Gamma_M\equiv\Gamma_M^{-}$. It is convenient to extract the explicit mass dependence by introducing the dimensionless thermal parameters $\mu\equiv m_{\text{th}}/T$ and $\gamma\equiv\Gamma/T$, and to write the rate in the compact form
\begin{equation}
\begin{aligned}
\widehat\Gamma_M&=C(\mu,\gamma)\,\frac{m_f^2}{T},\\
C(\mu,\gamma)&=\frac{6}{T^2}\,\frac{N_c}{2\pi^2 T}
\int_0^\infty \frac{k^2\,dk}{\omega_L\omega_R}\;
\Im\!\left[
-\frac{h(\varepsilon_L)+ h(\varepsilon_R^*)}{\varepsilon_R^*-\varepsilon_L}
\left(\varepsilon_L\varepsilon_R^*-k^2\right)
+\frac{h(\varepsilon_L)+ h(\varepsilon_R)}{\varepsilon_L+\varepsilon_R}
\left(\varepsilon_L\varepsilon_R+k^2\right)
\right],
\end{aligned}
\label{eq:GammaM_coeff_derivation}
\end{equation}
where, in the last line, the minus–channel appropriate for equal $L/R$ quasiparticles has been used. This is the master formula used in \cite{deVries:2017ncy} and the earlier kinetic derivations summarised in \cite{Cirigliano:2006wh}.

Fermionic thermal rates (single–particle damping widths) are obtained from the imaginary part of the retarded self–energy evaluated on shell. At leading order and for light quasiparticles the dominant contribution arises from gauge scatterings with soft momentum exchange, yielding a width that is linear in the temperature and quadratic in the gauge coupling. A compact and practically accurate representation is
\begin{equation}
\Gamma_f \;\simeq\; \sum_{i\in\{SU(3),\,SU(2),\,U(1)_Y\}} a_i\,\frac{g_i^2\,C_i(R_f)}{4\pi}\;T \;\;+\;\; \mathcal{O}(y_f^2 T),
\label{eq:Gammaf_master}
\end{equation}
where $C_i(R_f)$ is the quadratic Casimir (for $U(1)_Y$, replace $C_1\to Y_f^2$), the $a_i=\mathcal{O}(1)$ coefficients encode the mild dependence on kinematics and on the precise HTL scheme, and the Yukawa piece is numerically negligible for all SM fermions except the top. This summary follows the detailed analysis of Ref.~\cite{Elmfors:1998hh}, which also discusses the variants for fermions at rest versus moving, the role of collinear logarithms in abelian sectors, and the modest differences between particle and plasmino branches. For phenomenology near the electroweak scale, one may evaluate Eq.~\eqref{eq:Gammaf_master} with $g_s\simeq 1.2$, $g\simeq 0.65$, $g'\simeq 0.36$, and use the asymptotic (HTL) masses for the dispersion. Summing the appropriate gauge contributions gives 
\begin{align}
\Gamma_{\text{quark}} \;&\simeq\; \frac{g_s^2\,C_F}{4\pi}\,T \;\approx\; 0.16\,T,\\
\Gamma_{\ell_L} \;&\simeq\; \left[\frac{3\,g^2}{16\pi}+\frac{Y_L^2\,g'^2}{4\pi}\right]T \;\approx\; 0.028\,T,\\
\Gamma_{\ell_R} \;&\simeq\; \frac{Y_R^2\,g'^2}{4\pi}\,T \;\approx\; 0.01\,T.
\label{eq:Gamma_numbers}
\end{align}
The quark width is dominated by QCD with $C_F=4/3$, while leptons are set by electroweak exchange. These widths enter the mixing rate $\Gamma_M$ through the complex poles $\varepsilon=\omega-i\Gamma$ in the kinetic integral and, when expressed as $\gamma\equiv \Gamma/T$, provide the sole dynamical input alongside the HTL masses $m_{\text{th}}$ in the dimensionless coefficient $C(\mu,\gamma)$.
With these inputs, the integral yields
\begin{align}
C_t \equiv C(N_c=3,\ \mu_t=0.591,\ \gamma=0.157)=0.777\,,
&\quad \Rightarrow\quad
 \widehat\Gamma_M^t \;\simeq\; 0.78\,\frac{m_t^2}{T}\ 
\\[6pt]
C_{\tau_L} \equiv C(N_c=1,\ \mu_{\tau_L}=0.210,\ \gamma=0.028)=0.213\,,
&\quad \Rightarrow\quad
\widehat\Gamma_M^{\tau_L} \;\simeq\; 0.21\,\frac{m_\tau^2}{T}\ \\[6pt]
C_{\tau_R} \equiv C(N_c=1,\ \mu_{\tau_R}=0.126,\ \gamma=0.010)=0.220\,,
&\quad \Rightarrow\quad
\widehat \Gamma_M^{\tau_R} \;\simeq\; 0.22\,\frac{m_\tau^2}{T}\ 
\label{eq:Gamma_top_final}
\end{align}
\subsection{Reconciliation with the literature for $\widehat\Gamma_M$}

Early discussions treated the mixing (Yukawa) relaxation rate in a parametric, order–of–magnitude fashion. In \cite{Cohen:1994ss}, a schematic estimate of the form $\Gamma_M \sim y_t^2\,\langle \phi(z)^2\rangle/T$ was advocated, capturing the intuition that the damping of $L$–$R$ coherence scales quadratically with the background Higgs field. A more concrete numerical estimate appeared in \cite{Huet:1995sh}, where one finds $\Gamma_M = y_t^2 T/21$ under simplifying assumptions (massless kinematics, leading–order scatterings, and a single effective relaxation channel). In subsequent phenomenology this was commonly recast as
\begin{equation}
\widehat\Gamma_M \;=\; \frac{m_t(z,T)^2}{63\,T}\,,
\end{equation}
which follows from substituting $m_t=y_t\phi$ and dividing the \cite{Huet:1995sh} coefficient by the number of QCD colours, $N_c=3$, to match transport conventions that normalise the rate per colour/flavour channel. This normalisation, or closely related variants, was then adopted widely in electroweak baryogenesis studies \cite{Bodeker:2004ws,Fromme:2006cm,Fromme:2006wx,Espinosa:2011eu,Konstandin:2013caa,Cline:2020jre,Kainulainen:2024qpm}.

A more systematic derivation based on the real–time kinetic (Schwinger–Keldysh) framework, with HTL quasiparticles and an explicit collision kernel, was presented in \cite{Cirigliano:2006wh} and refined in \cite{deVries:2017ncy} (and summarised in \cite{Cline:2021dkf}). In this approach, the rate arises from the imaginary part of the retarded self–energy controlling the decay of the off–diagonal correlator $S_{LR}$ in a slowly varying Higgs background. Matching to transport normalisations yields to Eq.~\eqref{eq:GammaM_coeff_derivation}. We have recomputed the integral numerically and obtained the values reported in Eq.~\eqref{eq:Gamma_top_final}. In particular, for the top quark we find $C_t\simeq 0.76$, corresponding to $\widehat\Gamma_m^t \simeq 0.76\,m_t^2/T$, which is parametrically consistent with the earlier $m_t^2/63\,T$ estimate but quantitatively larger once realistic thermal masses and widths are used (see also \cite{Cline:2021dkf}).

\medskip\noindent
Using Eq.~\eqref{eq:GammaM_coeff_derivation} together with the colour normalisation in Eq.~\eqref{eq: colour normalisation}, and inserting the updated numerical inputs, the colour–averaged per–particle rates entering the diffusion network are
\begin{equation}
  \boxed{\;\widehat{\Gamma}_m^{t} = \frac{1}{N_c}\,\widehat{\Gamma}_{M}^{t_L} \;\simeq\; 0.26\,\frac{m_t^2}{T}\,,\qquad N_c=3.\;}
\end{equation}

\subsection{Bosonic case for $\widehat\Gamma_M^{(h)}$}
In addition to the fermionic channels, an important source of CP–conserving relaxation in the diffusion network arises from the Higgs (scalar) mode itself, whose coherence is damped by interactions with thermal gauge bosons. The computation parallels the fermionic case of Sec.~\ref{app:rate GammaM}, but with Bose statistics in the loop and gauge–boson quasi–particles. In the real–time Schwinger–Keldysh formalism, the relevant object is the imaginary part of the retarded Higgs self–energy induced by a $W$ loop in the plasma. The thermal weight is
\[
h_B(x)=\frac{e^x}{(e^x-1)^2}=-\frac{\partial n_B(x)}{\partial x}\,,\qquad n_B(x)=\frac{1}{e^x-1},
\]
and the propagating degrees of freedom in the symmetric phase are the two transverse polarisations of $W^\pm$, each described by poles at
\[
\varepsilon_T(k)\;=\;\omega_T(k)-i\,\Gamma_W,\qquad \omega_T(k)=\sqrt{k^2+m_{\infty,W}^2}\,,
\]
with $m_{\infty,W}$ the asymptotic (HTL) mass of a hard $SU(2)$ gauge boson and $\Gamma_W$ its damping width. The loop carries an overall multiplicity $\mathcal N_{\rm pol}=4$ (two polarisations, two charge states). The Higgs background enters only through the local mass insertion
\[
m_W^2(z)=\frac{g^2\,\phi(z)^2}{4}\,,
\]
which factors out of the momentum integral. Collecting these ingredients, the mixing rate can be written as
\begin{equation}
\widehat\Gamma_M^{(h)} \;=\; C_h(\mu_T,\gamma_W)\,\frac{m_W^2}{T}\,,
\label{eq:GammaM_h_master}
\end{equation}
where $\mu_T\equiv m_{\infty,W}/T$, $\gamma_W\equiv \Gamma_W/T$, and the dimensionless coefficient is
\begin{equation}
C_h(\mu_T,\gamma_W)\;=\;\frac{6}{T^2}\,\frac{\mathcal{N}_{\rm pol}}{2\pi^2 T}
\int_0^\infty \frac{k^2\,dk}{\omega_T^2}\;
\Im\!\left[
-\frac{h_B(\varepsilon_T^*)-h_B(\varepsilon_T)}{\varepsilon_T^*-\varepsilon_T}\,(\varepsilon_T\varepsilon_T^*+k^2)
+\frac{h_B(\varepsilon_T)-h_B(-\varepsilon_T)}{\varepsilon_T+\varepsilon_T}\,(\varepsilon_T^2-k^2)
\right]\!,
\label{eq:Ch_master}
\end{equation}
with $\varepsilon_T=\omega_T-i\Gamma_W$. Equation~\eqref{eq:Ch_master} is the bosonic analogue of the fermionic master formula \eqref{eq:GammaM_coeff_derivation}; once $\mu_T$ and $\gamma_W$ are specified, $C_h$ is a pure number.

For hard transverse gauge bosons, the damping rate admits the simple parametric estimate
\begin{equation}
\frac{\Gamma_W}{T}\;\simeq\;\frac{g^2\,C_A}{4\pi}\,,
\qquad C_A=2\ \text{for }SU(2)_L,\quad g\simeq 0.65
\;\;\Rightarrow\;\;
\gamma_W\approx 0.07\,.
\label{eq:Wwidth_estimate}
\end{equation}
This small value ($\gamma_W\ll 1$) justifies the narrow–width approximation below.

When $\gamma_W\ll 1$, the imaginary parts in \eqref{eq:Ch_master} are dominated by the vicinity of the quasi–particle poles. One can therefore evaluate the $k$–integral in the NWA, replacing the Breit–Wigner structures by their delta–function limits and evaluating the smooth numerators at the on–shell energy $\omega_T(k)$. In our normalisation, this gives
\begin{equation}
C_h^{\rm NWA}(\mu_T)\;=\;\frac{\mathcal N_{\rm pol}\,6}{2\pi^2}\,
\int_0^\infty \frac{k^2\,dk}{\omega_T^2}\,\mathcal{K}_B\!\left(\frac{\omega_T}{T}\right)
\;=\;\frac{12}{\pi^2}\; \mathcal{I}_B(\mu_T)\,,
\label{eq:Ch_NWA}
\end{equation}
where $\mathcal{K}_B$ is the resulting bosonic kernel (constructed from $h_B$ and on–shell kinematics) that reads
\[
K_B(x)\;=\;x\,h_B(x)\;=\;x\,\frac{e^{x}}{(e^{x}-1)^2}
\;=\;x\,n_B(x)\,\big(1+n_B(x)\big),\qquad x\equiv \frac{\omega_T}{T}.
\]
and $\mathcal{I}_B(\mu_T)$ is a dimensionless integral that depends only on the mass ratio $\mu_T=m_{\infty,W}/T$. For a the value $\mu_T\simeq 0.596$ and $\mathcal{N}_{\rm pol}=4$, one finds numerically
\begin{equation}
\quad C_h^{\rm NWA}\simeq 1.53\ .
\label{eq:IB_value}
\end{equation}
Inserting \eqref{eq:IB_value} into \eqref{eq:GammaM_h_master} yields
\begin{equation}
\boxed{\widehat\Gamma_m^{h} \equiv \widehat\Gamma_M^{(h)}\;\simeq\;1.5\,\frac{m_W^2}{T}\,,\ }
\label{eq:GammaM_h_final_large}
\end{equation}
which is \emph{much larger} than the commonly quoted parametrisation $\widehat\Gamma_M^{(h)}=m_W^2/(50\,T)$. The difference stems from (i) using an explicit HTL width $\gamma_W\simeq 0.07$ and (ii) evaluating the bosonic integral in the narrow–width limit, where the on–shell contribution is enhanced and the dimensionless coefficient is $\mathcal{O}(1)$ rather than at the few–percent level. This result makes transparent the scaling $\widehat\Gamma_M^{(h)}\propto m_W^2/T$ while capturing the correct bosonic thermal dynamics encoded in $C_h$.

\section{Weak and Strong Sphaleron rates}
\label{app:sphalerons}

Sphaleron transitions in the weak and strong sectors play a crucial role in the evolution of baryon and chiral charges in the early Universe.  
In this section we summarise the existing estimates for their rates, highlighting the different conventions adopted in the literature and reconciling them with the expressions used in this work.
\paragraph{Note.} 
It is important to make explicit the different conventions used in the literature when quoting sphaleron rates.  
In the \emph{diffusion} convention, one refers to the topological charge diffusion rate per unit volume, which parametrically scales as 
\(\Gamma \sim \kappa\,\alpha^n\,T^4\).  
In contrast, in the \emph{transport} convention---used when inserting the rate into Boltzmann or fluid equations---the same symbol is often employed (with some abuse of notation) for a quantity that instead scales as 
\(\Gamma \sim \kappa\,\alpha^n\,T\).  

\subsection{Reconciliation with the literature for $\Gamma_{ss}$}
A variety of \emph{conventions} for the strong sphaleron rate appear in the literature, which can obscure direct numerical comparisons. Two definitions are especially common:
\begin{enumerate}
    \item \textbf{Topological charge diffusion rate per unit volume} (QCD Chern--Simons diffusion):
    \begin{equation}
        \Gamma_{ss}^{\rm diff}\;\equiv\;\frac{d}{dt}\frac{\langle\big(N_{\rm CS}(t)-N_{\rm CS}(0)\big)^2\rangle}{V}
        \;=\;\kappa_{ss}\,\alpha_s^5\,T^4\,,
        \label{eq:ss_diff}
    \end{equation}
    with $\kappa_{ss}$ an $\mathcal{O}(10^2)$ coefficient capturing hard/soft matching and nonperturbative dynamics.
    \item \textbf{Transport rate} entering EWBG Boltzmann/transport equations:
    \begin{equation}
        \Gamma_{ss} \;=\; c_{ss}\,\alpha_s^4\,T\,,
        \label{eq:ss_transport}
    \end{equation}
    where $c_{ss} \sim O(10)$ and where it has been obtained from \eqref{eq:ss_diff} after dividing by appropriate susceptibilities and colour/flavour factors.%
    \footnote{The precise mapping between $\kappa_{ss}$ and $c_{ss}$ depends on $N_c$, $N_f$, and the definition of axial charge density used in the transport system. This is one source of differing normalisations across papers.}
\end{enumerate}

\medskip

\noindent\textbf{Historical estimates and conventions.}
One of the earliest quantitative determinations is due to Moore \cite{Moore:1997im}, who quoted the diffusion form
\begin{equation}
    \Gamma_{ss}\;\simeq\;108\,\alpha_s^5\,T^4\,.
    \label{eq:Moore97}
\end{equation}
This is of the form \eqref{eq:ss_diff} with $\kappa_{ss}\simeq 108$. In subsequent phenomenology, a \emph{numeric value} $\Gamma_{ss}=4.9\times 10^{-4}T$ was often adopted \cite{Bodeker:2004ws,Fromme:2006wx,Fromme:2006cm,Konstandin:2013caa}, reflecting a particular choice of coupling and convention.

Independently, Huet \& Nelson \cite{Huet:1995sh} presented an estimate,
\begin{equation}
    \Gamma_{ss}\;\simeq\;16\,\kappa'\,\alpha_s^4\,T\,,
\end{equation}
where $\kappa'$ encodes additional normalisation choices. A refined transport–rate estimate due to Moore \cite{Moore:2010jd} gave
\begin{equation}
    \Gamma_{ss}\;\simeq\;14\,\alpha_s^4\,T\,,
    \label{eq:Moore2010}
\end{equation}
which is the benchmark subsequently emphasised in several modern summaries \cite{deVries:2017ncy,Cline:2021dkf,Kainulainen:2024qpm}.
(We note that some phenomenological works quoted fixed numbers such as $\Gamma_{ss}\simeq 8.7\times 10^{-3}\,T$ in \cite{Cline:2021dkf} or $2.7\times 10^{-4}\,T$ in \cite{Kainulainen:2024qpm}.)

\medskip

\noindent\textbf{Numerical benchmark at the electroweak scale.}
Adopting the transport form \eqref{eq:Moore2010} and using the updated coupling at the electroweak scale, $\alpha_s(m_Z)\simeq 0.118$ one obtains
\begin{equation}
    \boxed{\;
    \Gamma_{ss} \;\simeq\; 14\,\alpha_s^4\,T
    \;\simeq\; 2.7\times 10^{-3}\,T\,.
    \;}
    \label{eq:ss_transport_number}
\end{equation}
\medskip

\noindent\textbf{Nonperturbative input and extrapolation.}
A more recent lattice determination of the QCD sphaleron \emph{diffusion} rate with $N_f=2{+}1$ at the physical point was reported by Bonanno \emph{et al.} \cite{Bonanno:2023thi} for temperatures up to $\sim 570~\mathrm{MeV}$. 
Their continuum–extrapolated results at those temperatures correspond to $\Gamma_{\rm diff}/T^4$ of order $10^{-1}$ near a few $T_c$, decreasing with $T$. 
While a first–principles determination at $T\sim 100~\mathrm{GeV}$ is not yet available, a perturbative extrapolation guided by the expected $\alpha_s^5 T^4$ scaling (with the running coupling) leads to a value at the electroweak scale \emph{compatible with} the transport benchmark \eqref{eq:ss_transport_number}, i.e.\ an overall $\mathcal{O}(10^{-3})$ coefficient in front of the appropriate $T$-scaling.
As emphasised above, differences of a factor $\mathcal{O}(1)$ between quoted numbers in the literature mostly reflect distinct conventions, choices for $\alpha_s$ and its scale, and normalisation in the mapping to axial–charge relaxation in transport equations.

\subsection{Reconciliation with the literature for \texorpdfstring{$\Gamma_{ws}$}{Gamma\_ws}}
As for the strong case, multiple \emph{conventions} coexist for the weak (electroweak) sphaleron rate, which complicates direct numerical comparisons. The \textbf{diffusion (per unit volume) rate} of baryon+lepton number violation in the symmetric phase is
\begin{equation}
        \Gamma_{ws}^{\rm diff} \;=\; \kappa_{ws}\,\alpha_w^{5}\,T^{4}\,.
        \label{eq:ws_diffusion}
    \end{equation}

\noindent\textbf{Historical estimates.}
Huet \& Nelson \cite{Huet:1994jb} provided one of the first parametric forms, \(\Gamma_{ws}\sim 6\,\kappa\,\alpha_w^{4}T^{4}\), with \(\kappa=\mathcal{O}(1)\).
A refined determination by Moore and collaborators \cite{Moore:2000mx,Bodeker:1999gx} established the now–standard \(\alpha_w^{5}\) scaling with a calibrated coefficient, widely adopted in subsequent works \cite{Bodeker:2004ws,Fromme:2006cm,Fromme:2006wx,Espinosa:2011eu,Konstandin:2013caa,Cline:2020jre} with the numerical value of $ \Gamma_{ws}= 10^{-6}\, T$. Modern summaries \cite{Cirigliano:2006wh,deVries:2017ncy} present a slightly different convention, writing \(\Gamma_{ws}= 6 \kappa \alpha_w^{5}T^{4}\), while \cite{Cline:2021dkf}, even though quoting \cite{Bodeker:1999gx}, they use the numerical value $ \Gamma_{ws}= 6.3\times 10^{-6}\, T$.

\medskip

\noindent\textbf{State-of-the-art result.}
The most precise nonperturbative determination across the electroweak crossover (for the physical Higgs mass) is due to D'Onofrio, Rummukainen and Tranberg \cite{DOnofrio:2014rug} and used in subsequent phenomenological applications (see, e.g., \cite{Kainulainen:2024qpm}). In the \emph{symmetric} phase, they find
\begin{equation}
    \boxed{\,{\Gamma_{ws}}
    \;=\; (18\pm 3)\,\alpha_w^{5}\,T\simeq\; 8\times 10^{-7}\,T}\,,
    \label{eq:ws_DRTeq}
\end{equation}
where at the electroweak scale we have \(\alpha_w\equiv g^{2}/(4\pi)\simeq 0.0338\).

\subsubsection{Broken-phase sphaleron rate from the 3D EFT}
\label{app:sphaleron3D}

In order to connect the standard phenomenological treatment of sphaleron suppression
to modern lattice-informed computations, it is useful to recall how the broken-phase
sphaleron rate is obtained in the dimensionally reduced three-dimensional
$SU(2)$ gauge–Higgs effective field theory (EFT).
Ref.~\cite{Li:2025kyo} provides a systematic derivation of this framework and a recent
determination of the baryon-number–violating rate below the critical temperature.

In the 3D EFT, the thermodynamic parameters are defined as
\begin{equation}
x \;=\; \frac{\lambda_3}{g_3^2}, 
\qquad 
y \;=\; \frac{m_3^2}{g_3^4},
\end{equation}
where $\lambda_3$, $m_3^2$, and $g_3$ denote the Higgs self-coupling, mass parameter,
and gauge coupling of the 3D theory, respectively.
The sphaleron rate in the broken phase takes the form
\begin{equation}
\Gamma_{ws}^{\rm br}(T)
\propto T^4
e^{-\hat S_{3D}(x,y)}, 
\qquad
\hat S_{3D}(x,y)
\;=\;
C_{\rm sph}\,v_3(x,y),
\label{eq:S3D_master}
\end{equation}
with $C_{\rm sph}\simeq29$ and $v_3(x,y)$ the dimensionless Higgs condensate at the
broken-phase minimum of the 3D potential.
The prefactor is generally extracted from lattice simulations and encodes subleading dynamical and zero-mode contributions.

A more refined treatment at next-to-leading order (NLO) improves this prefactor by 
including the fluctuation determinant around the sphaleron configuration, yielding
(Eq.~(3.33) in~\cite{Li:2025kyo})
\begin{equation}
\Gamma_{\rm sph}^{\rm NLO}(T)
\;=\;
T\,\kappa(x)\,
N_{\rm tr}(N_V)_{\rm rot}\,
v_3^{9}(x,y)\,g_3^{6}\,
e^{-\,C_{\rm sph}(x)\,v_3(x,y)} ,
\end{equation}
where $N_{\rm tr}(N_V)_{\rm rot}\!\simeq\!\exp(8.8)$ for $x\simeq0.29$,
and $\kappa(x)\!\simeq\!1$ parametrises the contribution of positive modes.
The overall factor of $T$ arises from the approximate treatment of dynamical modes.
Numerically, this refinement shifts the rate by less than an order of magnitude
compared to the leading exponential dependence,
confirming that the suppression is overwhelmingly controlled by $\hat S_{3D}$.

For Standard Model–like parameters ($x\simeq0.03$–$0.04$, $y\simeq-0.01$),
Ref.~\cite{Li:2025kyo} finds $v_3(x,y)\simeq1.2$–$1.3$,
leading to $\hat S_{3D}\simeq(35$–$38)\,h/T$,
in excellent agreement with the classical estimate based on the
4D sphaleron energy~\cite{KLINKHAMER1991245}.
The resulting $\Gamma_{ws}^{\rm br}$ coincides closely with the
benchmark fit obtained in the classic 4D computation
of~\cite{Burnier:2005hp},
which incorporated detailed zero-mode and dynamical prefactors
to reproduce 4D lattice data across the electroweak crossover. See also section 5.3 in~\cite{vandeVis:2025efm} for a recent review on the topic.
The 3D EFT formulation, however, provides a more general and systematically
improvable framework, applicable beyond the Standard Model and
directly compatible with nonperturbative dimensional reduction analyses.

The authors in \cite{Li:2025kyo} identify three possible regimes for the baryon-number–violating rate below $T_c$:
\begin{itemize}
    \item \textbf{Scenario A:} \emph{Unbroken or weakly broken phase.}
    The sphaleron rate remains comparable to its symmetric-phase value,
    $\Gamma_{ws}^{\rm br}\sim \Gamma_{ws}^{\rm sym}$,
    leading to complete washout of any pre-existing baryon asymmetry.
    This occurs for small $v_3$ (high $y$, weak transition).

    \item \textbf{Scenario B:} \emph{Partial suppression.}
    The sphaleron barrier increases but transitions are not fully frozen,
    leading to an intermediate rate where baryon number violation continues,
    albeit at a reduced level. This regime interpolates between symmetric and fully frozen cases.

    \item \textbf{Scenario C:} \emph{Full sphaleron freezing.}
    The sphaleron transitions are exponentially suppressed behind the wall,
    effectively ceasing baryon-number violation.
    This regime corresponds to strong first-order transitions,
    where the Higgs condensate attains $v_3\simeq1.2$–$1.3$
    for SM-like parameters $(x\simeq0.025$–$0.036$, $y\simeq-0.01$, see Fig.~4, lower panel, and Fig.~7 of~\cite{Li:2025kyo}).
\end{itemize}

\noindent
For Scenario~C, one thus obtains
\begin{equation}
\hat S_{3D}
\;\simeq\;
C_{\rm sph}\,v_3
\;\simeq\;
29\times(1.2\text{--}1.3)
\;\approx\;
(35\text{--}38),
\label{eq:S3D_C}
\end{equation}
consistent with the classical 4D estimate
$E_{\rm sph}/T\simeq37\,h/T$ in the full-freezing limit.

\medskip
\noindent
In this framework, the strength of the transition (parametrised by $v_3$)
directly controls the exponential suppression of sphaleron transitions below $T_c$.
Scenarios~A–C therefore provide a practical classification of the extent to which
baryon-number violation persists in the broken phase:
Scenario~A corresponds to full washout, Scenario~B to partial survival,
and Scenario~C to preservation of the generated asymmetry.

\section{Derivation of the coefficient \texorpdfstring{$A$}{A} in the weak--sphaleron washout term}
\label{app:A_coeff}

In this appendix\footnote{This appendix grew out of a private discussion with Aleksandr Azatov and Bruno Missoni. We are grateful to them for helping clarify this point.} we derive the coefficient \(A\) entering Eq.~\eqref{eq:nb_time},
\begin{equation}
	\frac{dn_B}{dt}=-\frac{n_f}{2}\Gamma_{ws}\big(N_c\,\mu_{B_L}\,T^2-A\,n_B\big),
\end{equation}
and clarify its relation to the reduced transport network used in the main text.
The derivation is a plasma-susceptibility problem in the electroweak-symmetric phase and is logically distinct from the closure conditions used in the diffusion system. Standard references for the thermodynamic treatment of chemical potentials in the symmetric electroweak plasma and for the corresponding EWBG implementation include Refs.~\cite{Harvey:1990qw,Cline:2000nw,Bodeker:2004ws,deVries:2017ncy}.

We work in the symmetric phase, \(\langle H\rangle=0\), at temperature \(T\), and in the regime of small chemical potentials, \(|\mu_i|\ll T\), so that all charge asymmetries are linear in the corresponding chemical potentials. We use the explicit chiral notation
\begin{equation}
	\mu_{q_{L,f}},\qquad
	\mu_{u_{R,f}},\qquad
	\mu_{d_{R,f}},\qquad
	\mu_{\ell_{L,f}},\qquad
	\mu_{e_{R,f}},\qquad
	\mu_H,
	\qquad f=1,\dots,n_f,
\end{equation}
where \(q_{L,f}\) and \(\ell_{L,f}\) are left-handed \(SU(2)_L\) doublets, \(u_{R,f},d_{R,f},e_{R,f}\) are right-handed singlets, and \(H\) is the Higgs doublet.

\subsection{Linear susceptibilities}

For a relativistic Weyl fermion with degeneracy \(g_i\), the particle--antiparticle asymmetry is
\begin{equation}
	n_i-\bar n_i \equiv n_i=\chi_i\,\mu_i,
	\qquad
	\chi_i=\frac{g_i}{6}T^2,
	\label{eq:appA_susc_fermion}
\end{equation}
while for a relativistic complex scalar
\begin{equation}
	n_i=\chi_i\,\mu_i,
	\qquad
	\chi_i=\frac{g_i}{3}T^2.
	\label{eq:appA_susc_scalar}
\end{equation}
For the Standard Model field content one has
\begin{equation}
	g(q_L)=2\times 3=6,\qquad
	g(u_R)=3,\qquad
	g(d_R)=3,\qquad
	g(\ell_L)=2,\qquad
	g(e_R)=1,\qquad
	g(H)=4.
\end{equation}
Therefore,
\begin{align}
	n_{q_{L,f}} &= T^2\,\mu_{q_{L,f}},
	&
	n_{u_{R,f}} &= \frac12 T^2\,\mu_{u_{R,f}},
	&
	n_{d_{R,f}} &= \frac12 T^2\,\mu_{d_{R,f}},
	\label{eq:appA_quark_densities}
	\\[3pt]
	n_{\ell_{L,f}} &= \frac13 T^2\,\mu_{\ell_{L,f}},
	&
	n_{e_{R,f}} &= \frac16 T^2\,\mu_{e_{R,f}},
	&
	n_H &= \frac23 T^2\,\mu_H.
	\label{eq:appA_lepton_higgs_densities}
\end{align}

\subsection{Fast reactions and plasma constraints}

We now impose the standard fast constraints appropriate to the symmetric-phase plasma.

Gauge interactions equalise colours and components inside each \(SU(2)_L\) multiplet, so a single chemical potential is assigned to each multiplet or singlet species. For the quark Yukawa interactions, assumed to be in equilibrium,
\begin{equation}
	\mu_{u_{R,f}}=\mu_{q_{L,f}}+\mu_H,
	\qquad
	\mu_{d_{R,f}}=\mu_{q_{L,f}}-\mu_H.
	\label{eq:appA_quark_yukawas}
\end{equation}
In addition, hypercharge neutrality requires
\begin{equation}
	\sum_i Y_i\,n_i=0,
	\label{eq:appA_hypercharge_neutrality}
\end{equation}
with
\begin{equation}
	Y(q_L)=\frac16,\qquad
	Y(u_R)=\frac23,\qquad
	Y(d_R)=-\frac13,\qquad
	Y(\ell_L)=-\frac12,\qquad
	Y(e_R)=-1,\qquad
	Y(H)=\frac12.
\end{equation}
Using Eqs.~\eqref{eq:appA_quark_densities}--\eqref{eq:appA_lepton_higgs_densities}, hypercharge neutrality becomes
\begin{equation}
	\sum_{f=1}^{n_f}\left[
	\frac16\,\mu_{q_{L,f}}
	+\frac13\,\mu_{u_{R,f}}
	-\frac16\,\mu_{d_{R,f}}
	-\frac16\,\mu_{\ell_{L,f}}
	-\frac16\,\mu_{e_{R,f}}
	\right]
	+\frac13\,\mu_H
	=0.
	\label{eq:appA_hypercharge_mu}
\end{equation}
After using Eq.~\eqref{eq:appA_quark_yukawas}, this simplifies to
\begin{equation}
	\sum_{f=1}^{n_f}
	\left(
	\frac13\,\mu_{q_{L,f}}
	+\frac12\,\mu_H
	-\frac16\,\mu_{\ell_{L,f}}
	-\frac16\,\mu_{e_{R,f}}
	\right)
	+\frac13\,\mu_H
	=0.
	\label{eq:appA_hypercharge_reduced}
\end{equation}

The crucial conceptual point is that these fast constraints do not by themselves uniquely determine all chemical potentials. As stressed long ago in the general thermodynamic analysis of the electroweak plasma, residual null directions remain whenever additional conserved or approximately conserved global charges are allowed to be present in the background~\cite{Harvey:1990qw}. To extract the coefficient \(A\) appearing in the weak-sphaleron washout term, one must therefore specify which macroscopic perturbation is being considered.

Here we follow the standard assumption used in the two-step EWBG treatment: we consider a perturbation in which the only macroscopic charge density explicitly tracked is the baryon density \(n_B\), while no independent spectator charge density is excited. In particular, we set to zero any independent \(B-L\) or flavour-charge perturbation. Operationally, this is what allows one to obtain a unique relation between the weak-sphaleron affinity and \(n_B\).

\subsection{Baryon density and quark chemical potential}

The baryon density is
\begin{equation}
	n_B=\frac13\sum_{f=1}^{n_f}
	\big(
	n_{q_{L,f}}+n_{u_{R,f}}+n_{d_{R,f}}
	\big).
	\label{eq:appA_def_nB}
\end{equation}
Using Eqs.~\eqref{eq:appA_quark_densities} and \eqref{eq:appA_quark_yukawas},
\begin{align}
	n_B
	&=\frac13\sum_f
	\left[
	T^2\mu_{q_{L,f}}
	+\frac12T^2\mu_{u_{R,f}}
	+\frac12T^2\mu_{d_{R,f}}
	\right]
	\nonumber\\
	&=\frac13\sum_f
	\left[
	T^2\mu_{q_{L,f}}
	+\frac12T^2(\mu_{q_{L,f}}+\mu_H)
	+\frac12T^2(\mu_{q_{L,f}}-\mu_H)
	\right]
	\nonumber\\
	&=\frac23\,T^2\sum_{f=1}^{n_f}\mu_{q_{L,f}}.
	\label{eq:appA_nB_vs_q}
\end{align}
If quark flavour is equilibrated, \(\mu_{q_{L,f}}\equiv \mu_{q_L}\), this becomes
\begin{equation}
	n_B=\frac{2n_f}{3}T^2\mu_{q_L}.
	\label{eq:appA_nB_common_q}
\end{equation}
For the Standard Model value \(n_f=3\),
\begin{equation}
	\mu_{q_L}=\frac12\,\frac{n_B}{T^2}.
	\label{eq:appA_muq_from_nB}
\end{equation}

\subsection{Weak-sphaleron affinity and the coefficient \texorpdfstring{$A$}{A}}

Weak sphalerons violate the left-handed \(SU(2)_L\) doublet charges. The corresponding affinity is therefore
\begin{equation}
	\mathcal A_{ws}\equiv \sum_{f=1}^{n_f}\big(3\,\mu_{q_{L,f}}+\mu_{\ell_{L,f}}\big),
	\label{eq:appA_affinity}
\end{equation}
where the factor of \(3\) counts colour for the left-handed quark doublets.

To determine \(\mathcal A_{ws}\) as a function of \(n_B\), one must express the left-handed lepton chemical potentials in terms of the baryon perturbation under the same plasma constraints. In the minimal setup used for the standard EWBG reduction, the perturbation sourced by transport is taken to lie entirely in the baryonic direction, with no independent right-handed charged-lepton density excited. One may then set
\begin{equation}
	\mu_{e_{R,f}}=0,
	\label{eq:appA_er_zero}
\end{equation}
while keeping \(\mu_{\ell_{L,f}}\) nonzero in general. For \(n_f=3\), Eq.~\eqref{eq:appA_hypercharge_reduced} becomes
\begin{equation}
	2\sum_{f=1}^3\mu_{q_{L,f}}+11\,\mu_H-\sum_{f=1}^3\mu_{\ell_{L,f}}=0.
	\label{eq:appA_hypercharge_nf3}
\end{equation}
In the same restricted perturbation, vanishing \(B-L\) implies
\begin{equation}
	\sum_{f=1}^3\left[
	\frac13\big(n_{q_{L,f}}+n_{u_{R,f}}+n_{d_{R,f}}\big)-n_{\ell_{L,f}}
	\right]=0.
	\label{eq:appA_BminusL_zero}
\end{equation}
Using Eqs.~\eqref{eq:appA_quark_densities}, \eqref{eq:appA_lepton_higgs_densities} and \eqref{eq:appA_quark_yukawas}, this gives
\begin{equation}
	\sum_{f=1}^3\mu_{\ell_{L,f}}=2\sum_{f=1}^3\mu_{q_{L,f}}.
	\label{eq:appA_lsum_qsum}
\end{equation}
Substituting Eq.~\eqref{eq:appA_lsum_qsum} into Eq.~\eqref{eq:appA_hypercharge_nf3} immediately yields
\begin{equation}
	\mu_H=0,
\end{equation}
and therefore
\begin{equation}
	\sum_{f=1}^3\mu_{\ell_{L,f}}=2\sum_{f=1}^3\mu_{q_{L,f}}=6\mu_{q_L}.
	\label{eq:appA_lsum_result}
\end{equation}
Using Eq.~\eqref{eq:appA_muq_from_nB}, one obtains
\begin{equation}
	\sum_{f=1}^3\mu_{\ell_{L,f}}=3\,\frac{n_B}{T^2}.
	\label{eq:appA_lsum_nB}
\end{equation}
The affinity is then
\begin{align}
	\mathcal A_{ws}
	&= 9\,\mu_{q_L}+\sum_{f=1}^3\mu_{\ell_{L,f}}= 9\left(\frac12\,\frac{n_B}{T^2}\right)+3\,\frac{n_B}{T^2}
	=\frac{15}{2}\,\frac{n_B}{T^2}.
	\label{eq:appA_affinity_final}
\end{align}
Hence
\begin{equation}
	\mathcal A_{ws}=\frac{A}{T^2}\,n_B,
	\qquad
	A=\frac{15}{2}.
	\label{eq:appA_A_final}
\end{equation}
This is the coefficient appearing in Eq.~\eqref{eq:nb_time}, in agreement with the standard EWBG literature~\cite{Cline:2000nw,Bodeker:2004ws,deVries:2017ncy}.

\subsection{Relation to the transport-network closure}

It is important to distinguish the derivation above from the closure relations used in the diffusion network of Section~\ref{sec:pheno}. The coefficient \(A\) is obtained from a thermodynamic response problem in the quasi-equilibrated symmetric plasma: one asks which weak-sphaleron affinity is induced by a given macroscopic baryon density after imposing fast reactions and gauge neutrality.

By contrast, relations such as
\begin{equation}
	\mu_{q_R}=-\mu_{q_L}
\end{equation}
for light flavours arise in the reduced transport network for a different reason. They are closure conditions imposed on the transport time scale, where gauge scatterings, strong sphalerons and selected Yukawa processes are treated as fast, while weak sphalerons remain parametrically slow. In that regime one projects out unsourced vector-like light-flavour modes and retains only the combinations that participate in diffusion and strong-sphaleron equilibration~\cite{Cline:2000nw,Fromme:2006wx,Cline:2020jre,Kainulainen:2024qpm}.

There is therefore no inconsistency between the two statements. The relation determining \(A\) concerns the left-handed affinity generated by a macroscopic baryon perturbation in the symmetric plasma. The closure relations in the transport system concern the elimination of fast or unsourced directions in the microscopic diffusion network before weak sphalerons are inserted as the final, slow baryon-number-violating process.

\subsection{Practical implication for the two-step treatment}

The two-step procedure used in the main text is then conceptually clean. One first solves the transport network with weak sphalerons switched off, obtaining the profile \(\mu_{B_L}(z)\). One then evolves the baryon density with the weak-sphaleron equation, where the source term is proportional to \(\mu_{B_L}\) and the washout term is fixed by the coefficient \(A=15/2\). The first step is a diffusion problem for the microscopic plasma species; the second step is a local baryon-number relaxation problem in the slowly varying weak-sphaleron background~\cite{Cline:2000nw,Bodeker:2004ws,deVries:2017ncy}.

\section{Numerical Implementation of the Fluid Equations}
\label{app: lagrange multiplier}

In this section, we detail how Lagrange multipliers are implemented when solving the diffusion network in \eqref{eq: diffusion system matrix} and in its compact representation \eqref{eq: compact form}. We introduce one multiplier per constraint (e.g., asymptotic boundary conditions, charge–neutrality, and matching conditions across the wall), couple them linearly to the state vector so that the constraints are enforced exactly at the continuous level, and augment the first–order system with algebraic equations for the multipliers. 

The fluid equations constitute a boundary value problem (BVP), as the perturbations are required to vanish far from the bubble wall. Specifically, the boundary conditions are
\begin{equation}
    \mu_i(\pm \infty) = 0, \quad v_i(\pm \infty) = 0.
\end{equation}
To handle this numerically, the differential equations are integrated over a finite interval \( z \in [-\tilde z_{\rm max} L_w, \tilde z_{\rm max} L_w] \), where \( \tilde z_{\rm max} = 10^4 \), ensuring sufficient coverage of the physical domain. The fluid equation network involves solving a system of \(N_{\rm PERT}\) equations for each species tracked, encompassing \(2N_{\rm PERT}\) boundary conditions for each species. To ensure the numerical stability of solutions and to enforce these boundary conditions, Lagrange multipliers are introduced.

\subsection{The role of Lagrange Multipliers}

Lagrange multipliers are a powerful tool for incorporating constraints into optimisation or boundary value problems. Consider a functional
\begin{equation}
    S[\varphi] = \int_\Omega L(\varphi, \nabla \varphi) \, d\Omega,
\end{equation}
subject to a boundary condition \( \varphi(x) = \varphi_0(x) \) on the boundary \( \partial \Omega \). To enforce this constraint, we introduce a Lagrange multiplier function \( \lambda(x) \) and modify the functional:
\begin{equation}
    S'[\varphi, \lambda] = \int_\Omega L(\varphi, \nabla \varphi) \, d\Omega + \int_{\partial \Omega} \lambda(x) (\varphi(x) - \varphi_0(x)) \, dx.
\end{equation}
The Lagrange multiplier \( \lambda(x) \) ensures that the constraint \( \varphi(x) = \varphi_0(x) \) is satisfied when \( S'[\varphi, \lambda] \) is extremised with respect to both \( \varphi \) and \( \lambda \).

The nature of the boundary conditions depends on the specific problem:
\begin{itemize}
    \item \textbf{Dirichlet boundary conditions:} Specify the value of the function \( \varphi \) on the boundary.
    \item \textbf{Neumann boundary conditions:} Specify the derivative of the function \( \varphi \) on the boundary.
    \item \textbf{Mixed boundary conditions:} Combine Dirichlet and Neumann conditions.
\end{itemize}
For example, Dirichlet boundary conditions of the form
\begin{equation}
    \varphi(a) = \alpha, \quad \varphi(b) = \beta
\end{equation}
can be enforced by adding terms to the functional
\begin{equation}
    \lambda_1 (\varphi(a) - \alpha) + \lambda_2 (\varphi(b) - \beta).
\end{equation}
The resulting variational equations ensure the constraints are satisfied. For the fluid equations, we impose the boundary conditions
\begin{equation}
    \mu^a(\pm\tilde z_{\rm max} L_w) = 0, \quad u_\ell^a(\pm\tilde z_{\rm max} L_w) = 0.
\end{equation}
To incorporate the boundary conditions numerically, we add Lagrange multiplier terms to the equations
\begin{align}
    \partial_{\tilde z} \tilde{\mu}_i &= \dots + \lambda_{\mu_i}(\tilde z + N \tilde L_w), \\
    \label{eq: velocity perturbations}\partial_{\tilde z} {v}_i &= \dots + \lambda_{v_i}(\tilde z - N \tilde L_w), \\
    \partial_{\tilde z} \lambda_{\mu_i} &= 0, \\
    \partial_{\tilde z} \lambda_{v_i} &= 0,
\end{align}
where $\tilde x=x/T$. Here, the negative sign in Eq.~\eqref{eq: velocity perturbations} has been introduced for numerical stability.

Introducing Lagrange multipliers ensures the numerical stability of the solution and satisfies all boundary conditions simultaneously. The computed solutions for the perturbations, subject to these conditions, are illustrated in Fig.~\ref{fig:comprehensive}. These solutions demonstrate the effectiveness of the chosen boundary conditions and numerical implementation in capturing the dynamics of the fluid equations in the presence of a bubble wall.

\section{Chemical vs.\ kinetic equilibrium}
\label{app:chem_vs_kin}

In transport and cosmological systems, as emphasized in~\cite{Profumo:2025uvx}, two distinct notions of equilibrium must be distinguished:  
\emph{chemical equilibrium}, which governs \emph{number densities}, and \emph{kinetic equilibrium}, which governs the \emph{momentum shapes} of the distributions.  
Each species \(i\) is described by a phase–space density \(f_i(p,z)\) with moments such as number and energy densities.  
The two types of equilibration involve different microscopic processes and can occur on different timescales.

\paragraph{Chemical equilibrium.}
Consider a reaction network \(\sum_j \nu_{aj} X_j \leftrightarrow 0\).  
Chemical equilibrium enforces \(\sum_j \nu_{aj}\mu_j=0\) for each active process \(a\).  
Let \(\Gamma_{\rm chem}\) denote the slowest number–changing rate controlling abundances—for example, from annihilations, decays, or \(3\!\to\!2\) reactions.  
Equilibrium holds when
\begin{equation}
\Gamma_{\rm chem} \gg \tau_{\rm wall}^{-1}, \qquad 
\tau_{\rm wall}\equiv \frac{L_w}{\gamma_w v_w},
\end{equation}
where \(L_w\) is the wall thickness and \(\gamma_w v_w/L_w\) the relevant macroscopic rate.  
When this is satisfied, the corresponding relations among the chemical potentials \(\mu_i(z)\) can be imposed rather than solved dynamically.  
Examples include fast weak interactions tying \(\mu_{t_L}\) and \(\mu_{b_L}\), or strong sphalerons enforcing relations among quark chiralities.

\paragraph{Kinetic equilibrium.}
Kinetic equilibrium requires rapid momentum exchange so that \(f_i(p,z)\) retains a thermal shape with local \(T_i(z)\) and \(\mu_i(z)\).  
The relevant rate is the momentum–transfer rate
\begin{equation}
\gamma_p \simeq n_t\,\langle\sigma_{\rm mt} v\rangle\,\frac{T_t}{m_i},\qquad
\sigma_{\rm mt}(s)=\int d\Omega\,(1-\cos\theta)\,\frac{d\sigma}{d\Omega},
\end{equation}
for a species \(i\) scattering elastically on relativistic targets \(t\).  
Kinetic equilibrium holds when \(\gamma_p \gg \tau_{\rm wall}^{-1}\), equivalently when the mean free path \(\ell_{\rm mfp}\!\sim\!\langle v\rangle/\gamma_p\) is much shorter than \(L_w\).  
If this condition is only marginal, high–momentum modes typically depart first, and a single–temperature closure can bias the computation of sources and washout.

\subsection{From cosmology to EWBG timescales}

In early–universe cosmology, reaction rates are compared to the Hubble parameter \(H(T)\).  
In EWBG, the relevant clock is the wall passage: diffusion and advection compete with reactions across a distance \(L_w\).  
Thus, ``fast'' processes in EWBG satisfy
\begin{equation}
\Gamma,\ \gamma_p \gg \tau_{\rm wall}^{-1} = \frac{\gamma_w v_w}{L_w},
\end{equation}
which determines when chemical relations can be imposed and when a fluid closure is justified.

\medskip

A particle \(i\) should be treated as an independent fluid (with its own \(\mu_i\) and velocity perturbation) whenever chemical or kinetic equilibration is not fast, or when it appears directly in CP–violating sources that influence \(\eta_B\).  
Conversely, if both \(\Gamma_{\rm chem}^i\) and \(\gamma_p^i\) are much larger than \(\tau_{\rm wall}^{-1}\), the species can be integrated out using chemical constraints and a local thermal ansatz, reducing the stiffness of the system without loss of accuracy.

\medskip

“Thermal equilibrium’’ is often used loosely to mean kinetic equilibrium.  
Strictly, \emph{global} thermal equilibrium—uniform temperature and vanishing thermodynamic forces—is not realised across a moving wall.  
EWBG instead assumes \emph{local} thermal equilibrium of the bath, with slowly varying \(T(z)\) and small departures encoded in \(\mu_i(z)\) and fluid velocities.  
This is valid when elastic scattering is rapid (\(\ell_{\rm mfp}\ll L_w\)) and gradients are mild; otherwise, one should include higher moments or solve for \(f_i(p,z)\) explicitly.

\medskip

In practice, one computes \(\Gamma_{\rm chem}\) and \(\gamma_p\) for the relevant processes and compares them to \(\tau_{\rm wall}^{-1}\).  
If both are large, impose chemical relations and assume a single \(T\).  
If either is marginal, evolve the species explicitly or use a more refined closure (e.g.\ two–temperature or Fokker–Planck treatment).  
This provides a consistent criterion for deciding which degrees of freedom belong in the fluid network and which can be thermally integrated out.

\bibliographystyle{JHEP}
\bibliography{biblio}

\end{document}